\pdfoutput=1

\documentclass{cup-elements_dat}

\usepackage{amsmath}
\usepackage{amssymb}
\usepackage{amsthm}
\usepackage{palatino,mathpazo} 
\usepackage[T1]{fontenc}
\usepackage[utf8]{inputenc}
\usepackage[mathscr]{eucal}
\usepackage{bm}
\usepackage[all,cmtip,2cell]{xy}
\usepackage{graphicx}
\usepackage{stmaryrd}
\usepackage{rotating}
\usepackage{enumitem}
\usepackage{algorithm}
\usepackage{algorithmic}
\usepackage{prooftree}
\usepackage{cmll}
\usepackage[braket]{qcircuit}
\usepackage{pdflscape}

\newtheorem{theorem}{Theorem}[section]
\newtheorem{corollary}[theorem]{Corollary}
\newtheorem{remark}[theorem]{Remark}
\newtheorem{definition}[theorem]{Definition}
\theoremstyle{definition}
\newtheorem{example}[theorem]{Example}

\allowdisplaybreaks

\usepackage{makecell}

\makeatletter
\renewcommand\subsubsection{\@startsection{subsubsection}{3}{\z@}%
  {3.25ex \@plus1ex \@minus.2ex}%
  {-1em}%
  {\normalfont\normalsize\bfseries}}

\newcounter{para}[subsubsection]

\usepackage{tikz}
\usetikzlibrary{arrows,snakes,automata,backgrounds,petri, positioning,calc,cd, automata}
\usetikzlibrary{shapes.geometric,shapes.misc,matrix,arrows.meta,decorations.markings}
\tikzset{x=0.9em, y=2ex, baseline=-0.5ex}
\tikzset{ihbase/.style={inner sep=0,circle,draw,fill=lightgray,minimum size=0.4em,node contents={}}}
\tikzset{ihblack/.style={ihbase,fill=black}}
\tikzset{ihwhite/.style={ihbase,fill=white}}
\tikzset{mat/.style={draw,fill=white,rectangle,node font=\scriptsize}}
\tikzset{ha/.style={mat,rounded rectangle,rounded rectangle left arc=none}}
\tikzset{haop/.style={mat,rounded rectangle,rounded rectangle right arc=none}}
\tikzset{blackha/.style={mat,rounded rectangle,rounded rectangle left arc=none,font=\color{white},fill=black}}
\tikzset{blackhaop/.style={mat,rounded rectangle,rounded rectangle right arc=none,font=\color{white},fill=black}}
\tikzset{anti/.style={inner sep=0,isosceles triangle,fill=black,draw=black, minimum width=0.75em, node contents={}}}
\tikzset{antiop/.style={anti,shape border rotate=180}}
\tikzset{antisq/.style={inner sep=0,rectangle,fill=black, minimum height=1em, minimum width=0.6em, node contents={}}}
\tikzset{count/.style={above,inner ysep=0.15em,font=\scriptsize}}
\tikzset{axiom/.style={above,font=\small}}
\tikzset{dir/.style={-Latex}}
\tikzset{st/.style={decoration={markings,
    mark={at position 0.5 with {\draw (0, 2pt) to (0, -2pt);}}},
    postaction=decorate}}

\newcommand{\sem}[1]{\left\llbracket{#1}\right\rrbracket}

}

\newcommand{\genericcomult}[2]{
  \begin{tikzpicture}
    \node at (1, 0) [ihbase,solid,name=copy,#1];
    \draw[#2] (copy) .. controls (1.25, 0.5) .. (2, 0.5);
    \draw[#2] (0, 0) -- (copy);
    \draw[#2] (copy) .. controls (1.25, -0.5) .. (2, -0.5);
  \end{tikzpicture}
}
\newcommand{\genericcomultn}[2]{
  \tikz {
    \draw (1, 0) node[ihbase,name=copy,#1] .. controls (1.25, 0.5) .. (1.5, 0.5)
    -- node[count] {$#2$} (2.25, 0.5);
    \draw (0, 0) -- node[count] {$#2$} (copy) .. controls (1.25, -0.5) .. (1.5, -0.5)
    -- node[count] {$#2$} (2.25, -0.5);
  }
}
\newcommand{\genericcounit}[2]{
  \tikz \draw[#2] (0, 0) -- (1, 0) node[ihbase,#1, solid];
}
\newcommand{\genericcounitn}[2]{
  \tikz \draw (0, 0) -- node[count] {$#2$} (1, 0) node[ihbase,#1];
}
\newcommand{\genericmult}[2]{
  \tikz {
    \node at (1,0) (copy) [ihbase,#1,solid];
    \draw[#2] (0,  0.5) .. controls (0.75,  0.5) .. (copy);
    \draw[#2] (0, -0.5) .. controls (0.75, -0.5) .. (copy);
    \draw[#2] (copy) -- (2, 0);
  }
}
\newcommand{\genericmultn}[2]{
  \tikz {
    \draw (0,  0.5) -- node[count] {$#2$} (0.75,  0.5)
    .. controls (1,  0.5) .. (1.25, 0) node[ihbase,name=copy,#1];
    \draw (0, -0.5) -- node[count] {$#2$} (0.75, -0.5)
    .. controls (1, -0.5) .. (copy) -- node[count] {$#2$} (2.25, 0);
  }
}
\newcommand{\genericunit}[2]{
  \tikz \draw[#2] (0, 0) node[ihbase,#1, solid] -- (1, 0);
}
\newcommand{\genericunitn}[2]{
  \tikz \draw (0, 0) node[ihbase,#1, solid] -- node[count] {$#2$} (1, 0);
}

\newcommand{\Bcomult}{\genericcomult{ihblack}{}}
\newcommand{\Bcomultn}[1]{\genericcomultn{ihblack}{#1}}
\newcommand{\Bcounit}{\genericcounit{ihblack}{}}
\newcommand{\Bcounitn}[1]{\genericcounitn{ihblack}{#1}}
\newcommand{\Bmult}{\genericmult{ihblack}{}}
\newcommand{\Bmultn}[1]{\genericmultn{ihblack}{#1}}
\newcommand{\Bunit}{\genericunit{ihblack}{}}
\newcommand{\Bunitn}[1]{\genericunitn{ihblack}{#1}}

\newcommand{\Wmult}{\genericmult{ihwhite}{}}
\newcommand{\Wmultn}[1]{\genericmultn{ihwhite}{#1}}

\newcommand{\Wunit}{\genericunit{ihwhite}{}}
\newcommand{\Wunitn}[1]{\genericunitn{ihwhite}{#1}}

\newcommand{\Wcomult}{\genericcomult{ihwhite}{}}
\newcommand{\Wcomultn}[1]{\genericcomultn{ihwhite}{#1}}

\newcommand{\Wcounit}{\genericcounit{ihwhite}{}}
\newcommand{\Wcounitn}[1]{\genericcounitn{ihwhite}{#1}}

\newcommand\idzero{
\InputIfFileExists{empty-diag.tikz}{}{\input{./tikz/empty-diag.tikz}}
} 

\newcommand{\idone}{
  \tikz \draw (0, 0) -- (1, 0);
}

\newcommand{\idright}{\begin{tikzpicture}
	\begin{pgfonlayer}{nodelayer}
		\node [style=none] (0) at (-0.5, 0) {};
		\node [style=none] (1) at (0.2, 0) {};
		\node [style=none] (2) at (0.5, 0) {};
	\end{pgfonlayer}
	\begin{pgfonlayer}{edgelayer}
		\draw [->] (0.center) to (1.center);
		\draw (1.center) to (2.center);
	\end{pgfonlayer}
\end{tikzpicture}}

\newcommand{\idxright}[1]{\begin{tikzpicture}
	\begin{pgfonlayer}{nodelayer}
		\node [style=none] (0) at (0.5, 0) {};
		\node [style=none] (1) at (-0.4, 0) {};
		\node [style=none] (2) at (-1.5, 0) {};
		\node [style=none] (3) at (-0.5, 0.75) {\scriptsize $#1$};
	\end{pgfonlayer}
	\begin{pgfonlayer}{edgelayer}
		\draw (1.center) to (0.center);
		\draw [->] (2.center) to (1.center);
	\end{pgfonlayer}
\end{tikzpicture}
}

\newcommand{\idxleft}[1]{\begin{tikzpicture}
	\begin{pgfonlayer}{nodelayer}
		\node [style=none] (0) at (-1.5, 0) {};
		\node [style=none] (1) at (-0.6, 0) {};
		\node [style=none] (2) at (0.5, 0) {};
		\node [style=none] (3) at (-0.5, 0.75) {\scriptsize $#1$};
	\end{pgfonlayer}
	\begin{pgfonlayer}{edgelayer}
		\draw (1.center) to (0.center);
		\draw [->] (2.center) to (1.center);
	\end{pgfonlayer}
\end{tikzpicture}
}

\newcommand{\idx}[1]{
  \tikz \draw (0, 0) -- (1.5, 0) node [midway, above] {\scriptsize $#1$};
}

\newcommand{\sym}{
  \tikz {
    \draw (0,  0.5) .. controls (0.5,  0.5) and (0.5, -0.5) .. (1, -0.5);
    \draw (0, -0.5) .. controls (0.5, -0.5) and (0.5,  0.5) .. (1,  0.5);
  }
}

\newcommand{\braidxundery}[2]{\begin{tikzpicture}
	\begin{pgfonlayer}{nodelayer}
		\node [style=none] (0) at (-0.5, -0.25) {};
		\node [style=none] (1) at (0.75, 0.5) {};
		\node [style=none] (2) at (-0.5, 0.5) {};
		\node [style=none] (3) at (0, 0.25) {};
		\node [style=none] (4) at (0.25, 0) {};
		\node [style=none] (5) at (0.75, -0.25) {};
		\node [style=none] (6) at (1, -0.25) {\scriptsize $#1$};
		\node [style=none] (6) at (1, 0.5) {\scriptsize $#2$};
	\end{pgfonlayer}
	\begin{pgfonlayer}{edgelayer}
		\draw [in=-180, out=0] (0.center) to (1.center);
		\draw [in=150, out=0] (2.center) to (3.center);
		\draw [in=-180, out=0] (0.center) to (1.center);
		\draw [in=180, out=-30, looseness=0.75] (4.center) to (5.center);
	\end{pgfonlayer}
\end{tikzpicture}}

\newcommand\scalar[1]{
  \tikz {
    \node[ha] (ha) {$#1$};
    \draw (ha.west) -- ++(-0.75, 0);
    \draw (ha.east) -- ++(0.75, 0);
  }
}

\pgfdeclarelayer{edgelayer}
\pgfdeclarelayer{nodelayer}
\pgfsetlayers{edgelayer,nodelayer,main}

\definecolor{light-gray}{gray}{.5}

\tikzset{
BWmatrix/.pic={
    \coordinate (center) at (0,0);
    \filldraw[fill=white, draw=black, line width=1pt] (.5,0) 
        [rounded corners=14pt] -- (1,0) 
        [rounded corners=14pt] -- (1,1)
        [rounded corners=0pt] -- (.5,1) 
        [rounded corners=0pt] -- cycle;
    \filldraw[fill=black, draw=black, line width=1pt] (0,0) 
        -- (.5,0) 
        -- (.5,1)
        -- (0,1) 
        -- cycle;
   },
pics/BWmatrix/.default=0.2
}

\tikzstyle{pl}=[circle,thick,draw=black!75,fill=white,minimum size=10pt]
\tikzstyle{port}=[circle, fill,inner sep=1.2pt]
\tikzstyle{transition}=[rectangle,thick,draw=black,
  			  fill=black,minimum size=6pt]

\tikzstyle{arrow}=[->]

\newcommand{\diagbox}[3]{
\begin{tikzpicture}
	\begin{pgfonlayer}{nodelayer}
		\node [style=basic box] (0) at (0, 0) {$#1$};
		\node [style=none] (1) at (1.5, 0) {};
		\node [style=none] (2) at (-1.5, 0) {};
		\node [style=none] (3) at (1.5, 0.5) {\scriptsize $#3$};
		\node [style=none] (4) at (-1.5, 0.5) {\scriptsize $#2$};
	\end{pgfonlayer}
	\begin{pgfonlayer}{edgelayer}
		\draw (2.center) to (0);
		\draw (0) to (1.center);
	\end{pgfonlayer}
\end{tikzpicture}
}

\newcommand{\diagstate}[4]{
\begin{tikzpicture}
	\begin{pgfonlayer}{nodelayer}
		\node [style=none] (6) at (1.75, -0.5) {};
		\node [style=none] (7) at (0.5, -0.5) {};
		\node [style=basic box] (10) at (0, 0) {$#1$};
		\node [style=none] (11) at (1.75, 0) {\scriptsize $#3$};
		\node [style=none] (12) at (-1.75, 0) {\scriptsize $#2$};
		\node [style=none] (13) at (-1.75, -0.5) {};
		\node [style=none] (14) at (-0.5, -0.5) {};
		\node [style=none] (15) at (0.5, 0.5) {};
		\node [style=none] (16) at (1.75, 0.5) {};
		\node [style=none] (17) at (-1.75, 0.5) {};
		\node [style=none] (18) at (-0.5, 0.5) {};
		\node [style=none] (19) at (-1.75, 1) {\scriptsize $#4$};
		\node [style=none] (20) at (1.75, 1) {\scriptsize $#4$};
	\end{pgfonlayer}
	\begin{pgfonlayer}{edgelayer}
		\draw [in=180, out=0, looseness=1.25] (7.center) to (6.center);
		\draw [in=180, out=0, looseness=1.25] (13.center) to (14.center);
		\draw [in=180, out=0, looseness=1.25] (15.center) to (16.center);
		\draw [in=180, out=0, looseness=1.25] (17.center) to (18.center);
	\end{pgfonlayer}
\end{tikzpicture}
}

\newcommand{\traceform}[4]{
\begin{tikzpicture}
	\begin{pgfonlayer}{nodelayer}
		\node [style=none] (1) at (0.5, 1) {};
		\node [style=none] (6) at (-0.5, 1) {};
		\node [style=none] (8) at (2.25, 0) {};
		\node [style=none] (9) at (0.5, 0) {};
		\node [style=basic box] (12) at (0, 0.5) {$#1$};
		\node [style=none] (13) at (2, 0.5) {\scriptsize $#3$};
		\node [style=none] (14) at (1.25, 2.25) {};
		\node [style=none] (15) at (-1.25, 2.25) {};
		\node [style=none] (16) at (-1.75, 0.5) {\scriptsize $#2$};
		\node [style=none] (17) at (-2, 0) {};
		\node [style=none] (18) at (-0.5, 0) {};
		\node [style=none] (20) at (1.25, 1) {};
		\node [style=none] (21) at (2.25, 1.85) {\scriptsize $#4$};
		\node [style=none] (22) at (-1.25, 1) {};
	\end{pgfonlayer}
	\begin{pgfonlayer}{edgelayer}
		\draw [in=180, out=0, looseness=1.25] (9.center) to (8.center);
		\draw (15.center) to (14.center);
		\draw [in=180, out=0, looseness=1.25] (17.center) to (18.center);
		\draw (6.center) to (22.center);
		\draw [bend right=90, looseness=1.75] (20.center) to (14.center);
		\draw [bend left=90, looseness=1.75] (22.center) to (15.center);
		\draw (1.center) to (20.center);
	\end{pgfonlayer}
\end{tikzpicture}
}

\newcommand{\capx}[1]{
\begin{tikzpicture}
	\begin{pgfonlayer}{nodelayer}
		\node [style=none] (43) at (0.5, -0.5) {};
		\node [style=none] (44) at (0.5, 0.5) {};
		\node [style=none] (45) at (0.25, 0.5) {};
		\node [style=none] (46) at (0.25, -0.5) {};
		\node [style=none] (47) at (-0.15, -0.5) {\scriptsize $#1$};
		\node [style=none] (47) at (-0.15, 0.5) {\scriptsize $#1$};
	\end{pgfonlayer}
	\begin{pgfonlayer}{edgelayer}
		\draw [->, bend right=90, looseness=2.00] (43.center) to (44.center);
		\draw (45.center) to (44.center);
		\draw (43.center) to (46.center);
	\end{pgfonlayer}
\end{tikzpicture}
}

\newcommand{\sdcapx}[1]{\begin{tikzpicture}
	\begin{pgfonlayer}{nodelayer}
		\node [style=none] (43) at (0.5, 0.5) {};
		\node [style=none] (44) at (0.5, -0.5) {};
		\node [style=none] (45) at (0.25, -0.5) {};
		\node [style=none] (46) at (0.25, 0.5) {};
		\node [style=none] (47) at (-0.15, 0.5) {\scriptsize $#1$};
		\node [style=none] (47) at (-0.15, -0.5) {\scriptsize $#1$};
	\end{pgfonlayer}
	\begin{pgfonlayer}{edgelayer}
		\draw [bend left=90, looseness=2.00] (43.center) to (44.center);
		\draw (45.center) to (44.center);
		\draw (43.center) to (46.center);
	\end{pgfonlayer}
\end{tikzpicture}}

\newcommand{\cupx}[1]{\begin{tikzpicture}
	\begin{pgfonlayer}{nodelayer}
		\node [style=none] (43) at (0.25, 0.5) {};
		\node [style=none] (44) at (0.25, -0.5) {};
		\node [style=none] (45) at (0.5, -0.5) {};
		\node [style=none] (46) at (0.5, 0.5) {};
		\node [style=none] (47) at (0.8, 0.5) {\scriptsize $#1$};
		\node [style=none] (47) at (0.8, -0.5) {\scriptsize $#1$};
	\end{pgfonlayer}
	\begin{pgfonlayer}{edgelayer}
		\draw [->, bend left=90, looseness=2.00] (44.center) to (43.center);
		\draw (45.center) to (44.center);
		\draw (43.center) to (46.center);
	\end{pgfonlayer}
\end{tikzpicture}}

\newcommand{\sdcupx}[1]{\begin{tikzpicture}
	\begin{pgfonlayer}{nodelayer}
		\node [style=none] (43) at (0.25, 0.5) {};
		\node [style=none] (44) at (0.25, -0.5) {};
		\node [style=none] (45) at (0.5, -0.5) {};
		\node [style=none] (46) at (0.5, 0.5) {};
		\node [style=none] (47) at (0.8, 0.5) {\scriptsize $#1$};
		\node [style=none] (47) at (0.8, -0.5) {\scriptsize $#1$};
	\end{pgfonlayer}
	\begin{pgfonlayer}{edgelayer}
		\draw [bend left=90, looseness=2.00] (44.center) to (43.center);
		\draw (45.center) to (44.center);
		\draw (43.center) to (46.center);
	\end{pgfonlayer}
\end{tikzpicture}}

\newcommand{\myeq}[1]{\mathrel{\overset{\makebox[0pt]{\mbox{\normalfont\tiny\sffamily #1}}}{=}}}

\newcommand{\R}{\mathbb{R}}
\newcommand{\N}{\mathbb{N}}

\newcommand{\Field}{\mathbb{K}}
\newcommand{\Rig}{\mathsf{R}}
\newcommand{\catC}{\mathsf{C}} 

\newcommand{\Free}[2]{\mathsf{Free}_{#1}(#2)}
\newcommand{\FreeX}[1]{\Free{X}{#1}}

\newcommand{\FreeSMC}[1]{\mathsf{Free}_{SMC}(#1)}

\newcommand{\scFrob}{\mathsf{scFrob}}
\newcommand{\Span}[1]{\mathbf{Span}(#1)}
\newcommand{\Cospan}[1]{\mathbf{Cospan}(#1)}
\newcommand{\Mat}[1]{\mathsf{Mat}_{#1}}

\newcommand{\id}{\mathrm{id}}

\newcommand{\from}{\mathrel{:}\,}

\newcommand{\poi}{\,;\,}

\newcommand{\adjto}{\,\lower1pt\hbox{$\dashv$}\,}

\newcommand{\Set}{\mathsf{Set}}
\newcommand{\Bij}{\mathsf{Bij}}

\newcommand{\LinRel}{\mathsf{LinRel}_\Field}

\newcommand{\Rel}{\mathsf{Rel}}

\newcommand{\fSet}{\mathsf{fSet}}

\newcommand{\fVect}{\mathsf{fVect}}

\newcommand{\Tr}{\mathsf{Tr}}

\newcommand{\Sem}{\mathsf{Sem}}
\newcommand{\eqE}[1]{\mathrel{\overset{\makebox[0pt]{\mbox{\normalfont\tiny\sffamily #1}}}{=}}}

\newcommand{\Iff}{\Leftrightarrow}

\newcommand{\Hyp}[1]{\mathsf{Hyp}_{\scriptscriptstyle{#1}}} 
\newcommand{\MonHyp}[1]{\mathsf{MHyp}_{\scriptscriptstyle{#1}}} 

\newcommand{\CspGraph}[1]{\sem{#1}}
\newcommand{\cospn}[5]{#1 \xrightarrow{#2} #3 \xleftarrow{#4} #5}
\newcommand{\spn}[5]{#1 \xleftarrow{#2} #3 \xrightarrow{#4} #5}

\newcommand{\ruleLabel}[1]{#1}

\makeatletter
\def\moverlay{\mathpalette\mov@rlay}
\def\mov@rlay#1#2{\leavevmode\vtop{%
\baselineskip\z@skip \lineskiplimit-\maxdimen
\ialign{\hfil$#1##$\hfil\cr#2\crcr}}}
\makeatother

 \newcommand{\derivationRule}[3]{{\prooftree{ #1}\justifies{ #2}\thickness=0.15em\using\ruleLabel{#3}\endprooftree}}

\newcommand\twarr[2]{%
\mathrel{\mathop{\moverlay{\scriptstyle\xrightarrow{\,#1\,}\cr{\lower.2em\hbox{$\scriptstyle{}_{#2}$}}}}}}
\newcommand\twarrw[2]{%
\mathrel{\mathop{\moverlay{\scriptstyle\Longrightarrow\cr{\lower-.6em\hbox{$\scriptstyle{}_{#1}$}}
\cr{\lower.3em\hbox{$\scriptstyle{}_{#2}$}}}}}}

\newcommand{\dtransw}[2]{\raise1pt\hbox{$\;\twarrw{#1}{#2}\;$}}

\newcommand{\diagregexp}[1]{
\begin{tikzpicture}
	\begin{pgfonlayer}{nodelayer}
		\node [style=none] (0) at (1.5, 0) {};
		\node [style=rcoreg] (1) at (0, 0) {{\color{red} $e$}};
	\end{pgfonlayer}
	\begin{pgfonlayer}{edgelayer}
		\draw [red] (1) to (0.center);
	\end{pgfonlayer}
\end{tikzpicture}}

\usepackage{siunitx}
\usepackage[american,siunitx]{circuitikz}
\ctikzset{
    csources/scale=0.5,
    diodes/scale=0.5,
    instruments/scale=0.5,
    resistors/scale=0.5,
    sources/scale=0.5,
    transistors/scale=0.5,
    bipoles/border margin=1,
    bipoles/thickness=1,
    tripoles/thickness=1,
    current arrow scale=24,
}
\tikzstyle{vsource}=[rmeter, t={\textsf{\tiny -- +}}] 
\tikzstyle{ammeter}=[rmeter, t={\textsf{A}}] 
\tikzstyle{vmeter}=[rmeter, t={\textsf{V}}] 
\tikzstyle{elecdot}=[blue,circle,fill,inner sep=0.85pt]

\tikzstyle{none}=[inner sep=0pt]
\tikzstyle{plain}=[inner sep=0pt]
\tikzstyle{black}=[circle, draw=black, fill=black, inner sep=0pt, minimum size=4pt]
\tikzstyle{black-faded}=[circle, draw=light-gray, fill=light-gray, inner sep=0pt, minimum size=4pt]
\tikzstyle{white}=[circle, draw=black, fill=white, inner sep=0pt, minimum size=4.5pt]
\tikzstyle{z spider}=[circle, draw=black, fill=white, inner sep=2pt, minimum size=4.5pt]
\tikzstyle{x spider}=[circle, draw=black, fill=light-gray, inner sep=2pt, minimum size=4pt]
\tikzstyle{white-faded}=[circle, draw=light-gray, fill=white, inner sep=0pt, minimum size=4.5pt]
\tikzstyle{delay}=[fill=black, regular polygon, regular polygon sides=3,rotate=-90, scale=.55]
\tikzstyle{delay-op}=[fill=black, regular polygon, regular polygon sides=3,rotate=90, scale=.55]
\tikzstyle{reg}=[draw, fill=white, rounded rectangle, rounded rectangle left arc=none, minimum height=1.2em, minimum width=1.4em, node font={\scriptsize}]
\tikzstyle{effect}=[draw, fill=white, rectangle, minimum height=1.2em, minimum width=1em]
\tikzstyle{coreg}=[draw, fill=white, rounded rectangle, rounded rectangle right arc=none, minimum height=1.2em, minimum width=1.4em, node font={\scriptsize}]
\tikzstyle{state}=[draw, fill=white, rectangle, minimum height=1.2em, minimum width=1em]
\tikzstyle{hyperedge}=[draw, fill=white, rectangle, rounded corners, minimum height=1.2em, minimum width=1em]
\tikzstyle{box}=[shape=rectangle, text height=1.5ex, text depth=0.25ex, yshift=0.2mm, fill=white, draw=black, minimum height=3mm, minimum width=5mm, font={\small}]
\tikzstyle{basic box}=[draw, fill=white, rectangle, minimum height=1.2em, minimum width=1em]
\tikzstyle{small box}=[draw, fill=white, rectangle, rounded corners, minimum height=1.2em, minimum width=1.4em, node font={\scriptsize}]
\tikzstyle{rcoreg}=[draw=red, fill=white, rounded rectangle, rounded rectangle right arc=none, minimum height=1.2em, minimum width=1.4em, node font={\scriptsize}]
\tikzstyle{regb}=[draw, fill=black, rounded rectangle, rounded rectangle left arc=none, minimum height=1.2em, minimum width=1.4em, node font={\scriptsize}]
\tikzstyle{regbw}=[draw, left color=black, right color=white, middle color=white, rounded rectangle, rounded rectangle left arc=none, minimum height=1.2em, minimum width=1.4em, node font={\scriptsize}]
\tikzstyle{regwb}=[draw, left color=white, right color=black, middle color=white, rounded rectangle, rounded rectangle left arc=none, minimum height=1.2em, minimum width=1.4em, node font={\scriptsize}]
\tikzstyle{coregb}=[draw, fill=black, rounded rectangle, rounded rectangle right arc=none, minimum height=1.2em, minimum width=1.4em, node font={\scriptsize}]
\tikzstyle{coregbw}=[draw, left color=black, right color=white, middle color=white, rounded rectangle, rounded rectangle right arc=none, minimum height=1.2em, minimum width=1.4em, node font={\scriptsize}]
\tikzstyle{coregwb}=[draw, left color=white, right color=black, middle color=white, rounded rectangle, rounded rectangle right arc=none, minimum height=1.2em, minimum width=1.4em, node font={\scriptsize}]
\tikzstyle{rn}=[circle, draw=red, fill=red, inner sep=0pt, minimum size=4pt]
\tikzstyle{wrn}=[circle, draw=red, fill=white, inner sep=0pt, minimum size=4pt]
\tikzstyle{place}=[circle, draw=black, fill=white, inner sep=0pt, minimum size=9pt]
\tikzstyle{act}=[circle, draw=black, fill=white, inner sep=0pt, minimum size=4.5pt]
\tikzstyle{coact}=[draw, fill=white, rounded rectangle, rounded rectangle right arc=none, minimum height=.7em, minimum width=.9em, node font={\scriptsize}]
\tikzstyle{basic rounded box}=[draw, fill=white, rectangle, rounded corners, minimum height=1.2em, minimum width=1.4em]
\tikzstyle{small rounded box}=[draw, fill=white, rectangle, rounded corners, minimum height=1.2em, minimum width=1.4em, node font={\scriptsize}]
\tikzstyle{red dot}=[fill=red, draw=black, shape=circle, scale=0.3]
\tikzstyle{blue dot}=[fill=blue, draw=black, shape=circle, scale=0.3]
\tikzstyle{green dot}=[fill=green, draw=black, shape=circle, scale=0.6]
\tikzstyle{medium box}=[fill=white, draw=black, shape=rectangle, minimum width=0.3cm, minimum height=0.5cm]
\tikzstyle{white square}=[fill=white, draw, inner sep=0.6mm, minimum height=1.5mm, minimum width=1.5mm, shape=rectangle, shape=rectangle]
\tikzstyle{vertex}=[inner sep=0mm, minimum size=1mm, shape=circle, draw=black, fill=black]
\tikzstyle{vertex set}=[inner sep=0mm, minimum size=1mm, shape=circle, draw=black, fill=white, font={\footnotesize\boldmath}]
\tikzstyle{s flat}=[fill=white, draw=black, shape=rectangle, minimum width=8mm, minimum height=5mm]
\tikzstyle{black dot}=[fill=black, draw=black, shape=circle, scale=0.3]
\tikzstyle{empty dot}=[fill=none, draw=black, shape=circle, scale=0.3]
\tikzstyle{l flat}=[fill=white, draw=black, shape=rectangle, minimum width=1.8cm, minimum height=0.3cm]
\tikzstyle{s rect}=[fill=white, draw=black, shape=rectangle, minimum width=0.1cm, minimum height=0.1cm]
\tikzstyle{s vert}=[fill=white, draw=black, shape=rectangle, minimum width=5mm, minimum height=8mm]
\tikzstyle{m vert}=[fill=white, draw=black, shape=rectangle, minimum width=5mm, minimum height=12mm]
\tikzstyle{m flat}=[fill=white, draw=black, shape=rectangle, minimum width=6mm, minimum height=5mm]
\tikzstyle{mm flat}=[fill=white, draw=black, shape=rectangle, minimum height=5mm, minimum width=10mm]
\tikzstyle{mmm flat}=[fill=white, draw=black, shape=rectangle, minimum height=5mm, minimum width=12mm]
\tikzstyle{grey dot}=[fill={rgb,255: red,191; green,191; blue,191}, draw={rgb,255: red,191; green,191; blue,191}, shape=circle, scale=0.3]
\tikzstyle{mm vert}=[fill=white, draw=black, shape=rectangle, minimum width=5mm, minimum height=14mm]
\tikzstyle{20mm vert}=[fill=white, draw=black, shape=rectangle, minimum width=5mm, minimum height=20mm]
\tikzstyle{16mm vert}=[fill=white, draw=black, shape=rectangle, minimum width=5mm, minimum height=16mm]
\tikzstyle{18mm vert}=[fill=white, draw=black, shape=rectangle, minimum width=5mm, minimum height=18mm]

\tikzstyle{dashes}=[-, dashed]
\tikzstyle{right arrow}=[->]
\tikzstyle{left arrow}=[<-]
\tikzstyle{grey fill}=[-, fill={rgb,255: red,191; green,191; blue,191}, draw={rgb,255: red,191; green,191; blue,191}]
\tikzstyle{blue fill}=[-, fill=cyan, draw=cyan]
\tikzstyle{yellow fill}=[-, fill=yellow, draw=yellow]
\tikzstyle{green fill}=[-, fill=green, draw=green]
\tikzstyle{red wire}=[-, draw=red]
\tikzstyle{blue wire}=[-, draw=blue]
\tikzstyle{grey wire}=[-, draw={rgb,255: red,128; green,128; blue,128}]
\tikzstyle{green wire}=[-, draw=green]
\tikzstyle{yellow wire}=[-, draw=yellow]
\tikzstyle{black fill}=[-, fill=black]
\tikzstyle{white thick}=[-, draw=white, line width=4.5pt]

\CUPseries{}
\CUPelements{}

\title{An Introduction to String Diagrams \\ for Computer Scientists}
\author{Robin Piedeleu}
\affil{University College London}
\author{Fabio Zanasi}
\affil{University College London and University of Bologna}

\begin{document}
\frontmatter

\maketitle


\copyrightauthor{Robin Piedeleu and Fabio Zanasi, 2023}


\newpage
\mbox{ }
\vspace{1ex}
\begin{center}
\includegraphics[scale=.15]{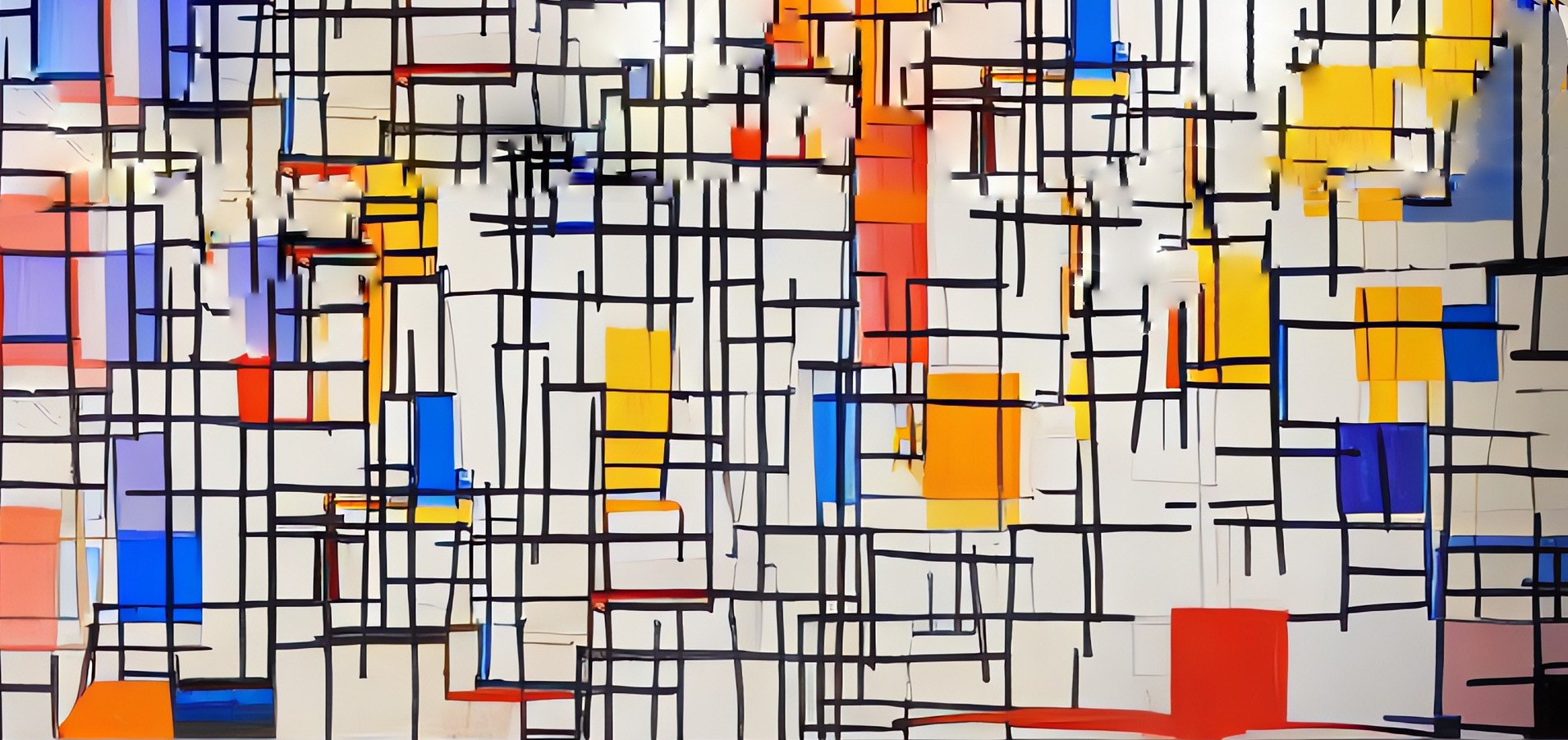}
\end{center}
\vspace{10ex}

\setcounter{tocdepth}{3}
\mainmatter

\section{The Case for String Diagrams}\label{sec:intro}

\paragraph{The algebraic structure of programs} When learning a programming language, one of the most basic tasks is understanding how to correctly write programs in the language \emph{syntax}. This syntax is often specified inductively, as a context-free grammar. For instance, the following grammar defines the syntax of a very elementary imperative programming language, where variables $x, y, \dots$ and natural numbers $n \in \mathbb{N}$ may occur:
\begin{equation}\label{eq:proggrammar}
\begin{gathered}
 \hspace{-.5cm} b \quad ::= \quad True \ \mid \ x = y \ \mid \ x = n \ \mid \ \neg b \ \mid \ b \wedge b  \ \mid \  b \vee b \\
  \hspace{-.3cm}  p \ ::= \ skip \ \mid \  x \! := \! n \ \mid \  x \! :=  \! y \ \mid \  x \! := \! y\!+\!1 \ \mid  \ \textit{while b do p} \ \mid \ p \, ; \, p
	\end{gathered}
\end{equation}
With the second row of the grammar, we can write arbitrary programs $p$ featuring assignment of value to a variable, while loops, and program concatenation. In particular, while loops will depend upon a boolean expression $b$, whose construction is dictated by the first row of the grammar. For practitioners, this information is essential to correctly write code in the given language: an interpreter will only execute programs that are written according to to the grammar. For computer scientists, interested in formal analysis of programs, this information has deeper consequences: it gives us a powerful tool to prove \emph{mathematical properties} of the language, by \emph{induction} over the syntax. This principle is a generalisation of how we are used to reason about the natural numbers. Indeed, the set $\mathbb{N}$ of natural numbers can also be specified via a grammar:
\begin{equation}\label{eq:natgrammar}
	n \quad ::= \quad 0 \ \mid \ n+1
\end{equation}
When proving properties of $\mathbb{N}$ by induction, what we are really doing is reasoning by case analysis on the clauses of grammar~\eqref{eq:natgrammar}. For instance, suppose to prove by induction that, for each $n \in \mathbb{N}$, $n +1 \leq 2^n$. In the base case, we assume that $n$ is $0$; we can verify that $0+1 \leq 2^0 = 1$. In the inductive step, we consider the case that $n$ is $n' +1$ for some $n'$. If we assume $n' +1 \leq 2^{n'}$, then we can show the statement for $n = n'+1$, as follows: $(n'+1) + 1 \leq 2^{n'} +1 \leq 2^{n'} + 2^{n'} = 2^{n'+1}$.

In the same way, we can reason by induction on programs, whenever their syntax is specified by a grammar such as~\eqref{eq:proggrammar}. For example, we can prove that a certain property $P$ holds for any program $p$ defined by~\eqref{eq:proggrammar}, as follows: first, we need to show that $P$ holds for $skip$, $ x:= n$, $x := y$, and $x := y\!+\!1$. Then, assuming $P$ holds for $p$, we show that it holds for $\textit{while b to p}$. Finally, assuming $P$ holds for $p$ and $p'$, we show that it holds for $p \, ; \, p'$.

This style of reasoning is extremely useful for a number of tasks. For instance, we may prove by induction important properties of our program, such as its \emph{correctness}, \emph{safety}, or \emph{liveness}, as studied in the research area of \emph{formal verification}. We may also define the \emph{semantics} by induction, \emph{i.e.}, assign programs their behaviour in a way that respects their structure. In programming language theory, there are usually two different ways of defining the semantics of a language: operational and denotational. The former specifies directly \emph{how} to execute every expression, while the latter specifies \emph{what} an expression means by assigning it a mathematical objects that abstracts its intended behaviour. An inductively defined semantics is particularly important because it enables \emph{compositional} (or \emph{modular}) reasoning: the meaning of a complex program may be entirely understood in terms of the semantics of its more elementary expressions. For instance, if our semantics associates a function $[p]$ to each program $p$, and associates to $p \ ; \ p'$ the composite function $[p'] \circ [p]$, that means that the semantics of the expression $p \ ; \ p'$ \emph{exclusively} depends on the semantics of simpler expressions $p$ and $p'$.

Moreover, the description of a language as a syntax equipped with a compositional semantics informs us about the \emph{algebraic} structures underpinning program behaviour. For instance, in any sensible semantics, the program constructs $;$ and $skip$ of the  grammar~\eqref{eq:proggrammar} acquire a \emph{monoid} structure, with the binary operation $;$ as its multiplication and the constant $skip$ as its identity element. Indeed, the laws of monoids, namely that $[(p;q);r] = [p;(q;r)]$ (associativity) and $[p ; skip] = [p] = [skip ; p]$ (unitality), will usually hold for the semantics of these  operations.

\paragraph{Graphical Models of Computation} As we have seen, defining a formal language via an inductively defined syntax brings clear benefits. However, not all computational phenomena may be adequately captured via this kind of formalism. Think for instance about data flowing through a digital controller. In this model, information propagates through components in complex ways, requiring constraints on how resources are processed. For example, a gate may only receive a certain quantity of data at a time, or a deadlock could occur. Sophisticated forms of interaction, such as entanglement in quantum processes, or conditional (in)dependence between random variables in a probabilistic systems, also require a language capable of capturing resource-exchange between components in a clear and expressive manner.

Historically, scientists have adopted \emph{graphical} formalisms to properly visualise and reason about these phenomena. Graphs provide a simple pictorial representation of how information flows through a component-based system, which would otherwise be difficult to encode into a conventional textual notation. Notable examples of these formalisms include electrical and digital circuits, quantum circuits, signal flow graphs (used in control theory), Petri nets (used in concurrency theory), probabilistic graphical models like Bayesian networks and factor graphs, and neural networks.
\begin{figure}[H]
\begin{align*}
 \raisebox{2em}{\Qcircuit @C=1em @R=.7em {
  \lstick{}   & \ctrl{1} & \gate{H} & \meter                  & \control \cw \\
  \lstick{} & \targ    & \meter   & \control \cw \cwx[1] \\
  \lstick{} & \qw	     & \qw      & \targ                   & \control \cwx[-2] \qw & \rstick{} \qw
 }} \qquad\qquad
 \raisebox{-2.5em}{\includegraphics[height=2cm]{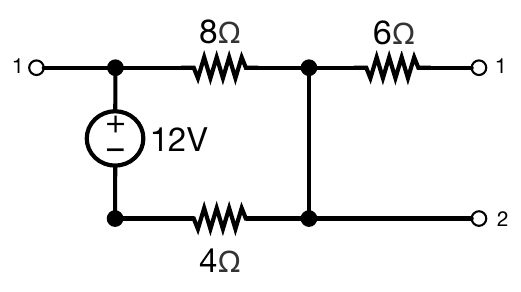}}\\
 
\InputIfFileExists{petri-ex.tikz}{}{\input{./tikz/petri-ex.tikz}}

 \qquad\qquad 
\InputIfFileExists{bayesian-net-ex.tikz}{}{\input{./tikz/bayesian-net-ex.tikz}}
 \qquad 
\InputIfFileExists{nn-ex.tikz}{}{\input{./tikz/nn-ex.tikz}}

\end{align*}
{\small 	\caption{\emph{Some examples of graphical formalisms: a quantum circuit, an electric circuit, a Petri net, a Bayesian network, and a neural network.}}
}
\end{figure}

On the other hand, graphical models have clear drawbacks compared to syntactically defined formal languages. Our ability to reason mathematically about combinatorial, graph-like structures is limited. We typically miss a formal theory of how to \emph{decompose} these models into simpler components, and also of how to \emph{compose} them together to create more complex models. In short, graphical models are often treated as monolithic rather than modular entities. In turn, this means that we cannot use induction on the model structure to prove their properties, as we would with a standard program syntax. Crucial features of program analysis, such as the definition of a compositional semantics, and the investigation of  algebraic structures underpinning model behaviour, face significant obstacles when adapted to graphical formalisms. 

\paragraph{String Diagrams: The Best of Both Worlds} String diagrams originate in the abstract mathematical framework of \emph{category theory}, as a pictorial notation to describe the morphisms in a monoidal category. However, over the past three decades their use has expanded significantly in computer science and related fields, extending way beyond their initial purpose.

What makes string diagrams so appealing is their dual nature. Just like graphical models, they are a \emph{pictorial} formalism: we can specify and reason about a string diagram as if it was a graph, with nodes and edges. However, just like programming languages, string diagrams may be also regarded as a formal \emph{syntax}; we can think of them as made of elementary components (akin to the gates of a circuit, but a lot more general than that), composed via syntactically defined operations.

Remarkably, understanding string diagrams as syntactically defined objects does not require switching to a different (textual) formalism---the graphical representation itself \emph{is} made of syntax. The theory of monoidal categories provides a rigorous formalisation of how to switch between the combinatorial and the syntactic perspective on string diagrams, as well as a rich framework to investigate their semantics and algebraic properties. Indeed, like programming languages, we can assign a semantics to string diagrams compositionally, as a \emph{functor} between categories. This gives a modular way to specify and reason about the behaviour of the models that they represent.

\begin{figure}[H]
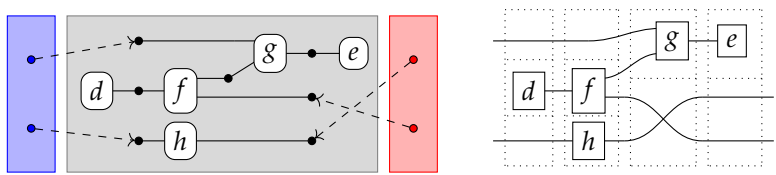

\[
\InputIfFileExists{ex-hypergraph-diag-2.tikz}{}{\input{./tikz/ex-hypergraph-diag-2.tikz}}
\qquad  
\InputIfFileExists{ex-diag-framed.tikz}{}{\input{./tikz/ex-diag-framed.tikz}}
\]
{\small 	\caption{\emph{An example of a string diagram regarded as a (hyper)graph (left), with blue and red boxes signalling the interfaces for composing with other string diagrams, and the same string diagram regarded as a piece of syntax (right), with dotted boxes placed to emphasise where elementary components compose, vertically and horizontally. 
}
}
}
\end{figure}

\paragraph{String Diagrams in Contemporary Research} Thanks to their versatility, string diagrams are increasingly adopted as a reasoning tool by scientists across various research fields. We may identify two major trends in the use of string diagrams: as a way to reason about graphical models syntactically, and as a way to reason about (textual) formal languages in a more visual, resource-sensitive manner.

Within the first trend, string diagrammatic approaches have enabled the adoption of compositional semantics and algebraic reasoning for graphical formalisms that previously lacked these features. Examples include Petri nets~\cite{BHPSZ-popl19}, linear dynamical systems~\cite{Bonchi2015,Fong2015,BaezErbele-CategoriesInControl}, quantum circuits~\cite{zxIntro}, electrical circuits~\cite{boisseau2021string}, Bayesian networks~\cite{fritz2020synthetic,Fong12,jacobs2021causal}, amongst others. Besides providing a unifying mathematical perspective on these models, string diagrams have also demonstrated the ability to produce tangible outcomes, employable at industrial scale. A convincing example is the ZX-calculus, a diagrammatic language that generalises quantum circuits. It now serves as the basis for the development of state-of-the-art quantum circuit optimisation algorithms~\cite{duncan2020graph}, and is seeing widespread adoption by companies dealing with quantum computing.

\begin{figure}[H]
\[\raisebox{-.45\height}{\includegraphics[scale=0.4]{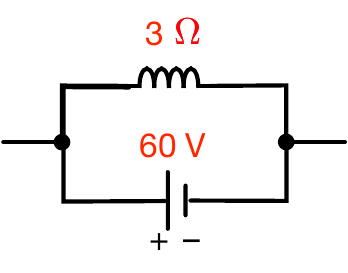}}\quad \mapsto\quad  
\InputIfFileExists{ex-circuit-diagram-translation.tikz}{}{\input{./tikz/ex-circuit-diagram-translation.tikz}}
\]
\[\raisebox{.2\height}{
\InputIfFileExists{ex-petri-net.tikz}{}{\input{./tikz/ex-petri-net.tikz}}
} \quad \mapsto \quad
\InputIfFileExists{ex-petri-net-diagram-translation.tikz}{}{\input{./tikz/ex-petri-net-diagram-translation.tikz}}
\]
{\small \caption{\emph{String diagrams representing  the behaviour of an electrical circuit (left) and a Petri net (right). The abstract perspective offered by the diagrammatic approach reveals that seemingly very different phenomena may be captured via the same set of elementary components.}}}
\end{figure}

As examples of the second trend, string diagrams have been instrumental in the development of compilers~\cite{muroya2017dynamic} for higher-order functional languages, and in a provably sound algorithm for reverse-mode automatic differentiation~\cite{Alvarez-Picallo23}. In both these examples, string diagrams serve as an intermediate formalism that sits between high-level programming languages and lower-level implementations. As the former, they can be manipulated syntactically. As the latter, they explicitly represent information propagation and other structural properties of systems.

\begin{figure}[H]
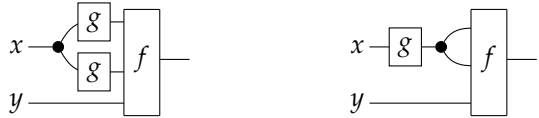

\[
\InputIfFileExists{ex-copy.tikz}{}{\input{./tikz/ex-copy.tikz}}
\qquad \qquad \qquad 
\InputIfFileExists{ex-copy-1.tikz}{}{\input{./tikz/ex-copy-1.tikz}}
\]
{\small 	\caption{\emph{A major appeal of string diagrams is \emph{resource-sensitivity}: they uncover any implicit assumption on how resources are handled during a computation. For instance, in the string diagrams above resources $x$ and $y$ are being fed to processes $f$ and $g$. Suppose applying $g$ to $x$ is an expensive computation. In the scenario where $f$ receives the value $g(x)$ twice, we are able to distinguish the case where we duplicate $x$ and then feed it to $g$ (left), and the more efficient way, where we duplicate $g(x)$ (right). Note traditional algebraic syntax would represent both cases as the same term, $f(g(x), g(x),y)$.}
}}
\end{figure}


\paragraph{Outline} This introduction provides a basic overview of string diagrams and their applications. Section~\ref{sec:string-diagram-syntax} introduces the formal syntax (for the most common variant) of string diagrams, the rules to manipulate them, and equational theories. Section~\ref{sec:graphs} shows how string diagrams may be also thought of as certain (hyper)graphs, thus providing an equivalent combinatorial perspective on these objects. In Section~\ref{sec:thstringdiag}, we consider other flavours of string diagrams, which correspond to a different syntax and can be manipulated in more permissive or restrictive ways. Section~\ref{sec:semantics} explains how to assign semantics to string diagrams; we cover common examples of semantics and their equational properties. Section~\ref{sec:other-trends} contains pointers to different trends that we do not cover in detail in this introduction. Finally, Section~\ref{sec:diagrams-science} is a non-exhaustive lists of applications of string diagrams, both in and outside of computer science.

The use of category theory is kept to a bare minimum, and we have prioritised intuition over technicalities whenever possible. For the reader's convenience, we have included an appendix containing the most rudimentary definitions of category theory. However, it should not be treated as an introduction to the topic, for which we recommend~\cite{leinster2014basic}.

\section{String Diagrams as Syntax}
\label{sec:string-diagram-syntax}

We have seen a couple of examples,~\eqref{eq:proggrammar} and \eqref{eq:natgrammar}, of how to specify expressions of a formal language via a grammar. In order to generalise this technique to diagrammatic expressions, it is best viewed through the lens of abstract algebra. From an algebraic viewpoint, a grammar is a means of presenting the \emph{signature} of our language: the list of \emph{operations} which we may use as the building blocks to construct more complex expressions. Each operation comes with its  \emph{type}, which remained implicit in~\eqref{eq:proggrammar} and \eqref{eq:natgrammar}. For instance, we may regard $;$ (program composition) in~\eqref{eq:proggrammar} as a \emph{binary} operation, which takes as inputs two programs $p$ and $p'$ as arguments and returns as output a program $p ; p'$ as value. The type of this operation is thus $program \times program \to program$. Analogously, $skip$, $x := y$, $x:= y+1$ and $x := n$ may be seen as constants (operations with no inputs) of type $program$, and the while loop yields an operation of type $boolean \times program \to program$: give a boolean expression $b$ and a program $p$, we obtain a program $\textit{while b do p}$.

This example suggests that, more generally, a signature $\Sigma$ should consist of two pieces of information: a set $\Sigma_1$ of generating operations, and a set $\Sigma_0$ of generating objects (e.g. $program$, $boolean$), which may be used to indicate the type of operations. Once we fix $\Sigma$, we may construct the expressions over $\Sigma$ the same way we would build the valid programs out of the grammar~\eqref{eq:proggrammar}. In algebra, such expressions are usually called \emph{$\Sigma$-terms}.

This process works in a fairly similar way for string diagrams, with some key differences. String diagrams will be built from signatures, except that now the type of operations may feature multiple outputs as well as multiple inputs, as displayed pictorially in~\eqref{eq:exoperation} below. We will also see that \emph{variables}, usually a building block of $\Sigma$-terms, are not a native concept, but rather something that may encoded in the diagrammatic representation.
More on this point in Example~\ref{ex:comonoids}, and Remarks~\ref{rmk:linear-properties} and~\ref{rmk:algebraic-Cartesian} below.

\paragraph{Signatures} A string diagrammatic syntax may be specified starting from a \emph{signature} $\Sigma = (\Sigma_0, \Sigma_1)$, with a set $\Sigma_0$ of \emph{generating objects} and a set $\Sigma_1$ of \emph{generating operations}. We will refer to either simply as generators when it is clear from the context whether we mean a generating object or operation. Each generating operation $d$ has a type $v \to w$, where $v \in \Sigma_0^{\star}$ (the set of words on alphabet $\Sigma_0$) is called the \emph{arity}, and $w \in \Sigma_0^{\star}$ the \emph{coarity} of $d$. Pictorially, an operation $d$ with arity $v = a_1 \dots a_m$ and coarity $w = b_1 \dots b_n$ will be represented as \begin{equation}\label{eq:exoperation}

\InputIfFileExists{smc/ex-operation.tikz}{}{\input{./tikz/smc/ex-operation.tikz}}

\end{equation}
or simply $\diagbox{d}{}{}$ when we do not need to name the list of generating objects in the arity and coarity.


\begin{example}~\label{ex:stack-signature}
The following forms a signature where we can think of the string diagrams as operations of a simple stack machine which can perform simple arithmetic on integers:
\[\Sigma_0 = \{stack,int\} \]
and
\[\Sigma_1 =\left\{
\InputIfFileExists{pop.tikz}{}{\input{./tikz/pop.tikz}}
, 
\InputIfFileExists{push.tikz}{}{\input{./tikz/push.tikz}}
, 
\InputIfFileExists{int-add.tikz}{}{\input{./tikz/int-add.tikz}}
, 
\begin{tikzpicture}
	\begin{pgfonlayer}{nodelayer}
		\node [style=none] (63) at (-1.75, 0) {};
		\node [style=none] (65) at (0.5, 0) {};
		\node [style=basic box] (70) at (1, 0) {\scriptsize $del$};
		\node [style=none] (74) at (-1.5, 0.5) {\scriptsize $int$};
	\end{pgfonlayer}
	\begin{pgfonlayer}{edgelayer}
		\draw [in=180, out=0] (63.center) to (65.center);
	\end{pgfonlayer}
\end{tikzpicture}
}
, 
\begin{tikzpicture}
	\begin{pgfonlayer}{nodelayer}
		\node [style=none] (87) at (-0.75, 0) {};
		\node [style=none] (88) at (1.25, 0) {};
		\node [style=basic box] (95) at (-1, 0) {\scriptsize $1$};
		\node [style=none] (139) at (1, 0.5) {\scriptsize $int$};
	\end{pgfonlayer}
	\begin{pgfonlayer}{edgelayer}
		\draw [in=180, out=0] (87.center) to (88.center);
	\end{pgfonlayer}
\end{tikzpicture}
}
 \right\}\]
\end{example}

\paragraph{Terms} Terms are generated by combining the generators of the signature in a certain way. Once again, let us look first at how terms would be specified in traditional algebra. One would start with a set $Var$ of variables and a signature $\Sigma$ of operations, and define terms inductively as:
\begin{itemize}
\item For each $x \in Var$, $x$ is a term.
\item For each $f \in \Sigma$, say of arity $n$, if $t_1, \dots, t_n$ are terms, then $f(t_1,\dots, t_n)$ is a term.
\end{itemize}
For string diagrammatic syntax, terms are generated in a similar fashion, with two important differences: (i) it is a \emph{variable-free} approach, and (ii) the way operations in $\Sigma$ are combined in the inductive step depends on the richness of the graphical structure we want to express.

A standard choice for string diagrams is to rely on \emph{symmetric monoidal} structure. This means that the generating operations in $\Sigma_1$ will be augmented with some `built-in' operations (`identity', `symmetry', and `null'), and combined via two forms of \emph{composition} (`sequential' and `parallel'). As a preliminary intuition, we may think of the built-in operations as the minimal structure needed to express graphical manipulations of terms, such as `stretching' a wire or crossing two wires. Fixing a signature $\Sigma = (\Sigma_0,\Sigma_1)$, the $\Sigma$-terms are generated by a few simple derivation rules or term formation rules: these are written in the form
\[
\derivationRule{\text{list of terms}}{\text{term}}{\scriptstyle\text{condition}}
\]
Here, we may regard the list of terms above the line as hypotheses needed to form the term in conclusion of the rule, below the line, provided that the side-condition is satisfied. For symmetric monoidal diagrams, the rules are as follows.
\begin{enumerate}
\item First, we have that every generating operation in $\Sigma_1$ yields a term:
\begin{gather*}
\derivationRule{}{\quad\diagbox{d}{v}{w}\quad}{\scriptstyle d\in\Sigma_1}
\end{gather*}
\item Next, each built-in operation (from left to right below: identity, symmetry, and null) also yields a term:
\[\derivationRule{}{\quad
\InputIfFileExists{id-x-with-frame.tikz}{}{\input{./tikz/id-x-with-frame.tikz}}
\quad}{\scriptstyle{x\in \Sigma_0}} \qquad \derivationRule{}{\quad ^{y}_x \ \sym \ ^x_{y}\quad}{\scriptstyle{x,y\in \Sigma_0}}\qquad \derivationRule{}{\quad 
\InputIfFileExists{empty-diag.tikz}{}{\input{./tikz/empty-diag.tikz}}
\quad}{}
 \]
\item For the inductive step, a new term may be built by combining two terms, either sequentially (left) or in parallel (right). Note that, for sequential composition, the output of the leftmost term needs to match the input of the rightmost term. For parallel composition, there is no such requirement, and the resulting term has input (output) the concatenation of the words forming the inputs (outputs) of the starting terms.
\[\derivationRule{\diagbox{c}{u}{v}\quad \diagbox{d}{v}{w}}{
\InputIfFileExists{horizontal-comp-framed.tikz}{}{\input{./tikz/horizontal-comp-framed.tikz}}
}{}\qquad\qquad  \derivationRule{\diagbox{\scriptstyle{d_1}}{v_1}{w_1}\quad \diagbox{\scriptstyle{d_2}}{v_2}{w_2}}{
\InputIfFileExists{vertical-comp-framed.tikz}{}{\input{./tikz/vertical-comp-framed.tikz}}
}{}\]
\end{enumerate}
Using the composition rules, we can define by induction, `identities'
\[
\InputIfFileExists{id-vx.tikz}{}{\input{./tikz/id-vx.tikz}}
:=
\InputIfFileExists{id-vx-def.tikz}{}{\input{./tikz/id-vx-def.tikz}}
\]
and `symmetries'
\[
\InputIfFileExists{smc/sym-vxw.tikz}{}{\input{./tikz/smc/sym-vxw.tikz}}
\::=\:
\InputIfFileExists{smc/sym-vxw-def.tikz}{}{\input{./tikz/smc/sym-vxw-def.tikz}}
 \qquad \quad 
\InputIfFileExists{smc/sym-vwx.tikz}{}{\input{./tikz/smc/sym-vwx.tikz}}
\::=\:
\InputIfFileExists{smc/sym-vwx-def.tikz}{}{\input{./tikz/smc/sym-vwx-def.tikz}}
\]
for arbitrary words $v,w$ in $\Sigma_0^*$.

Varying the set of built-in terms (second clause) and the ways of combining terms (last two clauses) will capture structures different from symmetric monoidal, as illustrated in Section~\ref{sec:thstringdiag} below.

Another important point: notice that \emph{null}, the identity over the empty word $\epsilon$ is not depicted, (or depicted as the \emph{empty diagram}), which is shown above as an empty dotted box. Furthermore, since the type of terms is a pair of words over some generating alphabet, they can have the empty word $\epsilon$ as arity or coarity. A term of type $d\from\epsilon \to w$, sometimes called a \emph{state}, has an empty left boundary
\[
\InputIfFileExists{state-d.tikz}{}{\input{./tikz/state-d.tikz}}
\]
while a term of type $d\from v\to \epsilon$, sometimes called an \emph{effect}\footnote{The names `state' and `effect' originated from the role played by string diagrams of this type in quantum theory~\cite{PQP}.}, has an empty right boundary
\[
\InputIfFileExists{effect-d.tikz}{}{\input{./tikz/effect-d.tikz}}
\]
Consequently, a term of type $d\from\epsilon\to \epsilon$, which is sometimes called a \emph{scalar}, or a \emph{closed} term, by analogy with the corresponding algebraic notion of terms containing no free variables, has no boundary at all; it is thus depicted as just a box, with no wires:
\[
\begin{tikzpicture}
	\begin{pgfonlayer}{nodelayer}
		\node [style=state] (36) at (0, 0) {$d$};
	\end{pgfonlayer}
\end{tikzpicture}
}
\]

\paragraph{From Terms to String Diagrams} Terms are not quite the same as string diagrams. Indeed, as soon as we consider more elaborate terms, we realise that the above definition require us to decorate pictures with extra notation, in order to keep track of the order in which we have applied the different forms of composition. For instance, we may construct the following term from the signature in Example~\ref{ex:stack-signature}:
\begin{equation}	\label{eq:exsigmaterm}

\InputIfFileExists{ex-stack-framed.tikz}{}{\input{./tikz/ex-stack-framed.tikz}}

\end{equation}
It denotes a very simple protocol, which pops two values of the stack, deletes the second one and increments the first by one, before pushing it back onto the stack.
We have only kept outer object labels for readability. Its full derivation tree is given in Figure~\ref{fig:stack-derivation-tree}. Notice that a bracketing by dotted frames fully specifies the corresponding derivation tree.
\begin{landscape}
\begin{figure}
 \begin{equation*}
 {\tiny
\derivationRule{\derivationRule{
\derivationRule{\derivationRule{}{
\InputIfFileExists{pop.tikz}{}{\input{./tikz/pop.tikz}}
}{}\derivationRule{\derivationRule{
\derivationRule{}{
\InputIfFileExists{pop.tikz}{}{\input{./tikz/pop.tikz}}
}{}\derivationRule{}{
\InputIfFileExists{int-id.tikz}{}{\input{./tikz/int-id.tikz}}
}{}}{
\InputIfFileExists{ex-stack-framed-llr0.tikz}{}{\input{./tikz/ex-stack-framed-llr0.tikz}}
}{}\derivationRule{}{
\InputIfFileExists{1-state.tikz}{}{\input{./tikz/1-state.tikz}}
}{}}{
\InputIfFileExists{ex-stack-framed-llr.tikz}{}{\input{./tikz/ex-stack-framed-llr.tikz}}
}{}}{
\InputIfFileExists{ex-stack-framed-ll.tikz}{}{\input{./tikz/ex-stack-framed-ll.tikz}}
}{} \derivationRule{\derivationRule{}{
\InputIfFileExists{int-add.tikz}{}{\input{./tikz/int-add.tikz}}
}{}\derivationRule{\derivationRule{}{
\InputIfFileExists{stack-id.tikz}{}{\input{./tikz/stack-id.tikz}}
}{}\derivationRule{}{
\InputIfFileExists{del.tikz}{}{\input{./tikz/del.tikz}}
}{}}{
\InputIfFileExists{ex-stack-framed-lr0.tikz}{}{\input{./tikz/ex-stack-framed-lr0.tikz}}
}{}}{
\InputIfFileExists{ex-stack-framed-lr.tikz}{}{\input{./tikz/ex-stack-framed-lr.tikz}}
}{}}{
\InputIfFileExists{ex-stack-framed-l.tikz}{}{\input{./tikz/ex-stack-framed-l.tikz}}
}{}\derivationRule{}{
\InputIfFileExists{push.tikz}{}{\input{./tikz/push.tikz}}
}{}}{
\InputIfFileExists{ex-stack-framed.tikz}{}{\input{./tikz/ex-stack-framed.tikz}}
}{}
}
\end{equation*}
\caption{An example derivation tree.}\label{fig:stack-derivation-tree}
\end{figure}
\end{landscape}
This example makes apparent that $\Sigma$-terms come with lots of extra information on how the graphical representation has been put together: the dotted boxes keep track of the order in which sequential and parallel composition have been applied. The move to string diagrams allows us to \emph{abstract away} this information, and focus solely on how the term components are wired together. More formally, a string diagram on $\Sigma$ is defined as an equivalence class of $\Sigma$-terms, where the quotient is taken with respect to the reflexive, symmetric and transitive closure of the following equations (where object labels are omitted for readability, and $c, c_i$, $d_i$ range over $\Sigma$-terms of the appropriate arity/coarity):
\begin{equation}\label{fig:smc-axioms}
\begin{gathered}
{
\InputIfFileExists{smc/sequential-associativity.tikz}{}{\input{./tikz/smc/sequential-associativity.tikz}}
 = 
\InputIfFileExists{smc/sequential-associativity-1.tikz}{}{\input{./tikz/smc/sequential-associativity-1.tikz}}
}
 \\
\scalebox{1}{
\InputIfFileExists{smc/unit-right.tikz}{}{\input{./tikz/smc/unit-right.tikz}}
 = \diagbox{c}{}{} = 
\InputIfFileExists{smc/unit-left.tikz}{}{\input{./tikz/smc/unit-left.tikz}}
}
 \\
\scalebox{1}{
\InputIfFileExists{smc/parallel-associativity.tikz}{}{\input{./tikz/smc/parallel-associativity.tikz}}
 = 
\InputIfFileExists{smc/parallel-associativity-1.tikz}{}{\input{./tikz/smc/parallel-associativity-1.tikz}}
}
\qquad 
  \scalebox{1}{ 
\InputIfFileExists{smc/parallel-unit-above.tikz}{}{\input{./tikz/smc/parallel-unit-above.tikz}}
 = \diagbox{c}{}{} =  
\InputIfFileExists{smc/parallel-unit-below.tikz}{}{\input{./tikz/smc/parallel-unit-below.tikz}}
}
\\
 \scalebox{1}{
\InputIfFileExists{smc/interchange-law.tikz}{}{\input{./tikz/smc/interchange-law.tikz}}
 = 
\InputIfFileExists{smc/interchange-law-1.tikz}{}{\input{./tikz/smc/interchange-law-1.tikz}}
 }
 \\
\scalebox{1}{
\InputIfFileExists{smc/sym-natural.tikz}{}{\input{./tikz/smc/sym-natural.tikz}}
= 
\InputIfFileExists{smc/sym-natural-1.tikz}{}{\input{./tikz/smc/sym-natural-1.tikz}}
}
\qquad\quad
\scalebox{1}{
\InputIfFileExists{smc/sym-iso.tikz}{}{\input{./tikz/smc/sym-iso.tikz}}
 = 
\begin{tikzpicture}
	\begin{pgfonlayer}{nodelayer}
		\node [style=none] (0) at (2, -0.75) {};
		\node [style=none] (1) at (-2, -0.75) {};
		\node [style=none] (2) at (-2, 0.5) {};
		\node [style=none] (3) at (2, 0.5) {};
	\end{pgfonlayer}
	\begin{pgfonlayer}{edgelayer}
		\draw (0.center) to (1.center);
		\draw (3.center) to (2.center);
	\end{pgfonlayer}
\end{tikzpicture}}
}
\end{gathered}
\end{equation}
If we think of the dotted frames as two-dimensional brackets, these laws tell us that the specific bracketing of a term does not matter. This is similar to how, in algebra, $(a\cdot b)\cdot c= a\cdot (b\cdot c)$ for an associative operation, justifying the use of the unbracketed expression $a\cdot b\cdot c$. In fact, there's an even better notation: when dealing with a single associative binary operation, we can simply forget it and write any product as a concatenation $abc$! This is a simple instance of the same key insight that allows us to draw string diagrams. It is helpful to think of these diagrammatic rules as a higher-dimensional version of associativity\footnote{In fact there is a sense in which this is precisely true, \emph{cf.} Section~\ref{sec:higher-dimensions} for a short discussion of this point.}. The first three lines above encode the associativity and unitality of the two forms of composition. On the fourth line, the \emph{interchange} law concerns the interplay between the two forms of composition: we can take the parallel composition of two sequentially composed terms, or vice-versa; the resulting string diagram is the same. The last line contains two axioms involving wire crossings $\sym$: the first, called the \emph{naturality} of $\sym$, tells us that boxes can be pulled across wires; the second, that the wire crossing is self-inverse. As a result, wires can be entangled or disentangled as long as we do not modify how the boxes are connected. We call these axioms the laws of symmetric monoidal categories (SMC).

Thanks to the laws of SMC, we can safely remove the brackets from the term in~\eqref{eq:exsigmaterm} to obtain the corresponding string diagram:
\begin{equation}\label{eq:exstringdiagram}
	
\InputIfFileExists{ex-stack.tikz}{}{\input{./tikz/ex-stack.tikz}}

\end{equation}
This representation is now unambiguous because the axioms in~\eqref{fig:smc-axioms} imply that any way of placing dotted frames around components of~\eqref{eq:exstringdiagram} lead to equivalent $\Sigma$-terms. For instance, the following two bracketed terms are equivalent as string diagrams:
\[
\begin{gathered}

\InputIfFileExists{ex-stack-framed.tikz}{}{\input{./tikz/ex-stack-framed.tikz}}

\\

\InputIfFileExists{ex-stack-framed-alt.tikz}{}{\input{./tikz/ex-stack-framed-alt.tikz}}

\end{gathered}
\]
There is an important subtlety: if, formally, a string diagram is an equivalence class of terms quotiented by the laws of~\eqref{fig:smc-axioms}, there is not a unique way to depict a string diagram. In other words, the graphical representation (even without dotted frames) sits in between terms and string diagrams, as it does not distinguish certain equivalent terms. In some cases the depiction absorbs the laws of SMC, \emph{e.g.}, for the two sides of the interchange law:
\begin{equation} \label{eq:interchange-law}

\InputIfFileExists{smc/interchange-law.tikz}{}{\input{./tikz/smc/interchange-law.tikz}}
 = 
\InputIfFileExists{smc/interchange-law-1.tikz}{}{\input{./tikz/smc/interchange-law-1.tikz}}
\end{equation}
In other cases, \emph{the way we draw} them distinguishes string diagrams that are equivalent under the laws of~\eqref{fig:smc-axioms}. Consider for example
\[ 
\InputIfFileExists{smc/unit-interchange-law.tikz}{}{\input{./tikz/smc/unit-interchange-law.tikz}}
 = 
\InputIfFileExists{smc/unit-interchange-law-1.tikz}{}{\input{./tikz/smc/unit-interchange-law-1.tikz}}
\]
This equality can be seen as an instance of the interchange law~\eqref{eq:interchange-law} with identity wires or as a consequence of the unitality of identity wires, which allows us to stretch wires as much as we like.

A related point is that string diagrams do not distinguish different ways of braiding wires, even if our drawings do:
\[
\InputIfFileExists{smc/sym-iso.tikz}{}{\input{./tikz/smc/sym-iso.tikz}}
 = 
\begin{tikzpicture}
	\begin{pgfonlayer}{nodelayer}
		\node [style=none] (0) at (2, -0.75) {};
		\node [style=none] (1) at (-2, -0.75) {};
		\node [style=none] (2) at (-2, 0.5) {};
		\node [style=none] (3) at (2, 0.5) {};
	\end{pgfonlayer}
	\begin{pgfonlayer}{edgelayer}
		\draw (0.center) to (1.center);
		\draw (3.center) to (2.center);
	\end{pgfonlayer}
\end{tikzpicture}}
\]
The laws of~\eqref{fig:smc-axioms} guarantee that any two string diagrams made entirely of wire crossings over the same number of wires are equal when they define the same \emph{permutation} of the wires. If the other rules are two-dimensional versions of associativity, the wire-crossing axioms are two-dimensional generalisation of \emph{commutativity}. In ordinary algebra, when we have a commutative and associative binary operation, we can write products using any ordering of its elements: $abc=bac=acb$. For string diagrams, the vertical juxtaposition of boxes is not strictly commutative; nevertheless, we are allowed to move boxes across wires, which is the next best thing:
\[
\InputIfFileExists{cxd-swap.tikz}{}{\input{./tikz/cxd-swap.tikz}}
 = 
\InputIfFileExists{swap-dxc.tikz}{}{\input{./tikz/swap-dxc.tikz}}
\]
(We invite the reader to show this, as their first exercise in diagrammatic reasoning).
Furthermore, we need to keep track of how boxes are wired, \emph{but} only the specific permutation of the wires matters, not how we have constructed it. Coming back to our stack-machine example, the following are equivalent string diagrams:
\[
\InputIfFileExists{ex-stack.tikz}{}{\input{./tikz/ex-stack.tikz}}
\;=\;
\InputIfFileExists{ex-stack-crossed.tikz}{}{\input{./tikz/ex-stack-crossed.tikz}}
\]

While this situation may appear slightly confusing at first, these example show that in practice the distinction between string diagrams (as equivalence classes) and how we depict them is harmless. The topological moves that are captured by the equations of~\eqref{fig:smc-axioms} are designed to be intuitive. They are often summarised by the following slogan: \emph{only the connectivity matters}. The rule of thumb is that any deformation that preserves the connectivity between the boxes and does not require us to bend the wires backwards will give two equivalent string diagrams.

Finally, keep in mind that the connection point from which we attach wires to boxes are ordered, so that the following two string diagrams are \emph{not} equivalent:
\[
\InputIfFileExists{pop.tikz}{}{\input{./tikz/pop.tikz}}
 \;\neq\; 
\InputIfFileExists{pop-sym.tikz}{}{\input{./tikz/pop-sym.tikz}}
\]

\begin{definition}[String diagrams over $\Sigma$] String diagrams over $\Sigma$ are $\Sigma$-terms quotiented by the equations in~\eqref{fig:smc-axioms}.
\end{definition}

\begin{example}\label{ex:diagram-equivalence}
Following the discussion above,
the reader should convince themselves that the two (unframed) terms below depict the same string diagram:
\[
\InputIfFileExists{ex-smc-diag.tikz}{}{\input{./tikz/ex-smc-diag.tikz}}
\;=\;
\InputIfFileExists{ex-smc-diag-1.tikz}{}{\input{./tikz/ex-smc-diag-1.tikz}}
\]
\end{example}


\paragraph{(Free) Symmetric Monoidal Categories}
In algebra, the collection of terms obtained from a signature, without any additional operations or equations, is often called the  \emph{free} structure over that signature. The diagrammatic language of string diagrams comes with an associated notion of free structure: the free \emph{symmetric monoidal category} (SMC) over a given signature\footnote{The notion of a free category generalises the same construction in algebra. It can be understood in terms of an adjunction, as explained for instance in~\cite[App. A.2]{BaezCoyaPropsNetworkTheory}. For the sake of our exposition, Definition~\ref{def:free-smc} below suffices.}.

At this point, the more mathematically-inclined reader might object that we still have not defined rigorously what a SMC is. Somewhat circularly, we could say that a SMC is a structure in which we can interpret string diagrams! Less tautologically, it is a category with an additional operation---the monoidal product---on objects and morphisms that satisfies the laws of Figure~\ref{fig:smc-axioms}. To state them without string diagrams, we need to introduce explicit notation for composition and the monoidal product. We do so in the following definition. Note that we assume basic knowledge of what a category is. The definition, along with related notions, can be found in Appendix.

\begin{definition}\label{def:smc}
A (strict) \emph{symmetric monoidal category} $(\catC, \otimes,I,\sigma)$ is a category $\catC$ equipped with a distinguished object $I$, a binary operation $\otimes$ on objects, an operation of type $\catC(X_1,Y_1)\times \catC(X_2,Y_2)\to \catC(X_1\otimes X_2,Y_1\otimes Y_2)$ on morphisms which we also write as $\otimes$,  such that $id_{X\otimes Y} = id_X\otimes id_Y$ and
\[c_1\otimes (c_2\otimes c_3) = (c_1\otimes c_2)\otimes c_3\qquad id_I\otimes c =c= c\otimes id_I\]
\[(c_1\otimes c_2)\poi (d_1\otimes d_2)=(c_1\poi d_1)\otimes (c_2\poi d_2)\]
and a family of morphisms $\sigma_X^Y$ for any two objects $X,Y$, such that
\[(id_X\otimes c)\poi\sigma_X^Z = \sigma_X^Y\poi (c\otimes id_X) \quad \text{for any } c\from Y\to Z\]
\[\sigma_X^Y\poi \sigma_Y^X = id_X\otimes id_Y\]
\end{definition}
\noindent Observe that these are exactly the laws of Figure~\ref{fig:smc-axioms} in symbolic form: we can simply replace `$\otimes$' by vertical composition and `$\poi$' by horizontal composition.

It is possible to translate any string diagrams into symbolic notation. For instance, the diagram of Example~\ref{ex:diagram-equivalence} can be written as $(d\otimes id\otimes f\otimes id);(id\otimes g\otimes \sigma);(e\otimes h\otimes id)$. This expression can be obtained by successively decomposing the diagrams into horizontal and vertical layers as follows:
\begin{align*}

\InputIfFileExists{ex-smc-diag-1-framed.tikz}{}{\input{./tikz/ex-smc-diag-1-framed.tikz}}
 
= 
\InputIfFileExists{ex-smc-diag-1-framed-horizontal.tikz}{}{\input{./tikz/ex-smc-diag-1-framed-horizontal.tikz}}
= 
\InputIfFileExists{ex-smc-diag-1-framed-decomposed.tikz}{}{\input{./tikz/ex-smc-diag-1-framed-decomposed.tikz}}

\end{align*}
As for string diagrams, there are multiple ways to write a given morphism in symbolic notation. In fact, because string diagrams absorb some of the laws of SMCs into the notation, there are usually more ways of writing a given morphism in symbolic notation than there are diagrammatic representations for it.

\begin{remark}[On strictness]\label{rmk:strictness}The last definition is \emph{not} the one that the reader is likely to encounter in the literature when looking up the terms ``symmetric monoidal category''. It defines what is called a \emph{strict} monoidal category; the usual notion is more general and allows for the equalities to be replaced by isomorphisms. We will not give a rigorous definition of this more general notion and refer the reader to any standard textbook on category theory for a general introduction to SMCs~\cite[Chapter XI]{maclane:71}. Our approach is nevertheless theoretically motivated by the following fundamental result: every SMC is equivalent (in a sense that we will not cover here in detail, but do recall in Appendix, at Definition~\ref{def:equivalence}) to a strict SMC. This fact is what allows us to draw string diagrams. It is known as the \emph{coherence theorem} for SMCs. Put differently, the coherence theorem allows us to forget explicit symbols for `$\otimes$' and `$\poi$', replacing them by vertical juxtaposition and horizontal composition without any brackets to denote the order of application\footnote{The coherence theorem is due to Mac Lane~\cite{maclane:71}. A recent exposition based on string diagrams can be found in~\cite{WilsonGZ22}.}. Once again, the reader is invited to think about this as a two-dimensional generalisation of well-known facts about monoids: just like we can we can simply concatenate elements of a monoid and omit the symbol for the multiplication and the parentheses to bracket its application, we can use string diagrams to represent morphisms of a SMC. It is then natural that more composition operations require more dimensions to represent. In fact, some of the earliest appearances of string diagrams\footnote{Though the difficulty of typesetting them at the time often meant that they did not appear as string diagrams in print!} occurred to construct free SMCs with additional structure and prove a coherence theorem for them~\cite{kelly1972many,kelly1980compactclosed}.
\end{remark}

\begin{definition}[Free SMC on a signature $\Sigma$]\label{def:free-smc} The symmetric monoidal category $\FreeSMC{\Sigma}$ is formed by letting objects be elements of $\Sigma_0^{*}$ and morphisms be string diagrams over $\Sigma$, \emph{i.e.}, $\Sigma$-terms quotiented by~\eqref{fig:smc-axioms}. The monoidal product is defined as word concatenation on objects. Composition and product of string diagrams are defined respectively by sequential and parallel composition of (some arbitrary representative of each equivalence class of) $\Sigma$-terms. 
\end{definition}



\begin{example}[Free SMC over a single object]\label{ex:free-smc-single-object}
The free SMC over the signature $\Sigma = (\{\bullet\},\varnothing)$ is easy to describe explicitly. Its string diagrams are generated by horizontal and vertical compositions of $\sym$ (where we omit labels for the single generating object $\bullet$), modulo the laws of SMCs. Here are a few examples:
\[
\InputIfFileExists{ex-permutations.tikz}{}{\input{./tikz/ex-permutations.tikz}}
\]
 In other words, they are permutations of the wires! If we write $\bullet^n$ for the concatenation of $n$ bullets, a string diagram $\bullet^n\to \bullet^n$ is a permutation of $n$ elements, and there are no diagrams $\bullet^n\to \bullet^m$ for $n\neq m$.

 Free SMCs on a single generating objects (and arbitrary generating operations) are usually called PROPs (\textbf{Pro}duct and \textbf{P}ermutation categories)~\cite{MacLane1965}. The PROP of permutations, which we just described, is the `simplest' possible PROP. More formally, it is the initial object in the category of PROPs. Sometimes, the notion of a `coloured' PROP is encountered: this is nothing but a (strict) SMC whose set of objects is freely generated from any set of generating objects, instead of just a single generating object as in the case of plain PROPs.

When we encounter PROPs, we will use natural numbers as objects, since all objects are of the form $\bullet^n$, and write the type of a string diagram $\bullet^n\to\bullet^m$ simply as $n\to m$.
\end{example}

\paragraph{Symmetric Monoidal Functors} Whenever we define a new mathematical structure, it is a good practice to introduce a corresponding notion of mapping between them. For SMCs, this is the notion of a \emph{symmetric monoidal functor}. We will need it when giving string diagrams a semantic interpretation, in Section~\ref{sec:semantics}.
\begin{definition}\label{def:symon-functor}
Let $(\mathsf{C},\otimes,I,\sigma)$ and  $(\mathsf{D},\boxtimes, J,\theta)$ be two SMCs.
A (strict) \emph{symmetric monoidal functor} $F: \mathsf{C}\to \mathsf{D}$
is a mapping from objects of $\mathsf{C}$ to those of $\mathsf{D}$ that satisfies
\[F(X_1\otimes X_2)=F(X_1)\boxtimes F(X_2) \qquad \text{ and }\qquad F(I) = J\]
and a mapping from morphisms of $\mathsf{C}$ to those of $\mathsf{D}$ that satisfies
\[F(c\poi d)=F(c)\poi F(d)\qquad F(id_X)=id_{F(Y)}\]
\[F(c_1\otimes c_2) = F(c_1)\boxtimes F(c_2) \qquad F(\sigma^Y_X) = \theta^{F(Y)}_{F(X)}\]
\end{definition}
In this introduction, for pedagogical reasons, we will mostly use \emph{strict} monoidal functors, that is, functors that preserve the monoidal structure on the nose. The reader should know that it is possible, and sometimes necessary, to relax this requirement, replacing the equalities $F(X_1\otimes X_2)=F(X_1)\boxtimes F(X_2)$ and $F(I) = J$ by \emph{isomorphisms} (which then have to satisfy certain compatibility conditions). See~\cite{maclane:71} for a standard treatment and~\cite{Selinger2009} for the connections with string diagrams.

If we have two such functors $F: \mathsf{C}\to \mathsf{D}$ and $G: \mathsf{D}\to \mathsf{C}$ that are inverses to each other---$FG$ and $GF$ are identity functors---we say that the two SMCs are \emph{isomorphic}. We will also use the notion \emph{equivalence} of SMCs. This is a more relaxed notion than that of isomorphism, where $FG$ and $GF$ are merely isomorphic to identity functors. It is more appropriate in some cases, in particular when the categories involves are not strict monoidal (see Remark~\ref{rmk:strictness}, for example). We will not dwell on equivalences of categories much in this text, but refer the reader to Definition~\ref{def:equivalence} and Remark~\ref{rmk:monoidal-equivalence} in Appendix.

\begin{remark}[On functors from free SMCs]\label{rmk:freeness}

When defining a functor $F$ out of a free SMC $\FreeSMC{\Sigma}$, there is a clear recipe to follow: we only need to specify to which object we want to map elements of $\Sigma_0$, and to which morphism $Fd\from Fu \to Fv$ we want to map each element of $d\from u\to v$ of $\Sigma_1$. This is because of the universal property of free constructions: if we have a mapping from the set of generating operations of some signature $\Sigma$ to morphisms of some SMC $\mathsf{C}$, there is a unique way of extending this mapping to a symmetric monoidal functor $\FreeSMC{\Sigma}\to \mathsf{C}$. This observation will come handy when defining the semantics of string diagrammatic theories, in Section~\ref{sec:semantics} below.
\end{remark}

\begin{example}\label{ex:permutations}
In Example~\ref{ex:free-smc-single-object}, we saw that the morphisms/string diagrams of the free SMC over a single object looked a lot like permutations. There is a way of making this precise, by establishing an isomorphism between this SMC and another whose morphisms are permutations of finite sets. Let $\Bij$ be the category whose objects are natural numbers, and morphisms $n\to n$ are permutations of $n= \{0,\dots, n-1\}$. We can equip it with a monoidal product, given by addition on objects, and on morphisms $\theta_1\from n_1\to n_1$ and $\theta_2\from n_2\to n_2$ by $\theta_1\otimes \theta_2(i) = \theta_1(i)$ if $i\leq n_1$ and $\theta_1\otimes \theta_2(i) = \theta_2(i)$ otherwise. The unit of the monoidal structure is the number $0$ and the symmetry is the permutation over two elements, which we write as $\sigma\from 2\to 2$, given by $\sigma(0) =1$ and $\sigma(1) =0$. The isomorphism is straightforward. In one direction, let $F\from \FreeSMC{\{\bullet\},\varnothing}\to \Bij$ be given by $F(\bullet^n)=n$ on objects and $F(\sym)=\sigma$. This is enough to describe $F$ fully because all string diagrams of $\FreeSMC{\{\bullet\},\varnothing}$ are vertical or horizontal composites of $\sym$ and $F$ has to preserve these two forms of composition, by Definition~\ref{def:symon-functor}. Furthermore, the required properties are immediately satisfied. To build its inverse, we need to know that we can factor any permutation into a composition of \emph{adjacent transpositions} (this fact is fairly intuitive and usually covered in introductory algebra courses, so we will not prove it here). Then, notice that the transposition $(i\: i+1)$ over $n$ elements should clearly be mapped to the string diagram that is the identity everywhere and $\sym$ at the $i$-th and $i+1$-th wires. Call this diagram $at_i$. Then, let $G\from \Bij\to \FreeSMC{\{\bullet\},\varnothing}$ be given by $G(n)=\bullet^n$ on objects and on morphisms by $G(\theta)=at_{i_1}at_{i_2}\dots at_{i_k}$ where $(i_1\: i_1+1)(i_2\: i_2+1)\dots (i_k\: i_k+1)=\theta$ is a decomposition of $\theta$ into adjacent transpositions. One can check that this is well-defined, and satisfies all the equations of Definition~\ref{def:symon-functor}. Moreover the two are inverses of each other. For example, to see that $GF(c)=c$ it is sufficient to check that equality for $\sym$. It holds clearly as $GF(\sym) = G(1 \: 2) = \sym$. The other direction is a bit more lengthy, but without any major difficulties.

Thus $(\Bij,+)$ gives a semantic account of the free SMC over a single object. Conversely, the latter can be seen as as diagrammatic syntax for the former.

In fact, this SMC is also equivalent to the non-strict SMC of finite sets and bijections between them, with the disjoint sum as monoidal product. The equivalence is also straightforward to establish, but requires us to fix a total ordering on every finite set. 

This example is just a taster of an idea that we will develop further in Section~\ref{sec:semantics}, dedicated to the \emph{semantics}---that is, the interpretation---of string diagrams.
\end{example}

\subsection{Adding Equations}\label{sec:eq-theories}

The equations of Figure~\ref{fig:smc-axioms} only capture a very basic notion of equivalence between string diagrams. When describing computational processes for example, it is useful to include more equations, specific to the domain of interest. In string diagram theory, these additional equations are encapsulated in the notion of \emph{symmetric monoidal theory}. More formally, a symmetric monoidal theory---or simply \emph{theory} when no ambiguity can arise---is a pair $(\Sigma, E)$, where $\Sigma$ is a signature and $E$ is a set of equalities $l = r$ between string diagrams of the same type over $\Sigma$. We write $\myeq{E}$ for the smallest congruence relation (w.r.t. sequential and parallel composition) containing $E$. We will see many examples of symmetric monoidal theories in Section~\ref{sec:common-theories}.

\begin{remark}[Equations and  diagrammatic rewriting.]\label{rmk:diagram-equations} It might be helpful to see equations as two-ways rewriting rules that can be applied in an arbitrary context. More precisely, assume that we have some equation of the form $l=r$, where $l,r$ have the same type; to apply it in context, we need to identify $l$ in a larger string diagram $c$, \emph{i.e.}, find $c_1$ and $c_2$ such that
\[\diagbox{c}{}{} = 
\InputIfFileExists{rewrite-l.tikz}{}{\input{./tikz/rewrite-l.tikz}}
\]
and simply replace $l$ by $r$, forming the new diagram $
\InputIfFileExists{rewrite-r.tikz}{}{\input{./tikz/rewrite-r.tikz}}
$. This is just the diagrammatic version of standard algebraic reasoning. We can summarise this process as follows:
\[\forall c_1,c_2\quad \diagbox{l}{}{} = \diagbox{r}{}{} \quad \Rightarrow \quad 
\InputIfFileExists{rewrite-l.tikz}{}{\input{./tikz/rewrite-l.tikz}}
 = 
\InputIfFileExists{rewrite-r.tikz}{}{\input{./tikz/rewrite-r.tikz}}
\]
For example,
\[
\InputIfFileExists{monoid-unitality-left.tikz}{}{\input{./tikz/monoid-unitality-left.tikz}}
 \quad \Rightarrow \quad 
\InputIfFileExists{monoid-unitality-ex-boxes.tikz}{}{\input{./tikz/monoid-unitality-ex-boxes.tikz}}
 \]
where the context is
\[
\InputIfFileExists{context-ex.tikz}{}{\input{./tikz/context-ex.tikz}}
\]
We will come back to this point, in the context of graph-rewriting, in Section~\ref{sec:rewriting}.
\end{remark}

In the same way that string diagrams corresponded to a free structure, the free symmetric monoidal category (SMC) over $\Sigma$, quotienting by further equations also determines a free structure: given a signature $\Sigma$ and a theory $E$ we can form the free SMC $\FreeSMC{\Sigma,E}$ obtained by quotienting the free SMC $\FreeSMC{\Sigma}$ by the equivalence relation over string diagrams given by $\eqE{E}$.

\begin{definition}[Free SMC over a theory $(\Sigma,E)$]\label{def:free-smc-equations} The symmetric monoidal category $\FreeSMC{\Sigma, E}$ is formed by letting objects be elements of $\Sigma_0^{\star}$ and morphisms be equivalence classes of string diagrams over $\Sigma$ quotiented by $\eqE{E}$. The monoidal product is defined as word concatenation on objects; composition and product of morphisms are defined respectively by sequential and parallel composition of arbitrary representatives of each equivalence class. 
\end{definition}

\subsection{Common Equational Theories}\label{sec:common-theories}

Some theories occur frequently in the literature. Many authors assume  familiarity with the axioms hiding behind the words ``monoids'', ``comonoids'', ``bimonoids/bialgebras'' or ``Frobenius monoid/algebra'', and how all of these theories relate to one another. For this reason it is valuable to know them well, especially when trying to distinguish routine moves from key steps in diagrammatic proofs. This section describes a few of the most commonly found examples.

\begin{example}[Monoids]\label{ex:monoids}
Let us begin with the deceptively easy example of monoids.
Many readers will undoubtedly be familiar with the algebraic theory of monoids, which can be presented by two generating operations, say $m(-,-)$ of arity $2$ and $u$ of arity $0$ (in other words, a constant) satisfying the following three axioms:
\[m(m(x,y),z) =m(x,m(y,z))\quad \text{ and }\quad m(u,x) = x = m(x,u)\]
Analogously, the symmetric monoidal theory of monoids can be presented by a signature $\Sigma = \left(\{\bullet\},\left\{\Wmultn{},\Wunitn{}\right\}\right)$, based on a single object-type $\bullet$, a \emph{multiplication} $\Wmult{} \colon 2 \to 1$ and a \emph{unit} $\Wunit{} \colon 0 \to 1$, 
and three axioms, for associativity and (two-sided) unitality:
\[
\InputIfFileExists{wmult-associative.tikz}{}{\input{./tikz/wmult-associative.tikz}}
\;\;\myeq{as}\;\; 
\InputIfFileExists{wmult-associative-1.tikz}{}{\input{./tikz/wmult-associative-1.tikz}}
\qquad  
\InputIfFileExists{wmult-unital-left.tikz}{}{\input{./tikz/wmult-unital-left.tikz}}
\;\;\myeq{unl}\;\;\idx{}
 \;\;\myeq{unr}\;\;
\InputIfFileExists{wmult-unital-right.tikz}{}{\input{./tikz/wmult-unital-right.tikz}}
\]
Observe that we have just replaced variables with wires and algebraic operations with diagrammatic generators. As in ordinary algebra,
two terms/diagrams are equal if they one can be obtained from the other by applying some sequence of these three equations (recall Remark~\ref{rmk:diagram-equations}).

We can also present commutative monoids in the same way. Recall that commutative monoids are those that satisfy $m(x,y)=m(y,x)$; diagrammatically, we can present the corresponding symmetric monoidal theory with the same signature and a single additional equality (note the use of the symmetry $\sym$):
\[
\InputIfFileExists{wmult-commutative.tikz}{}{\input{./tikz/wmult-commutative.tikz}}
\;\;\myeq{com}\;\;\,\Wmult\]
Of course, in the presence of commutativity, each of the unitality laws are derivable from the other. In the usual algebraic theory of monoids we would show this as follows: if $m(x,y)=m(y,x)$ then $m(u,x) = m(x,u) = x$ where the last step is right-unitality. The corresponding diagrammatic proof is very similar, with one additional step:
\[
\InputIfFileExists{wmult-unital-left.tikz}{}{\input{./tikz/wmult-unital-left.tikz}}
\;\;\myeq{com}\;\;
\InputIfFileExists{derived-left-unitality-1.tikz}{}{\input{./tikz/derived-left-unitality-1.tikz}}
\;\;\myeq{SMC}\;\;
\InputIfFileExists{derived-left-unitality-2.tikz}{}{\input{./tikz/derived-left-unitality-2.tikz}}
\;\;\myeq{unr}\;\;\idx{}\]
The second equality is a simple instance of the bottom left axiom of~\eqref{fig:smc-axioms}, for a string diagram with no wires on the left (that is, of type $\epsilon \to w$ for some $w$).

This makes an important point: two theories $(\Sigma,E)$ and $(\Sigma',E')$ might present the same structure, in the sense that the corresponding free SMCs $\FreeSMC{\Sigma,E}$ and $\FreeSMC{\Sigma',E'}$ might be isomorphic.
It is also a good place to mention that theories do not have to be minimal in any way; they can contain axioms that are derivable from the others.
There are various reasons one might prefer a theory that contains redundant axioms: to highlight some of the symmetries, to avoid having to re-derive some equalities as a lemma later on... 

When dealing with monoids, there are several straightforward syntactic simplifications that the reader is likely to encounter in the literature. First, a simple observation: in standard algebraic syntax, the associativity axiom $m(m(a, b),c) = m(a, m(b, c))$ implies that any two ways of applying monoid multiplication to the same list of elements are all equal. Therefore it is unambiguous to introduce a generalised monoid operation for any finite arity, \emph{e.g.} $m(a,b,c)$, to denote all possible ways of applying $m$ to these three elements, and avoid a flurry of parentheses. (Note that, with this syntactic sugar, the unit $e$ denotes the application of $m$ to zero elements.) The same trick works for an associative $\Wmult: 2\to 1$: we can define as a generalised $n$-ary operation as a dot with $n$-many wires
\[
\InputIfFileExists{n-ary-commutative-monoid.tikz}{}{\input{./tikz/n-ary-commutative-monoid.tikz}}
\]
as syntactic sugar to denote multiple applications of $\Wmult$. For this reason, the reader might also encounter diagrammatic proofs that identify different ways of applying a monoid operation to the same list of elements, much like a practitioner well-versed in ordinary algebra will usually omit parentheses where they can do so unambiguously.
\end{example}


\begin{example}[Comonoids]\label{ex:comonoids}
Unlike algebraic syntax, string diagrams allow for operations with co-arity different from $1$, manifested by multiple (or no) right boundary wires. It is therefore possible to flip the generators and axioms of the theory of monoids, to obtain the symmetric monoidal theory of \emph{comonoids}! Unsurprisingly, it is presented by a signature with a single object (which we therefore omit in diagrams), two generators, called comultiplication and counit,
\[\Bcomult\qquad \Bcounit\]
and the following three axioms:
\[
\InputIfFileExists{copy-associative.tikz}{}{\input{./tikz/copy-associative.tikz}}
\;\;\myeq{coas}\;\; 
\InputIfFileExists{copy-associative-1.tikz}{}{\input{./tikz/copy-associative-1.tikz}}
\qquad  
\InputIfFileExists{copy-unital-left.tikz}{}{\input{./tikz/copy-unital-left.tikz}}
\;\;\myeq{counl}\;\;\idx{}
 \;\;\myeq{counr}\;\;
\InputIfFileExists{copy-unital-right.tikz}{}{\input{./tikz/copy-unital-right.tikz}}
\]
called coassociativity and counitality. As one can see immediately, string diagrams for comonoids are just the mirrored version of those for monoids. Therefore, any diagrammatic statement involving only comonoids can be proved by simply flipping the corresponding proof about monoids along the vertical axis. For example, as we did for monoids, it is possible to reason silently modulo coassociativity and introduce syntactic sugar for a generalised comultiplication node with co-arity $n$ for any natural number:
\[

\InputIfFileExists{n-ary-cocommutative-comonoid.tikz}{}{\input{./tikz/n-ary-cocommutative-comonoid.tikz}}
\]
A comonoid is furthemore \emph{cocommutative} if
\[
\InputIfFileExists{copy-commutative.tikz}{}{\input{./tikz/copy-commutative.tikz}}
 \;\;\myeq{cocom}\;\; \Bcomult\]
As we will see, distinguished cocommutative comonoid structures play a special role in many theories: for example, they can be used to represent a form of copying and discarding, which allows us to interpret the wires of our diagrams as variables in standard algebraic syntax. The comultiplication allows us to reference a variable multiple times and the counit gives us the right to omit some variable in a string diagram. Following this intuition, we may for instance depict the term $f(g(x),g(x),y)$ in the context given by variables $x,y,z$, as follows:
\[x,y,z \vdash f(g(x),g(x), y)\qquad \mapsto \qquad 
\InputIfFileExists{ex-algebraic-term.tikz}{}{\input{./tikz/ex-algebraic-term.tikz}}
\]
For this reason, from the diagrammatic perspective, algebraic theories (or Lawvere theories, their categorical cousins, see~\cite{HylandP07}) always carry a chosen cocommutative comonoid structure~\cite{BonchiSZ18}, even though this structure does not appear in the usual symbolic notation for variables (which relies instead on an infinite supply of unique names to serve as identifiers for variables).  We will come back to this point in Remark~\ref{rmk:algebraic-Cartesian}.
\end{example}
\begin{remark}[Symmetric monoidal theories and linearity]\label{rmk:linear-properties}
Much like monoids in ordinary algebra, monoids or comonoids in symmetric monoidal theories can have additional properties. We have already encountered commutative monoids and cocommutative comonoids. However, the analogy between symmetric monoidal theories and algebraic theories hides an important subtlety: if, in the former, the wires play the role of variables, they have to be used precisely once. Unlike variables in ordinary algebra, we cannot use wires more than once or omit to use them at all! This restriction---termed \emph{resource-sensitivity}---is an important feature of diagrammatic syntax. Properties that do not involve multiple uses of variables can be specified completely analogously, as we saw for commutativity. Axioms that use each variable precisely once on each side of the equality sign are called \emph{linear} axioms. Non-linear axioms cannot be translated directly in the diagrammatic context, however. For example, it makes no sense to refer to the symmetric monoidal theory of idempotent monoids: those monoids that satisfy $m(x,x) = x$. Indeed, to even state the idempotency axiom one requires the ability to duplicate wires. As we will see, idempotency can also be expressed diagrammatically, but as a property of a more complex algebraic structure than a monoid; it can be stated as a property of a \emph{bimonoid}, which is our next example. This example is an instance of a more general pattern that allows us recover the resource-\emph{in}sensitivity of ordinary algebraic syntax. We will explore this correspondence more systematically in Section~\ref{sec:Cartesian}.
\end{remark}

The theory of monoids and comonoids can interact in different ways, as the next two examples illustrate. By `interact' in this context, we mean that there are different equations that one can impose when considering a signature that contains both the generators of monoids and those of comonoids with their respective theories.

\begin{example}[Co/commutative bimonoids]\label{ex:bimonoids}
One possible theory axiomatises a structure called a \emph{bimonoid}. It is presented by the generators of monoids and comonoids
\[\Wmult\quad \Wunit\quad \Bcomult \quad \Bcounit\]
together with their respective axioms, and the following additional four equations:
\begin{equation}\label{eq:bimonoids}
\begin{gathered}

\InputIfFileExists{wmult-bcomult.tikz}{}{\input{./tikz/wmult-bcomult.tikz}}
\:=\:
\InputIfFileExists{2-bcomult-sym-2-wmult.tikz}{}{\input{./tikz/2-bcomult-sym-2-wmult.tikz}}
\quad\qquad 
\begin{tikzpicture}
	\begin{pgfonlayer}{nodelayer}
		\node [style=white] (1) at (-0.75, 0) {};
		\node [style=black] (4) at (0.5, 0) {};
	\end{pgfonlayer}
	\begin{pgfonlayer}{edgelayer}
		\draw (1) to (4);
	\end{pgfonlayer}
\end{tikzpicture}
}
\:=\: 
\InputIfFileExists{empty-diag.tikz}{}{\input{./tikz/empty-diag.tikz}}

\\

\InputIfFileExists{wunit-bcomult.tikz}{}{\input{./tikz/wunit-bcomult.tikz}}
\:=\:
\begin{tikzpicture}
	\begin{pgfonlayer}{nodelayer}
		\node [style=none] (6) at (2, -0.5) {};
		\node [style=white] (8) at (0.75, -0.5) {};
		\node [style=none] (14) at (2, 0.5) {};
		\node [style=white] (16) at (0.75, 0.5) {};
	\end{pgfonlayer}
	\begin{pgfonlayer}{edgelayer}
		\draw (6.center) to (8);
		\draw (14.center) to (16);
	\end{pgfonlayer}
\end{tikzpicture}
}

\quad\qquad 

\InputIfFileExists{wmult-bcounit.tikz}{}{\input{./tikz/wmult-bcounit.tikz}}
\:=\:
\begin{tikzpicture}
	\begin{pgfonlayer}{nodelayer}
		\node [style=none] (6) at (0.75, 0.5) {};
		\node [style=black] (8) at (2, 0.5) {};
		\node [style=none] (14) at (0.75, -0.5) {};
		\node [style=black] (16) at (2, -0.5) {};
	\end{pgfonlayer}
	\begin{pgfonlayer}{edgelayer}
		\draw (6.center) to (8);
		\draw (14.center) to (16);
	\end{pgfonlayer}
\end{tikzpicture}
}

\end{gathered}
\end{equation}
Intuitively, these equalities can be seen as instances of the same general principle: whenever one of the monoid generators is composed horizontally with one of the comonoid generators, they pass through one another, producing multiple copies of each other. This is a two-dimensional form of \emph{distributivity}. For example, when the unit meets the comultiplication, the latter duplicates the former; when the multiplication meets the comultiplication, they duplicate each other (notice how this requires the symmetry, the ability to cross wires). Using the generalised monoid and comonoid operations introduced in the previous examples, we can formulate a generalised bimonoid axiom \emph{scheme} that captures all four axioms (and more):
\begin{equation}\label{eq:general-bimonoid-law}

\InputIfFileExists{b-w-general-bimonoid.tikz}{}{\input{./tikz/b-w-general-bimonoid.tikz}}

\end{equation}
\end{example}
\noindent Then, the four defining axioms can be recovered for the particular cases where the number of wires on each side is zero or two.

As we have already mentioned in Example~\ref{ex:comonoids}, comonoids can mimic the multiple use of variables in ordinary algebra. Thus, in the context of bimonoids, we can state ordinary equations that involve more or less than one occurrence of the same variable. For example, a bimonoid is \emph{idempotent} when it satisfies the following additional equality, which clearly translates the usual $m(x,x)=x$ into a diagrammatic axiom:
\[

\InputIfFileExists{bcomult-wmult.tikz}{}{\input{./tikz/bcomult-wmult.tikz}}
 = \idx{}\]
\begin{example}[Frobenius monoids]\label{ex:Frobenius}
Bimonoids are not the only way that monoids and comonoids can interact---there is another structure that frequently appear in the literature, under the name of \emph{Frobenius monoid}, or \emph{Frobenius algebra}\footnote{The term `Frobenius monoid' is due to monoids being traditionally more familiar than comonoids, even though both structures play an equally prominent role in a Frobenius monoid. As for the provenance of `Frobenius algebra', the term `algebra' usually refers to a monoid which is also a vector space (and whose multiplication is a linear map). This is the context in which Frobenius monoids were first studied.}. This structure is presented by the generators of monoids and comonoids. We will write them using nodes of the same colour, as this is how Frobenius monoids tend to appear in the literature, and will allow us to distinguish them from bimonoids in the rest of the paper:
\[
\Bcomult\quad \Bcounit \quad \Bunit\quad \Bmult
\]
together with their respective axioms, and the following additional axiom, called Frobenius' law:
\begin{equation}\label{eq:frobenius}

\InputIfFileExists{copy-Frobenius-left.tikz}{}{\input{./tikz/copy-Frobenius-left.tikz}}
\;\;\myeq{frob}\;\; 
\InputIfFileExists{copy-Frobenius-right.tikz}{}{\input{./tikz/copy-Frobenius-right.tikz}}

\end{equation}
This equality provides an alternative way for the multiplication and comultiplication of the monoid and comonoid structures to interact: unlike the case of bimonoids, this time they do not duplicate each other, but simply slide past one another, on either side. This is a fundamental difference which, in fact, turns out to be incompatible with the bimonoid axioms. We will examine this incompatibility more closely in Section~\ref{sec:mix}.

The reader might encounter other versions of this axiom in the literature, such as:
\[

\InputIfFileExists{copy-Frobenius-left.tikz}{}{\input{./tikz/copy-Frobenius-left.tikz}}
\;\;=\;\; 
\InputIfFileExists{copy-Frobenius.tikz}{}{\input{./tikz/copy-Frobenius.tikz}}
\;\;=\;\; 
\InputIfFileExists{copy-Frobenius-right.tikz}{}{\input{./tikz/copy-Frobenius-right.tikz}}

\]
In the presence of the other axioms (namely counitality and coassociativity), these two equalities are derivable from~\eqref{eq:frobenius}.
To get a feel for diagrammatic reasoning, let us prove it:
\begin{align*}

\InputIfFileExists{copy-Frobenius.tikz}{}{\input{./tikz/copy-Frobenius.tikz}}
 \;\;&\myeq{coun}\;\; 
\InputIfFileExists{derived-frobenius-1.tikz}{}{\input{./tikz/derived-frobenius-1.tikz}}
 \;\;\myeq{frob}\;\;
\InputIfFileExists{derived-frobenius-2.tikz}{}{\input{./tikz/derived-frobenius-2.tikz}}
\;\;\myeq{coas}\;\; 
\InputIfFileExists{derived-frobenius-3.tikz}{}{\input{./tikz/derived-frobenius-3.tikz}}
\\
&\myeq{frob}\;\;
\InputIfFileExists{derived-frobenius-4.tikz}{}{\input{./tikz/derived-frobenius-4.tikz}}
 \;\;\myeq{coun}\;\;
\InputIfFileExists{copy-Frobenius-left.tikz}{}{\input{./tikz/copy-Frobenius-left.tikz}}
 \end{align*}
At first, the string diagram novice may find it difficult to internalise all the laws that make up a theory such as that of Frobenius monoids. When proving some equality, it is not always clear which axiom to apply at which point to reach the desired goal and it is easy to get overwhelmed by all the choices available. However, in some nice cases, as we saw for (co)monoids, there are high level principles that allow us to simplify reasoning and see more clearly the key steps ahead. For example, reasoning up to associativity becomes second nature after enough practice and one no longer sees two different composites of (co)multiplication as different objects. 

In the same way, the Frobenius law can be thought of as a form of two-dimensional associativity; it simplifies reasoning about complex composites of monoid and comonoid operations even further and allows us to identify at a glance when any two string diagrams for this theory are equal. To explain this, it is helpful to think of string diagrams for the theory of Frobenius monoids as (undirected) graphs, whose vertices are any of the black dots, and edges are wires. We say that a string diagram is \emph{connected} if there is a path between any two vertices in the corresponding graph. It turns out that, for Frobenius monoids, any connected string diagram composed out of (finitely many) $\Bcomult$, $\Bcounit$, $\Bmult$ or $\Bunit$ using vertical or horizontal composition (without wire crossings) is equal to one of the following form~\cite[Section 5.2.1]{HeunenVicaryBook}:
\[

\InputIfFileExists{black-spider-normal-form.tikz}{}{\input{./tikz/black-spider-normal-form.tikz}}
\]
where we use ellipsis to represent an arbitrarily large composite following the same pattern.
In other words, the only relevant structure for a connected string diagram in the theory of Frobenius monoids is the number of left and right wires it has, and how many paths there are from any left leg to any right leg (how many loops it has in the normal form depicted above). This observation is sometimes called the \emph{spider theorem} and justifies introducing generalised vertices we call spiders as syntactic sugar:
\[
\InputIfFileExists{black-spider-k.tikz}{}{\input{./tikz/black-spider-k.tikz}}
\]
where the natural number $k$ represents the number of inner loops in the normal form above.
All the laws of Frobenius monoids can now be summarised into a single convenient axiom scheme:
\[
\InputIfFileExists{black-spider-fusion.tikz}{}{\input{./tikz/black-spider-fusion.tikz}}
\]
where $k$ is the number of middle wires that connect the two spiders on the left hand side of the equality. 
As a result, we need only keep track of the number of open wires and loops for any complicated string diagram; this greatly reduces the mental load to reason about this theory.

Frobenius monoids that satisfy the following idempotency axiom occur frequently in the literature:
\[

\InputIfFileExists{copy-special.tikz}{}{\input{./tikz/copy-special.tikz}}
=\idx{}
\]
 In this case, the Frobenius monoid is called \emph{special} (or sometimes, \emph{separable}). The normal form given by the spider theorem simplifies even further in this case, since we can now forget about the inner loops:
\[

\InputIfFileExists{black-spider.tikz}{}{\input{./tikz/black-spider.tikz}}
 \quad :=\quad 
\InputIfFileExists{black-special-spider-normal-form.tikz}{}{\input{./tikz/black-special-spider-normal-form.tikz}}

\]
The only relevant structure of any connected string diagram in the theory of special Frobenius monoids is the number of its left and right wires. We can thus introduce the same syntactic sugar, omitting the number of loops above the spider. The spider fusion scheme also simplifies further, as we no longer need to keep track of the number of legs that connect the two fusing spiders.
\end{example}
\begin{example}[Special and commutative Frobenius monoids]\label{ex:commutative-special-Frobenius}
 The commutative and special Frobenius monoids are very common in the literature, as they are an algebraic structure one finds naturally when reasoning about relations (as we will see when we study semantics of string diagrams in Section~\ref{sec:semantics}).
We summarise below the full equational theory for future reference:
\begin{equation}\label{eq:commutative-special-Frobenius}
\begin{gathered}

\InputIfFileExists{copy-associative.tikz}{}{\input{./tikz/copy-associative.tikz}}
 \;\myeq{coas}\; 
\InputIfFileExists{copy-associative-1.tikz}{}{\input{./tikz/copy-associative-1.tikz}}
\qquad   
\InputIfFileExists{copy-unital-left.tikz}{}{\input{./tikz/copy-unital-left.tikz}}
\;\myeq{counl}\;\idx{} \;\myeq{counr}\;
\InputIfFileExists{copy-unital-right.tikz}{}{\input{./tikz/copy-unital-right.tikz}}
\\ 

\InputIfFileExists{copy-commutative.tikz}{}{\input{./tikz/copy-commutative.tikz}}
\;\myeq{cocom}\; \Bcomult
\\

\InputIfFileExists{co-copy-associative.tikz}{}{\input{./tikz/co-copy-associative.tikz}}
\;\myeq{as}\; 
\InputIfFileExists{co-copy-associative-1.tikz}{}{\input{./tikz/co-copy-associative-1.tikz}}
\qquad  
\InputIfFileExists{co-copy-unital-left.tikz}{}{\input{./tikz/co-copy-unital-left.tikz}}
\;\myeq{unl}\;\idx{}
 \;\myeq{unr}\;
\InputIfFileExists{co-copy-unital-right.tikz}{}{\input{./tikz/co-copy-unital-right.tikz}}
\\
 
\InputIfFileExists{co-copy-commutative.tikz}{}{\input{./tikz/co-copy-commutative.tikz}}
\;\myeq{com}\;\,\Bmult
\\

\InputIfFileExists{copy-Frobenius-left.tikz}{}{\input{./tikz/copy-Frobenius-left.tikz}}
\;\;\myeq{frob}\;\; 
\InputIfFileExists{copy-Frobenius.tikz}{}{\input{./tikz/copy-Frobenius.tikz}}
\;\;\myeq{frob}\;\; 
\InputIfFileExists{copy-Frobenius-right.tikz}{}{\input{./tikz/copy-Frobenius-right.tikz}}

\\

\InputIfFileExists{copy-special.tikz}{}{\input{./tikz/copy-special.tikz}}
\;\myeq{spec}\;\idx{}
\end{gathered}
\end{equation}
In what follows, we will refer to this theory as $\scFrob$.

When adding commutativity in the picture, the spider theorem still holds, and includes string diagrams composed out of (finitely many) of $\Bcomult, \Bcounit, \Bunit, \Bmult$ \emph{and wire crossings}, using vertical or horizontal composition. Any string diagram in the free SMC over the theory of commutative and special Frobenius monoids is a fully determined by a list of spiders, and to where each of their respective legs are connected on the left and on the right boundary. In other words, string diagrams with $n$ left wires and $m$ right wires are in one-to-one correspondence with maps $n+m\to k$ for some $k$. This result will be the basis of a concrete description of the free SMC on the theory of a special commutative Frobenius monoid in  Example~\ref{ex:cospans}.
\end{example}
\begin{example}\label{eq:extraspecial-Frobenius}
A special Frobenius monoid that moreover satisfies the following axiom is sometimes called \emph{extra-special}
\begin{equation}
	
\begin{tikzpicture}
	\begin{pgfonlayer}{nodelayer}
		\node [style=black] (0) at (0.5, -0) {};
		\node [style=black] (1) at (-0.75, -0) {};
	\end{pgfonlayer}
	\begin{pgfonlayer}{edgelayer}
		\draw (0) to (1);
	\end{pgfonlayer}
\end{tikzpicture}}
\;\;=\;\;
\InputIfFileExists{empty-diag.tikz}{}{\input{./tikz/empty-diag.tikz}}
 \label{eq:extra-special}
\end{equation}
This means that we can forget about network of black nodes without any dangling wires---they can always be eliminated. In this case, string diagrams with $n$ left wires and $m$ right wires in the (free SMC over the) theory of a commutative extra-special Frobenius monoid are in one-to-one correspondence with partitions of $\{1,\dots,n+m\}$. Intriguingly, one may think of~\eqref{eq:extra-special} as a `garbage collector'; in the relational interpretation of string diagrams, it allows to capture \emph{equivalence relations}, as it eliminates empty equivalence classes. We will come back to the case of extra-special commutative Frobenius monoids in Example~\ref{ex:corelations}.
\end{example}

The reader will frequently encounter the theories we covered here as the building blocks of more complex diagrammatic calculi, designed to capture different kinds of phenomena. For example, the ZX-calculus, a theory that generalises quantum circuits (see Section~\ref{sec:diagrams-science}), contains not one, but two special commutative Frobenius monoids, often denoted by a red and green dot respectively. They interact together to form two bimonoids: the green monoid with the red comonoid forms a bimonoid and so do the red monoid with the green comonoid. Phew! At first, this seems like a lot of structure to absorb, but quickly, one learns to use the spider theorem to think of monochromatic string diagrams, so that most of the complexity comes from the interaction of the two colours. And even then, the generalised bimonoid law we saw in Example~\ref{ex:bimonoids} helps a lot. In fact, modern presentations of the ZX-calculus prefer to give the theory using spiders of arbitrary arity and co-arity as operations and the spider fusion rules as axioms. 
Strikingly, very similar equational theories can be found ubiquitously in a number of different applications, across different fields of science: it appears there is something fundamental to the interaction of monoid-comonoid pairs in the way we model computational phenomena. We will see several example applications in Section~\ref{sec:diagrams-science}.

\begin{remark}[Distributive Laws]\label{rmk:distributive-law}
It is noteworthy that both the equations of bimonoids (Example~\ref{ex:bimonoids}) and of Frobenius monoids (Example~\ref{ex:Frobenius}) describe the interaction of a monoid and a comonoid, even though they do it in different ways. A way to put it is in terms of factorisations: the bimonoid laws allow us to factorise any string diagram as one where all the comonoid generators precede the monoid generators, as in~\eqref{eq:general-bimonoid-law}; dually, the Frobenius laws yield a monoid-followed-by-comonoid factorisation, as in the spider theorem. More abstractly, the two equational theories can be described as different specifications of a \emph{distributive law} involving the monoid and the comonoid. Distributive laws are a familiar concept in algebra: the chief example is the one of a ring, whose equations describe the distributivity of a monoid over an abelian group. In the context of symmetric monoidal theories, distributive laws are even more powerful, as they can be used to study the interaction of theories with generators with arbitrary coarity, such as comonoids, Frobenius monoids, etc. The systematic study of distributive laws of symmetric monoidal theories has been initiated by Lack~\cite{Lack2004a}, and expanded in more recent works~\cite{ZanasiThesis, BonchiSZ-JPAA, BonchiSZ18}. Understanding a theory as the result a distributive law allows us to obtain a factorisation theorem for its string diagrams, such as~\eqref{eq:general-bimonoid-law} and the spider theorem. Moreover, it provides insights on a more concrete representation (a \emph{semantics}) for syntactically specified theories of string diagrams - a theme which we will explore in Section~\ref{sec:semantics}. For example, the phase-free fragment of the aforementioned ZX-calculus can be understood in terms of a distributive law between two bimonoids. This observation is instrumental in showing that the free model of the phase-free ZX-calculus is a category of linear subspaces~\cite{BialgAreFrob14}. We refer to~\cite{ZanasiThesis} for a more systematic introduction to distributive laws of symmetric monoidal theories, as well as on other ways of combining together theories of string diagrams.
\end{remark}

\section{String Diagrams as Graphs}\label{sec:graphs}

The previous section introduced string diagrams as a \emph{syntax}. However, a strength of the formalism is that string diagrams may be also treated as \emph{graphs}, with nodes and edges. This perspective is often convenient to investigate properties of string diagrams having to do with their combinatorial rather than syntactic structure, such as whether there is path between two components. 
Another important reason to explore a combinatorial perspective to string diagrams is that their graph representation `absorbs' the  laws of symmetric monoidal categories (Figure~\ref{fig:smc-axioms}). It is thus more adapted than the syntactc representation for certain computation tasks, such as rewriting (see Section~\ref{sec:rewriting} below). The goal of this section is to illustrate how string diagrams can be formally interpreted as graphs.

As a starting point, let us take for example a string diagram we have previously considered:
\[
\InputIfFileExists{ex-smc-diag.tikz}{}{\input{./tikz/ex-smc-diag.tikz}}
\]
If we forget about the term structure that underpins this representation, and try to understand it as a graph-like structure, the seemingly most natural approach is to think of boxes as \emph{nodes} and the wires as \emph{edges} of a graph. In fact, this is usually the intended interpretation adopted in the early days of string diagrams as a mathematical notation, see e.g.~\cite{JOYAL199155}. An immediate challenge for this approach is that `vanilla' graphs do not suffice: string diagrams present loose, open-ended edges, which only connect to a node on one side, or even on no side, as for instance the graph representation of the `identity' wires: $\idx{x}$. Historically, a solution to this problem has been to consider as interpretation a more sophisticated notion of graph, endowed with a topology from which one can define when edges are `loose', `half-loose', `pinned'... see~\cite{JOYAL199155}. Another, more recent approach understands string diagrams as graphs with two sorts of nodes, where the second sort just plays the bureaucratic role of giving an end to edges that otherwise would be drawn as loose~\cite{DixonKissinger13}.

The approach we present follows~\cite{BonchiGKSZ16}. We do not regard boxes as nodes, but rather as \emph{hyperedges}: edges that connect \emph{lists} of nodes, instead of individual nodes. This perspective allows us to work with a well-known data structure (simpler than the ones above) called a \emph{hypergraph}: the only entities appearing in a hypergraph are hyperedges and nodes: these interpret the boxes and the loose ends of wires in a string diagram, respectively. And the wires themselves? They are simply a depiction of how hyperedges connect with the associated nodes. Such an interpretation applies as follows to our leading example:
\[
\InputIfFileExists{ex-smc-diag.tikz}{}{\input{./tikz/ex-smc-diag.tikz}}
 \quad \mapsto \quad 
\InputIfFileExists{ex-smc-hypergraph.tikz}{}{\input{./tikz/ex-smc-hypergraph.tikz}}
\]
Note that, even though they are seemingly very close in shape, the two entities displayed above are of a very different nature. The one on the left is a syntactic object: the string diagram representing some term modulo the laws of SMCs. The one on the right is a combinatorial object: a hypergraph, with nodes indicated as dots and hyperedges indicated as boxes with round corners, labelled with $\Sigma$-operations.\footnote{The reader may wonder what happens when there is more than one generating object, so that string diagrams have non-trivial labels on wires. All of this section generalises to that more general setting: in the hypergraph interpretation, nodes may be labelled with the appropriate object, and constructions that merge nodes are disallowed unless the label matches. As this generalisation poses no significant conceptual difficulty, for the sake of clarity we opted for focussing our exposition on the case of theories with one generating object.}

In order to turn this mapping into a formal interpretation, we need an understanding of how to handle composition of string diagrams. Intuitively, parallel composition is simple: if we stack one hypergraph over the other, we still obtain a valid hypergraph. For instance:
\begin{gather*}

\InputIfFileExists{ex-smc-vertical-comp-1.tikz}{}{\input{./tikz/ex-smc-vertical-comp-1.tikz}}
\qquad
\InputIfFileExists{ex-smc-vertical-comp-2.tikz}{}{\input{./tikz/ex-smc-vertical-comp-2.tikz}}
\\
\rule{10cm}{0.15em}\\

\InputIfFileExists{ex-smc-vertical-comp.tikz}{}{\input{./tikz/ex-smc-vertical-comp.tikz}}

\end{gather*}

Sequential composition is subtler, as we need to formally specify how loose wires of one string diagram are `plugged in' loose wires of another diagram.
\begin{gather*}

\InputIfFileExists{ex-smc-horizontal-comp-1.tikz}{}{\input{./tikz/ex-smc-horizontal-comp-1.tikz}}
\qquad 
\InputIfFileExists{ex-smc-horizontal-comp-2.tikz}{}{\input{./tikz/ex-smc-horizontal-comp-2.tikz}}
\\
\rule{10cm}{0.15em}\\
\text{\Large{?}}
\end{gather*}
A proper definition of this composition operation is what leads to the notion of \emph{open} hypergraph: a hypergraph with a record of what nodes form its \emph{left} interface and what nodes form its \emph{right} interface. Note that one node can be in both, as in~\eqref{eq:openhyp-cospan} below. Pictorially, we will display the interfaces as separate discrete hypergraphs\footnote{A hypergraph is discrete when it has just nodes and no hyperedge.}, one on the left (framed in blue) and one on the right (framed in red), with dotted lines indicating which nodes of the actual hypergraph (framed in gray) lie on which interface. Our leading example corresponds to the open hypergraph on the right below.
\[\scalebox{0.85}{
\InputIfFileExists{ex-smc-diag.tikz}{}{\input{./tikz/ex-smc-diag.tikz}}
} \quad \mapsto \quad \scalebox{0.85}{
\InputIfFileExists{ex-smc-hypergraph-colour.tikz}{}{\input{./tikz/ex-smc-hypergraph-colour.tikz}}
}\]

Thanks to this additional information, open hypergraph come endowed with a built-in notion of sequential composition, mimicking the sequential composition of string diagrams. We are allowed to compose two open hypergraphs sequentially whenever the right interface of the first coincides with the left interface of the second.
\begin{gather*}

\InputIfFileExists{ex-smc-horizontal-comp-colour-1.tikz}{}{\input{./tikz/ex-smc-horizontal-comp-colour-1.tikz}}
\qquad
\InputIfFileExists{ex-smc-horizontal-comp-colour-2.tikz}{}{\input{./tikz/ex-smc-horizontal-comp-colour-2.tikz}}
\\
\rule{10cm}{0.15em}\\

\InputIfFileExists{ex-smc-hypergraph-colour.tikz}{}{\input{./tikz/ex-smc-hypergraph-colour.tikz}}

\end{gather*}
Equipped with this notion, we may define the interpretation of string diagrams as open hypergraphs, inductively on $\Sigma$-terms, as summarised in Figure~\ref{fig:hypergraph-sem}.\begin{figure}
\[\text{For all } g \in \Sigma_1\quad \diagbox{g}{v}{w}\;\mapsto\;
\InputIfFileExists{hypergraph-op.tikz}{}{\input{./tikz/hypergraph-op.tikz}}
\]
\[
\InputIfFileExists{hypergraph-id.tikz}{}{\input{./tikz/hypergraph-id.tikz}}
\quad\qquad 
\InputIfFileExists{hypergraph-sym.tikz}{}{\input{./tikz/hypergraph-sym.tikz}}
\]
\begin{equation*}
\begin{gathered}
\\[2em]

\InputIfFileExists{hypergraph-vertical-comp-1.tikz}{}{\input{./tikz/hypergraph-vertical-comp-1.tikz}}
\qquad
\InputIfFileExists{hypergraph-vertical-comp-2.tikz}{}{\input{./tikz/hypergraph-vertical-comp-2.tikz}}
\\
\rule{8cm}{0.15em}\\
{
\InputIfFileExists{hypergraph-vertical-comp.tikz}{}{\input{./tikz/hypergraph-vertical-comp.tikz}}
}
\vspace{1em}
\\[2em]

\InputIfFileExists{hypergraph-horizontal-comp-1.tikz}{}{\input{./tikz/hypergraph-horizontal-comp-1.tikz}}
\qquad
\InputIfFileExists{hypergraph-horizontal-comp-2.tikz}{}{\input{./tikz/hypergraph-horizontal-comp-2.tikz}}
\\
\rule{8cm}{0.15em}\\
{
\InputIfFileExists{hypergraph-horizontal-comp.tikz}{}{\input{./tikz/hypergraph-horizontal-comp.tikz}}
}
\end{gathered}
\end{equation*}
\caption{Interpretation of string diagrams as open hypergraphs.}\label{fig:hypergraph-sem}
\end{figure}
In words, the vertical composition takes the disjoint sum of each of the interfaces and hypergraphs, while the horizontal composition identifies the middle interface labels and includes them as nodes into the composite hypergraph.
Note this definition extends to an interpretation of string diagrams by verifying that it respects equality modulo the laws of SMCs---that is, if two $\Sigma$-terms are represented by the same string diagram, then they are mapped to the same open hypergraph.

As a side note, it is significant that moving from hypergraphs to open hypergraphs does not force us to complicate the notion of graph at hand, for instance by adding a different sort of nodes. From a mathematical viewpoint, an open hypergraph $G$ may be simply expressed as a structure consisting of two hypergraph homomorphisms $\langle p \colon G_L \to G ,q \colon G_R \to G\rangle$, where $G_L$ and $G_R$ are discrete hypergraphs, and the image $p[G_L]$ (resp. $q[G_R]$) identifies the nodes in the left (resp. right) interface of $G$. Here is a visualisation of such encoding: $G_L$ is the hypergraph in blue, $G_R$ is the one in red, and $G$ is the one in grey. The dotted lines, identifying the interfaces of $G$, now take formal meaning as the definition of functions $p \colon G_L \to G$ and $q \colon G_R \to G$.
\begin{equation}
	
\InputIfFileExists{ex-smc-hypergraph-cospan.tikz}{}{\input{./tikz/ex-smc-hypergraph-cospan.tikz}}
 \label{eq:openhyp-cospan}
\end{equation}
The structure $\langle p \colon G_L \to G ,q \colon G_R \to G\rangle$ is often called a \emph{cospan} of hypergraphs (see Example~\ref{ex:cospans} on the simpler cospans of sets), with carrier $G$. When referring to $G$ as an open hypergraph, we always implicitly refer to $G$ together to one such cospan structure. Reasoning with cospans is convenient as they come with a built-in notion of composition (by `pushout' in the category of hypergraphs) which is exactly how the informal composition of open hypergraphs given above is formally defined. We will come back to this point in Section~\ref{sec:rewriting}, as it plays a role in how we \emph{rewrite} with string diagrams.

\medskip

The interpretation of string diagrams as open hypergraphs given in Figure~\ref{fig:hypergraph-sem} above defines a monoidal functor $\sem{\cdot}$ from the free SMC over signature $\Sigma$ to the SMC of cospans of hypergraphs.

An important question stemming from the interpretation in Figure~\ref{fig:hypergraph-sem} is: to what extent the syntactic and the combinatorial perspective on string diagrams are interchangeable? First, one may show that $\sem{\cdot}$ is an \emph{injective} mapping: string diagrams that are distinct (modulo the laws of SMCs) are mapped to distinct open hypergraphs. However, it is clearly not surjective. Here are some examples of open hypergraphs over a signature $\Sigma$ which are not the interpretation of any $\Sigma$-string diagram.
\[
\begin{gathered}

\InputIfFileExists{ex-open-hypergraph-non-monogamous-1.tikz}{}{\input{./tikz/ex-open-hypergraph-non-monogamous-1.tikz}}
\qquad\qquad 
\InputIfFileExists{ex-open-hypergraph-non-monogamous-2.tikz}{}{\input{./tikz/ex-open-hypergraph-non-monogamous-2.tikz}}
 \\

\InputIfFileExists{ex-open-hypergraph-non-monogamous-3.tikz}{}{\input{./tikz/ex-open-hypergraph-non-monogamous-3.tikz}}

\end{gathered}
\]
These examples have something in common: nodes are allowed to behave more freely than in the image of interpretation of Figure~\ref{fig:hypergraph-sem}. For instance, in the first hypergraph there is an `internal' node (not on the interface) that has multiple outgoing links to hyperedges. In the second hypergraph, there is an internal node that has no incoming links. Finally, the third hypergraph features a node that can be plugged in twice on the left interface; when composing with another hypergraph on the left, it will have two incoming links.

We can prove that such features are forbidden in the image of $\sem{\cdot}$. The property that disallows them is called \emph{monogamy}~\cite{BonchiGKSZ16}.

\begin{definition}[Degree of a node]
The \emph{in-degree} of a node $v$ in a hypergraph $G$ is the number of pairs $(h,i)$ where $h$ is a hyperedge with $v$ as its $i$-th target.
Similarly, the \emph{out-degree} of $v$ in $G$ is the number of pairs $(h,i)$ where $h$ is a hyperedge with $v$ as its $i$-th source.
\end{definition}
\begin{definition}[Monogamy]
	An open hypergraph $\cospn{m}{f}{G}{g}{n}$ is monogamous if $f$ and $g$ are injective and for all nodes $v$ of $G$
\begin{itemize}
\item the in-degree of $v$ is $0$ if $v$ is in the image of $f$ and $1$ otherwise;
\item the out-degree of $v$ is $0$ if $v$ is in the image of $g$ and $1$ otherwise.
\end{itemize}
\end{definition}
Moreover, any hypergraph that corresponds to a string diagram is \emph{acyclic}: there are no directed paths containing the same node twice. These two properties are enough to characterise string diagrams.
\begin{theorem}\label{th:monogamicity-image}
	An open hypergraph is in the image of $\sem{\cdot}$ if and only if it is monogamous and acyclic.
\end{theorem}

\begin{corollary}
	String diagrams over $\Sigma$ are in 1-1 correspondence with $\Sigma$-labelled monogamous and acyclic open hypergraphs.
\end{corollary}

Theorem~\ref{th:monogamicity-image} settles the question of what kind of open hypergraphs correspond to `syntactically generated' string diagrams. We may also ask the converse question: what do we need to add to the algebraic specification of string diagrams in order to capture all the open hypergraphs?

Remarkably, the special and commutative \emph{Frobenius monoid} from Example~\ref{ex:commutative-special-Frobenius} is tailored to the role. Indeed we can give an interpretation to the operations of Example~\ref{ex:commutative-special-Frobenius}, as discrete open hypergraphs:
\[
\InputIfFileExists{hypergraph-copy.tikz}{}{\input{./tikz/hypergraph-copy.tikz}}
\quad\qquad 
\InputIfFileExists{hypergraph-del.tikz}{}{\input{./tikz/hypergraph-del.tikz}}
\]
\[
\InputIfFileExists{hypergraph-cocopy.tikz}{}{\input{./tikz/hypergraph-cocopy.tikz}}
\quad\qquad 
\InputIfFileExists{hypergraph-codel.tikz}{}{\input{./tikz/hypergraph-codel.tikz}}
\]
Intuitively, the Frobenius generators are modelling the possibility that a nodes has multiple or no ingoing/outgoing links, just as in the above examples. If we now consider string diagrams over $\Sigma$ augmented with the generating operations of a Frobenius monoid, we can infer the string diagrams for the above open hypergraphs:
\[
\begin{gathered}

\InputIfFileExists{ex-diagram-non-monogamous-1.tikz}{}{\input{./tikz/ex-diagram-non-monogamous-1.tikz}}
\;\mapsto\; 
\InputIfFileExists{ex-open-hypergraph-non-monogamous-1.tikz}{}{\input{./tikz/ex-open-hypergraph-non-monogamous-1.tikz}}

\\

\InputIfFileExists{ex-diagram-non-monogamous-2.tikz}{}{\input{./tikz/ex-diagram-non-monogamous-2.tikz}}
\;\mapsto\; 
\InputIfFileExists{ex-open-hypergraph-non-monogamous-2.tikz}{}{\input{./tikz/ex-open-hypergraph-non-monogamous-2.tikz}}

\\

\InputIfFileExists{ex-diagram-non-monogamous-3.tikz}{}{\input{./tikz/ex-diagram-non-monogamous-3.tikz}}
\;\mapsto\; 
\InputIfFileExists{ex-open-hypergraph-non-monogamous-3.tikz}{}{\input{./tikz/ex-open-hypergraph-non-monogamous-3.tikz}}

\end{gathered}
\]
These observations generalise to the following result, thus completing the picture of the correspondence between string diagrams and open hypergraphs. In stating it, we write $\Sigma+\scFrob$ for the signature given by the disjoint union of the generators of $\Sigma$ and of $\scFrob$, the theory of a special commutative Frobenius monoid given in~\eqref{eq:commutative-special-Frobenius}.
\begin{theorem}\label{thm:hypergraph-int-frob-iso}
	String diagrams on the signature $\Sigma + \scFrob$ modulo the axioms of special commutative Frobenius monoids given in~\eqref{eq:commutative-special-Frobenius} are in 1-1 correspondence with $\Sigma$-labelled open hypergraphs.
\end{theorem}
Note that it is not just the signature: the axioms of special commutative Frobenius monoids also play a role in the result, as they model precisely equivalence of open hypergraphs.

\begin{remark}\label{rmk:isoFrobHyp}
	Given a signature $\Sigma = (\Sigma_0,\Sigma_1)$, Open hypergraph with $\Sigma_0$-labelled nodes and $\Sigma_1$-labelled hyperedges form a symmetric monoidal category $\Hyp{\Sigma}$, whose morphisms are hypergraph homomorphisms respecting the labels. The monogamous and acyclic open hypergraphs form a subcategory $\MonHyp{\Sigma}$ of $\Hyp{\Sigma}$. One may phrase Theorem~\ref{th:monogamicity-image} and Theorem~\ref{thm:hypergraph-int-frob-iso} in terms of these categories, by saying that there is an isomorphism between $\FreeSMC{\Sigma}$ and $\MonHyp{\Sigma}$, and an isomorphism between $\FreeSMC{\Sigma+\scFrob}$ and $\Hyp{\Sigma}$.
\end{remark}

\section{Categories of String Diagrams}\label{sec:thstringdiag}


Manipulating string diagrams can be confusing to the newcomer because there are actually many flavours, each of which authorises or forbids different deformations and manipulations. To make matters worse, many papers will assume that the reader is comfortable with the rules of the game for the authors' specific flavour, and gloss over the basic transformations. This is not necessarily a bad thing, as the point of string diagrams is to serve as a useful computational tool, a syntax that empowers its users by absorbing irrelevant details into the topology of the notation itself. This section is here to convey the basic rules for the most common forms one is likely to encounter in the literature.
We will give some intuition about the manipulations that are authorised and those that are forbidden in each context, illustrating them through several examples.

In previous sections, we made the conscious choice of starting with string diagrams for \emph{symmetric monoidal categories}. Let us recall the rules of the game briefly:
we were allowed to compose boxes horizontally, as long as the types of the wires matched, and vertically, without restriction. In addition, we were allowed to cross wires however we wanted, and the only relevant structure of an arbitrary vertical or horizontal composite of multiple wire-crossings is the resulting permutation of the wires that it defines.

We will now see that there are various ways of strengthening or weakening these rules and the class of string diagrams under consideration.
For the reader willing to delve further into this subject, we recommend Selinger's extensive survey of diagrammatic languages for monoidal categories~\cite{Selinger2009}.

\subsection{Fewer Structural Laws}

\subsubsection{Monoidal Categories}\label{sec:mc}
What if we take away the ability to cross wires? Terms of the free monoidal category over a chosen signature $\Sigma = (\Sigma_0, \Sigma_1)$ are generated by the following derivation rules:
\[
\derivationRule{}{\quad\idx{x}\quad}{\scriptstyle{x\in \Sigma_0}}
\qquad\qquad
\derivationRule{}{\quad\diagbox{d}{v}{w}\quad }{\scriptstyle{d\in\Sigma_1}}
\]
\[
\derivationRule{\diagbox{c}{u}{v}\quad \diagbox{d}{v}{w}}{
\InputIfFileExists{horizontal-comp.tikz}{}{\input{./tikz/horizontal-comp.tikz}}
}{}
\qquad\qquad \derivationRule{\diagbox{\scriptstyle{d_1}}{v_1}{w_1}\qquad 
\diagbox{\scriptstyle{d_2}}{v_2}{w_2}}{
\InputIfFileExists{vertical-comp.tikz}{}{\input{./tikz/vertical-comp.tikz}}
}{}
\]
The difference with \emph{symmetric} monoidal categories is that we no longer have the built-in symmetry components $ \sym$ at our disposal. We consider two terms structurally equivalent when  they can be obtained from one another using the axioms of monoidal category, \emph{i.e.} the strict subset of those of symmetric monoidal categories that does not involve the symmetry/wire-crossing, as found in~\eqref{fig:smc-axioms}.
For example, we still have  the interchange law
\[
\InputIfFileExists{smc/interchange-law.tikz}{}{\input{./tikz/smc/interchange-law.tikz}}
 = 
\InputIfFileExists{smc/interchange-law-1.tikz}{}{\input{./tikz/smc/interchange-law-1.tikz}}
\]
but we do not have $\sym\sym^x_y \ = \ 
\InputIfFileExists{x+y.tikz}{}{\input{./tikz/x+y.tikz}}
$,
as $\sym $ is not even a term of our syntax.

Intuitively, they are the planar cousins of their symmetric counterpart, \emph{i.e.} the subset of string diagrams we can draw in the plane in a symmetric monoidal category without crossing any wires. In a way, the rules are simpler: we can only compose string diagrams horizontally (with the usual caveat that the right ports of the first have to match the left ports of the second) and vertically. That's it. Then, two string diagrams are equivalent if one can be deformed into the other \emph{without any intermediate steps that involve crossing wires}. For example,
\[
\InputIfFileExists{ex-plane-diag.tikz}{}{\input{./tikz/ex-plane-diag.tikz}}
\quad = \quad 
\InputIfFileExists{ex-plane-diag-deform.tikz}{}{\input{./tikz/ex-plane-diag-deform.tikz}}
\]
The last caveat is important, as two monoidal diagrams could be equivalent if we interpreted them (via the obvious embedding) as symmetric monoidal diagrams, but not equivalent as monoidal diagrams. This is the case for the two below:
\[
\InputIfFileExists{ex-plane-ineq.tikz}{}{\input{./tikz/ex-plane-ineq.tikz}}
\quad \neq \quad 
\InputIfFileExists{ex-plane-ineq-1.tikz}{}{\input{./tikz/ex-plane-ineq-1.tikz}}
\]
Thus, in the monoidal case, certain string diagrams can be trapped between some wires, without any way to move them on either side --- whereas, in the symmetric monoidal case, we could have just pulled the middle diagram out, past the surrounding wires.

\subsubsection{Braided Monoidal Categories}\label{sec:braided}
Braided monoidal categories~\cite{joyal1986braided} are one step up from monoidal categories, but are not symmetric. They allow a form of wire crossing that keeps track of which wire goes over which. To this effect, we introduce a  \emph{braiding} for each of the two possibilities depicted suggestively as follows:
\[
\InputIfFileExists{braiding-over.tikz}{}{\input{./tikz/braiding-over.tikz}}
 \qquad\qquad 
\InputIfFileExists{braiding-under.tikz}{}{\input{./tikz/braiding-under.tikz}}
\]
Term formation rules for the free braided monoidal category over a given signature $\Sigma$ are those of monoidal categories plus the following two:
\[\derivationRule{}{\quad
\InputIfFileExists{braiding-over.tikz}{}{\input{./tikz/braiding-over.tikz}}
\quad}{\scriptstyle{x,y\in \Sigma_0}}\qquad \quad \derivationRule{}{\quad
\InputIfFileExists{braiding-under.tikz}{}{\input{./tikz/braiding-under.tikz}}
\quad}{\scriptstyle{x,y\in \Sigma_0}}\]
The notion of structural equivalence is up to the axioms of braided monoidal categories, which we now give:
\[
\InputIfFileExists{braid-inverse.tikz}{}{\input{./tikz/braid-inverse.tikz}}
\]
\[
\InputIfFileExists{YB.tikz}{}{\input{./tikz/YB.tikz}}
\]
As the drawings suggest, the intuition for the braiding is that string diagrams now inhabit a three-dimensional space in which we are free to cross wires by moving them over or under each other.
The first two laws states that the two braidings are inverses of each other, and the third is an instance of the naturality of the wire crossings, called the \emph{Yang-Baxter} equation~\cite{Cheng_IteratedLaws}.
Notice that these axioms are similar to those for the symmetry in SMCs except that braidings are not self-inverse: $\sym\sym^x_y \ = \ 
\InputIfFileExists{x+y.tikz}{}{\input{./tikz/x+y.tikz}}
$ does not hold if we take $\braidxundery{x}{y}$ instead (see below). 
%
As a result, we can draw any string diagram we could draw in a symmetric monoidal category, but we have to pick which wire goes under and which goes over for each crossing.

Two string diagrams are equivalent if they can be deformed into each other without ever moving two wires through one another to magically disentangle them. Once more, this gives an equivalence that is finer\footnote{From a formal viewpoint, this statement is not entirely accurate, because the terms of free braided and symmetric monoidal categories are different. However, we can map those of the former to the latter by sending the braid to the symmetry. Our statement then amounts to saying that this mapping is not injective, \emph{i.e.}, that two different braided monoidal diagrams may be mapped to the same symmetric monoidal diagram.} than that of symmetric monoidal categories. A simple illustrative example of this phenomenon is the following twist:
\[
\InputIfFileExists{twist.tikz}{}{\input{./tikz/twist.tikz}}
\]
If we replaced the two braidings by the symmetry to obtain a term of a symmetric monoidal category, this string diagram would simply be the identity $
\InputIfFileExists{x+y.tikz}{}{\input{./tikz/x+y.tikz}}
$.
This serves as a reminder that the braiding is not self-inverse. One can find much more interesting examples---in fact, we can draw arbitrary braids:
\[
\InputIfFileExists{braid-ex.tikz}{}{\input{./tikz/braid-ex.tikz}}
\]
or string diagrams containing other generators with arbitrary braidings between them, which can be transformed like those of symmetric monoidal categories, as long as we do not move any of the wires through one another:
\[
\InputIfFileExists{ex-braided-diag.tikz}{}{\input{./tikz/ex-braided-diag.tikz}}
\;=\;
\InputIfFileExists{ex-braided-diag-1.tikz}{}{\input{./tikz/ex-braided-diag-1.tikz}}
\]

\begin{remark}
Contrary to the case of SMCs (see Section~\ref{sec:graphs}), there is no known representation of string diagrams for braided monoidal categories as graphs.
\end{remark}

\subsection{More Structural Laws}

Just like we can weaken the structure of symmetric monoidal categories and draw more restricted diagrams, we can also extend our diagrammatic powers. The following is a non-exhaustive list of the most common variations one might find in the literature.

\subsubsection{Traced Monoidal Categories}\label{sec:traced}
String diagrams in a (symmetric) monoidal category keep to a strict discipline of acyclicity: we can only connect the right and left ports of two boxes. One could imagine relaxing this requirement,  while keeping a clear correspondence between left ports as inputs and right ports as outputs.

Terms formation rules for traced monoidal categories are those of symmetric monoidal categories with the addition of an operation that allows us to form loops, called the \emph{partial trace}\footnote{The name comes from the usual linear algebraic notion of trace. We will examine this concrete case in Example~\ref{ex:linear-maps-tensor}.}:
\[\derivationRule{\quad\diagstate{d}{x}{y}{a}\quad}{\traceform{d}{x}{y}{a}}{}\]
The corresponding notion of structural equivalence is given by the following axioms (with object labels removed for clarity).
\begin{itemize}
\item Vanishing: \[
\InputIfFileExists{vanishing.tikz}{}{\input{./tikz/vanishing.tikz}}
=
\InputIfFileExists{vanishing-1.tikz}{}{\input{./tikz/vanishing-1.tikz}}
\qquad \qquad 
\InputIfFileExists{vanishing-product.tikz}{}{\input{./tikz/vanishing-product.tikz}}
 = 
\InputIfFileExists{vanishing-product-1.tikz}{}{\input{./tikz/vanishing-product-1.tikz}}
 \]
\item Superposing: \[
\InputIfFileExists{superposing.tikz}{}{\input{./tikz/superposing.tikz}}
= 
\InputIfFileExists{superposing-1.tikz}{}{\input{./tikz/superposing-1.tikz}}
\]
\item Yanking: \begin{equation}\label{eq:yanking}

\InputIfFileExists{yanking.tikz}{}{\input{./tikz/yanking.tikz}}
 =\idx{}
\end{equation}
\item Tightening:
\[
\InputIfFileExists{left-tightening.tikz}{}{\input{./tikz/left-tightening.tikz}}
=
\InputIfFileExists{left-tightening-1.tikz}{}{\input{./tikz/left-tightening-1.tikz}}
\qquad\quad 
\InputIfFileExists{right-tightening.tikz}{}{\input{./tikz/right-tightening.tikz}}
=
\InputIfFileExists{right-tightening-1.tikz}{}{\input{./tikz/right-tightening-1.tikz}}
\]
\item Sliding:
\[
\InputIfFileExists{sliding.tikz}{}{\input{./tikz/sliding.tikz}}
=
\InputIfFileExists{sliding-1.tikz}{}{\input{./tikz/sliding-1.tikz}}
\]
\end{itemize}
Here, we had to briefly go back to using dotted frames, because the axioms of traced monoidal categories are almost diagrammatic tautologies (which is the point of adopting a diagrammatic notation for them).

In short, string diagrams for traced monoidal categories include those of symmetric monoidal ones, but add the possibility of \emph{connecting any right port of any diagram to any left port of any other with the same type}, as in the following example:
\[
\InputIfFileExists{ex-trace-diag.tikz}{}{\input{./tikz/ex-trace-diag.tikz}}
\]
In essence, we have the ability to draw loops,  breaking free from the acyclicity requirement of plain symmetric monoidal diagrams. However, we cannot bend wires arbitrarily, or connect right (\emph{resp.} left) ports to right (\emph{resp.} left) ports. For example, the following is \emph{not} allowed:
\[
\InputIfFileExists{ex-self-dual-compact-diag.tikz}{}{\input{./tikz/ex-self-dual-compact-diag.tikz}}
\]
(Though we will soon introduce a syntax for which this kind of diagrams is allowed.)

The reader should convince themselves that the string diagram above is equivalent to the one on the right below:
\[
\InputIfFileExists{ex-trace-diag.tikz}{}{\input{./tikz/ex-trace-diag.tikz}}
\; =\; 
\InputIfFileExists{ex-trace-diag-deform.tikz}{}{\input{./tikz/ex-trace-diag-deform.tikz}}
\]
As for symmetric monoidal diagrams, the trick to check this equality lies in verifying that the connectivity of the different boxes is preserved.

\begin{remark}\label{rmk:feedback-categories} Traced string diagrams are often used when describing computational processes that feature some form of recursion or iteration. In this context, it is also natural to consider trace-like operations that do not satisfy the yanking axiom~\eqref{eq:yanking}. This makes sense when the trace-like operation is intended to represent a form of feedback which introduces a temporal delay. Examples abound in the theory of automata. Note that the sliding rule might also fail in this case. The associated graphical language generalises that of traced categories, and the associated structure is sometimes called a delayed or guarded traced category, or a category with feedback~\cite{katis2002feedback,di2021canonical,di2022monoidal}.
\end{remark}

\subsubsection{Compact Closed Categories}\label{sec:compact closed}
Compact closed (or more simply, compact) categories are special cases of traced monoidal categories where, rather than adding a global trace operation, we add ways of moving ports from left to right and vice-versa, using extra generators that represent wire-bending directly, as built-in operations.

The term formation rules are those of symmetric monoidal categories \emph{except} the identity introduction rules, with the following additions:
\[\derivationRule{}{\quad\raisebox{-0.5ex}{\idxright{x}}\quad}{\scriptstyle{x\in\Sigma_0}}\qquad \derivationRule{}{\quad\raisebox{-0.5ex}{\idxleft{x}}\quad}{\scriptstyle{x\in\Sigma_0}}\qquad \derivationRule{}{\quad\cupx{x}\quad}{\scriptstyle{x\in\Sigma_0}}\qquad \derivationRule{}{\quad\capx{x}\quad}{\scriptstyle{x\in\Sigma_0}}\]

In words, we introduce a new object $x^{*}$ (called the dual of $x$) for every object $x$ and write $\idxright{x}$ for the identity on $x$ and $\idxleft{x}$ for the identity on $x^*$. The objects $x$ and $x^*$ are related by two wire-bending diagrams $\cupx{x}$ and $\capx{x}$, called cup and cap respectively.

The corresponding notion of structural equivalence is defined by the axioms of symmetric monoidal category and the following two axioms, which capture the duality between inputs and outputs:
\begin{equation}\label{eq:snake}

\InputIfFileExists{z-yanking.tikz}{}{\input{./tikz/z-yanking.tikz}}
 \;\myeq{Z}\;\idxleft{x} \qquad\qquad 
\InputIfFileExists{s-yanking.tikz}{}{\input{./tikz/s-yanking.tikz}}
 \;\myeq{S}\;\idxright{x}
\end{equation}
These are sometimes called the \emph{snake} or \emph{yanking} equalities.

Using the symmetry, we can define syntactic sugar for two other cups and caps, bending wires in the other direction, which we write as:
\[
\InputIfFileExists{cup-down.tikz}{}{\input{./tikz/cup-down.tikz}}
 := 
\InputIfFileExists{cup-down-def.tikz}{}{\input{./tikz/cup-down-def.tikz}}
  \qquad\qquad 
\InputIfFileExists{cap-down.tikz}{}{\input{./tikz/cap-down.tikz}}
:= 
\InputIfFileExists{cap-down-def.tikz}{}{\input{./tikz/cap-down-def.tikz}}
 \]
From cups and caps, we also obtain a partial trace operation given simply by
\[\diagstate{d}{x}{y}{a} \quad \mapsto \quad 
\InputIfFileExists{trace-compact.tikz}{}{\input{./tikz/trace-compact.tikz}}
\]
This operation satisfies all the required axioms of traced monoidal categories (it is an instructive exercise to prove them). Importantly, what was a global operation before is now decomposed into smaller components that use the added generators. This is particularly helpful in applications, whenever we aim at reasoning  \emph{compositionally} about feedback loops in a system. Whereas the notion of trace is `native' to traced monoidal categories, it is a derived concept in compact closed categories.

For traced monoidal categories, we could draw loops directly, to connect any left port to any right port of a diagram. We could always read information in a given diagram as flowing from left to right, except in a looping wire, where it flowed backwards at the top of the loop, until it reaches its destination. Now that we can move left ports to the right of a diagram and right ports to the left, we have to be a bit more careful. This is why we have to annotate each wire with a direction.

We can understand this as layering a notion of input and output on top of those of left and right ports. We can call inputs those wires that flow into a diagram and outputs those that flow out of a diagram, whether they are on the left or right boundary. Then, in a compact closed category, we are allowed to \emph{connect any input to any output, i.e.,} we can assign a consistent direction to any wiring:
\[
\InputIfFileExists{ex-compact-diag.tikz}{}{\input{./tikz/ex-compact-diag.tikz}}
\]
Furthermore, as is now a leitmotiv, only the connectivity matters: we are allowed to straighten or bend wires at will, as long as we preserve the connections between the different sub-diagrams.
For example, the following two diagrams are equivalent:
\[
\InputIfFileExists{ex-compact-diag.tikz}{}{\input{./tikz/ex-compact-diag.tikz}}
\; =\; 
\InputIfFileExists{ex-compact-deform.tikz}{}{\input{./tikz/ex-compact-deform.tikz}}
\]

Finally, compact closed categories have the following very important property: diagrams of type $uv \to w$ are in one-to-one correspondence with diagrams  $u\to wv^*$. Diagrammatically, moving $v$ from the domain to the codomain is realised by bending the corresponding wire(s) using the cup $\cupx{v}$; moving in the other direction simply uses $\capx{v}$ to bend the $v$-wires back in place.
\[
\InputIfFileExists{map-state.tikz}{}{\input{./tikz/map-state.tikz}}
\qquad \quad 
\InputIfFileExists{state-map.tikz}{}{\input{./tikz/state-map.tikz}}
\]
The fact that these operations are inverses to each other is an easy consequence of the snake equalities~\eqref{eq:snake}, with which we can straighten the wires back into place. As a result, $wv^*$ can be seen as an internal analogue of the set of string diagrams $v\to w$. This property is found more generally in \emph{closed monoidal} categories, which we cover in Section~\ref{sec:closed}.

\subsubsection{Self-dual Compact Closed Categories}\label{sec:self-dual-compact}
There are significant instances of compact closed category where the distinction between inputs and outputs disappears completely: they are called \emph{self-dual}. The term formation rules are the same as those symmetric monoidal categories, with the following additions:
\[\derivationRule{}{\quad\sdcupx{x}\quad}{\scriptstyle{x\in\Sigma_0}}\qquad \derivationRule{}{\quad\sdcapx{x}\quad}{\scriptstyle{x\in\Sigma_0}}\]
They are also very close to those of compact closed categories, but we identify $x$ and its dual. As a result, there is no need to introduce a direction on wires. The added constants $\sdcupx{x}$ and $\sdcapx{x}$ satisfy the same equations as their directed cousins:
\[
\InputIfFileExists{self-dual-z-yanking.tikz}{}{\input{./tikz/self-dual-z-yanking.tikz}}
 \quad \myeq{Z} \quad \idx{x} \quad  \myeq{S} \quad 
\InputIfFileExists{self-dual-s-yanking-x.tikz}{}{\input{./tikz/self-dual-s-yanking-x.tikz}}
 \]
Without distinct duals there is no need to keep track of the directionality of wires, and the resulting diagrammatic calculus is even more permissive---there are no inputs or outputs, and we are now allowed to connect \emph{any two ports together}:
\[
\InputIfFileExists{ex-self-dual-compact-diag.tikz}{}{\input{./tikz/ex-self-dual-compact-diag.tikz}}
\]
As before the structural equivalence on diagrams allows us to identify any two diagrams where the same ports are connected:
\[
\InputIfFileExists{ex-self-dual-compact-diag.tikz}{}{\input{./tikz/ex-self-dual-compact-diag.tikz}}
\quad=\quad 
\InputIfFileExists{ex-self-dual-compact-deform.tikz}{}{\input{./tikz/ex-self-dual-compact-deform.tikz}}
\]
The previous three examples progressively relax which \emph{two} ports we can connect together; in the next examples, we relax the requirement that only two ports can be connected at a time by introducing different ways to split and end wires.

\subsubsection{Copy-Delete Monoidal Categories}\label{sec:CD}

The first of these adds the ability to split and end wires in order to connect some wire in the right boundary of a diagram to a (possibly empty) \emph{set} of wires in the left boundary of another. There are several names for these in the literature (Copy-Delete monoidal categories, gs-monoidal categories...) but we will call them CD categories for short.

The term formation rules for CD categories are the same as those symmetric monoidal categories, with the addition of wire splitting and ending for each generator:
\[\derivationRule{}{\quad\Bcomultn{x}\quad}{\scriptstyle{x\in\Sigma_0}}\qquad \derivationRule{}{\quad\Bcounitn{x}\quad}{\scriptstyle{x\in\Sigma_0}}\]
As anticipated, the copying and deleting operations allow us to connect a given right port of a diagram to a (possibly empty) \emph{set} of left ports of another:
\[
\InputIfFileExists{ex-cd-diag.tikz}{}{\input{./tikz/ex-cd-diag.tikz}}
\]
Notice however that there are several ways of connecting the same set of wires. For example, we could have connected the left boundary wire to $f,h$, and $c$ as follows:
\[
\InputIfFileExists{ex-cd-diag-1.tikz}{}{\input{./tikz/ex-cd-diag-1.tikz}}
\qquad \text{or}\qquad 
\InputIfFileExists{ex-cd-diag-2.tikz}{}{\input{./tikz/ex-cd-diag-2.tikz}}
\]
or any other way of connecting this wire to the same three boxes. To define a sensible notion of multi-wire connection, we need to add axioms that allow us to consider all these different ways of composing $\Bcomult$ and $\Bcounit$ equal if they connect the same wire to the same \emph{set} of wires. To achieve this, and obtain a suitable notion of structural equivalence for CD categories, we add the following equalities to those of SMCs:
\begin{equation}\label{eq:cd-cats-axioms}
\begin{gathered}
	\scalebox{1}{
\InputIfFileExists{copy-associative.tikz}{}{\input{./tikz/copy-associative.tikz}}
} \myeq{coas} \scalebox{1}{
\InputIfFileExists{copy-associative-1.tikz}{}{\input{./tikz/copy-associative-1.tikz}}
}\qquad  \qquad \scalebox{1}{
\InputIfFileExists{copy-commutative.tikz}{}{\input{./tikz/copy-commutative.tikz}}
}\myeq{cocom} \scalebox{1}{\Bcomult} \\[1em]
	\scalebox{1}{
\InputIfFileExists{copy-unital-left.tikz}{}{\input{./tikz/copy-unital-left.tikz}}
}\myeq{counl}\;\idx{}\; \myeq{counr}\scalebox{1}{
\InputIfFileExists{copy-unital-right.tikz}{}{\input{./tikz/copy-unital-right.tikz}}
}
	\end{gathered}
\end{equation}
The reader will recognise these laws from Example~\ref{ex:comonoids} as those of a \emph{commutative comonoid}: they tell us that there is only one way of splitting a single wire into $n$ wires, for any natural $n$.  

Using $\Bcomultn{x}$ and $\Bcounitn{x}$ for generating objects $x\in\Sigma_0$, we can define $\Bcomultn{w}$ and $\Bcounitn{w}$ for any word $w$ over $\Sigma_0$ or, more plainly, for arbitrarily many wires. We do so by induction:
\begin{equation}\label{eq:copy-delete-multiple-def}

\InputIfFileExists{bcomult-xv.tikz}{}{\input{./tikz/bcomult-xv.tikz}}
 \qquad \quad \qquad 
\InputIfFileExists{bcounit-xv.tikz}{}{\input{./tikz/bcounit-xv.tikz}}

\end{equation}
As mentioned in Example~\ref{ex:comonoids}, it is typical in applications that $\Bcomult$ and $\Bcounit$ are interpreted as gates that \emph{duplicating} and \emph{discard} a resource. With this perspective, CD categories are categories whose structure makes duplicating and discarding  of a resource explicit when it is used in some computation. This feature allows for a resource-sensitive analysis of processes. For instance, in CD categories we generally have that
\begin{equation}\label{eq:copy-delete-no}

\InputIfFileExists{Lawvere-distributive-copy.tikz}{}{\input{./tikz/Lawvere-distributive-copy.tikz}}
\; \neq\; 
\InputIfFileExists{Lawvere-distributive-copy-1.tikz}{}{\input{./tikz/Lawvere-distributive-copy-1.tikz}}

\end{equation}
Intuitively, this means that we distinguish the case of a process $d$ using resource of type $v$ once and then copying its output, from a process which duplicates it before letting two copies of $d$ consume it.

\subsubsection{Cartesian Categories}\label{sec:Cartesian}
Often, one would like to go further than having an operation that allows us to split and end wires---certain SMCs extend the capability of these operations to \emph{copy and delete boxes} too. This is the ability that cartesian categories\footnote{The reader familiar with category will know that a cartesian category is a category with finite cartesian products. Having cartesian products is a \emph{property} of a category. A monoidal product, on the other hand, is a \emph{structure} over a category. While cartesian products do define a monoidal product, a given category may be equipped with a monoidal product that is not its cartesian product.} give us.

The term formation rules for cartesian categories are the same as those of CD categories. The corresponding notion of structural equivalence further quotients that of CD categories with the following axioms, which capture the ability to copy and delete diagrams: for any $d:v\to w$ in $\Sigma_1$, we have
\begin{equation}\label{eq:copy-delete}

\InputIfFileExists{Lawvere-distributive-copy.tikz}{}{\input{./tikz/Lawvere-distributive-copy.tikz}}
\; \myeq{dup}\; 
\InputIfFileExists{Lawvere-distributive-copy-1.tikz}{}{\input{./tikz/Lawvere-distributive-copy-1.tikz}}
 \qquad \qquad 
\begin{tikzpicture}
	\begin{pgfonlayer}{nodelayer}
		\node [style=none] (6) at (-3.25, 0) {};
		\node [style=basic box] (10) at (-1.5, 0) {$d$};
		\node [style=black] (11) at (0, 0) {};
		\node [style=none] (15) at (-3, 0.5) {\scriptsize $v$};
	\end{pgfonlayer}
	\begin{pgfonlayer}{edgelayer}
		\draw (10) to (11);
		\draw (6.center) to (10);
	\end{pgfonlayer}
\end{tikzpicture}
}
\;\myeq{del}\;\Bcounitn{v}
\end{equation}
Note that the axiom scheme above applies to generating operations with potentially multiple wires. To instantiate it, recall what the definition of $\Bcomult$ and $\Bcounit$ for multiple wires given in~\eqref{eq:copy-delete-multiple-def}. Then, if we apply these to $g:x_1 x_2\to y_1 y_2$, we get
\[
\InputIfFileExists{dup-ex-two-wires.tikz}{}{\input{./tikz/dup-ex-two-wires.tikz}}
\qquad \qquad \quad 
\InputIfFileExists{del-ex-two-wires.tikz}{}{\input{./tikz/del-ex-two-wires.tikz}}
\]
where we omit object labels for clarity.
As for CD categories, we can now connect a given right port of a diagram to a (possibly empty) \emph{set} of left ports of another:
\[
\InputIfFileExists{ex-Cartesian-diag.tikz}{}{\input{./tikz/ex-Cartesian-diag.tikz}}
\]
But the structural notion of equivalence for cartesian categories is much coarser. For the first time in this paper, we encounter a diagrammatic language where the structural equivalence is not topological, and where equivalence cannot simply be checked by examining the connectivity of the different sub-diagrams. In practice, this can make it more difficult to identify when two diagrams are equivalent. As well as those that have the same connectivity between their different components, we can identify diagrams where one contains several copies of the same sub-diagram, connected by the same $\Bcomult$, or where one contains a sub-diagram connected to a $\Bcounit$ and the other does not, \emph{e.g.,}
\[
\InputIfFileExists{ex-Cartesian-diag.tikz}{}{\input{./tikz/ex-Cartesian-diag.tikz}}
\quad =\quad 
\InputIfFileExists{ex-Cartesian-deform.tikz}{}{\input{./tikz/ex-Cartesian-deform.tikz}}
\]
Above, from left to right, we have merged the two occurrences of $f$ and copied $d;(\Bcounit\otimes\id)$. It is helpful to break down the required equational steps:
\begin{align*}
\scalebox{0.80}{
\InputIfFileExists{ex-Cartesian-diag.tikz}{}{\input{./tikz/ex-Cartesian-diag.tikz}}
} & \myeq{dup}\; \scalebox{0.80}{
\InputIfFileExists{ex-Cartesian-diag-1.tikz}{}{\input{./tikz/ex-Cartesian-diag-1.tikz}}
}
\;\myeq{coun}\;\scalebox{0.80}{
\InputIfFileExists{ex-Cartesian-diag-2.tikz}{}{\input{./tikz/ex-Cartesian-diag-2.tikz}}
}\\
&\myeq{coun x 2}\;\scalebox{0.80}{
\InputIfFileExists{ex-Cartesian-diag-3.tikz}{}{\input{./tikz/ex-Cartesian-diag-3.tikz}}
}
\;\myeq{dup}\;\scalebox{0.80}{
\InputIfFileExists{ex-Cartesian-deform.tikz}{}{\input{./tikz/ex-Cartesian-deform.tikz}}
}
\end{align*}
In plain English, we first merge the two occurrences of $f$ using the $\mathsf{dup}$ equation (from right to left); apply the counitality axiom of the comonoid structure to get rid of the extra counit and leave a plain wire; apply the counitality axiom of the comonoid twice (from right to left this time) to produce a diagram from which the $\mathsf{dup}$ axiom applies to $d$, which is the last equality.

\begin{remark}[Cartesian categories and algebraic theories]\label{rmk:cartesian-decomposition}
The diagrammatic syntax of cartesian categories is the diagrammatic counterpart of the standard symbolic notation for algebraic theories. In this correspondence, wires take the place of variables. Since variables can be used arbitrarily many times, we need additional machinery in the diagrammatic setting to handle variable management: this is where $\Bcomult$ and $\Bcounit$ come in. Moreover,  just like we can substitute an arbitrary term for all occurrences of a give variable, we can copy and delete arbitrary diagrams using $\Bcomult$ and $\Bcounit$ with the axioms $\mathsf{dup}$-$\mathsf{del}$. This is what allows us to interpret composition as \emph{substitution}.

Let us examine the correspondence for a simple example; the general case is worked out (for the single-sorted case) in ~\cite{BonchiSZ18}. Consider the algebraic theory of monoids. It can be presented by two generating operations, $m(-,-)$ of arity $2$ and $e$ of arity $0$ (a constant) satisfying the following three axioms: $m(m(x,y),z) =m(x,m(y,z))$ and $m(e,x) = x = m(x,e)$. Terms of this algebraic theory are syntax trees whose leaves are labelled with variable names, as on the left below.
\[
\InputIfFileExists{ex-syntax-tree-monoid.tikz}{}{\input{./tikz/ex-syntax-tree-monoid.tikz}}
\qquad\leadsto \qquad 
\InputIfFileExists{ex-diagram-monoid.tikz}{}{\input{./tikz/ex-diagram-monoid.tikz}}
\]
By simply turning the tree on its side and gathering all leaves labelled by the same variable with $\Bcomult$ (or deleting those that we do not use with $\Bcounit$) we obtain the corresponding string diagram in the theory of cartesian categories, as on the right above.


We can also go in the other direction: from the diagrammatic syntax of a cartesian category over some signature, we can obtain an algebraic theory. This is done by noticing that every string diagram of such a category can be expressed as the composition of $\Bcomult$, $\Bcounit$ with diagrams that have a single outgoing wire. Indeed, the copying and deleting axioms imply that all string diagram $d\from v\to w$, with $w= x_1\dots x_n$, can be decomposed uniquely into $n$ diagrams $d_i\from v\to x_i$, $1\leq i\leq n$, where $x_i$ are generating objects. Let us see concretely what this decomposition looks like, and how to obtain the components for the case $w=x_1 x_2$: let
\[
\InputIfFileExists{Cartesian-proj.tikz}{}{\input{./tikz/Cartesian-proj.tikz}}
\]
We can then check that
\[
\InputIfFileExists{Cartesian-decompose.tikz}{}{\input{./tikz/Cartesian-decompose.tikz}}
\]
The general case is completely analogous.

In the single-sorted case, \emph{i.e.} when the set of objects contains a single generator, the connection is clear: the decomposition property above implies that every string diagram in a cartesian category can be seen as a composite of $\Bcomult$ and $\Bcounit$ with operations with \emph{arity} corresponding to the number of left wires  they have (and implicitly, co-arity one).

\end{remark}

Resuming our resource interpretation, observe that string diagrams in~\eqref{eq:copy-delete-no}, whose equality is not enforced in CD categories, are always equated in cartesian categories. This means that cartesian categories are \emph{resource-insensitive} by default, because they do not keep track of the interplay of processes and resources the same way CD categories do.

In applications, it is often interesting to enforce only a certain degree of (in)sensitivity, intermediate between CD and cartesian categories. A notable example is the one of \emph{Markov categories}. In a Markov category we can always discard string diagrams, but we cannot copy them at will. More formally, their structure is defined by dropping from the definition of cartesian category the leftmost equation in~\eqref{eq:copy-delete}. It turns out this setup is convenient to study probabilistic computation, as it provides a baseline for interpreting string diagrams as stochastic processes. See Section~\ref{sec:diagrams-science} for more pointers to the literature on the topic.

\subsubsection{Cocartesian Categories}\label{sec:cocartesian}
If we flip all diagram of the previous section along the vertical axis, we obtain the dual of a cartesian categories, namely \emph{cocartesian categories}. They extend the language of symmetric monoidal categories, not with a commutative  comonoid, but with a commutative monoid (\emph{cf.} Example~\ref{ex:monoids}) instead:
\[\derivationRule{}{\quad\Wmultn{x}\quad}{\scriptstyle{x\in\Sigma_0}}\qquad \derivationRule{}{\quad\Wunitn{x}\quad}{\scriptstyle{x\in\Sigma_0}}\]
Furthermore, we want these to \emph{merge} (or co-copy) and \emph{spawn} (or co-delete) any diagram as follows:
\begin{equation}\label{eq:cocopy-codelete}

\InputIfFileExists{coCartesian-cocopy.tikz}{}{\input{./tikz/coCartesian-cocopy.tikz}}
\;\; \myeq{codup}\;\; 
\InputIfFileExists{coCartesian-cocopy-1.tikz}{}{\input{./tikz/coCartesian-cocopy-1.tikz}}
 \qquad \qquad 
\begin{tikzpicture}
	\begin{pgfonlayer}{nodelayer}
		\node [style=none] (6) at (1.25, 0) {};
		\node [style=basic box] (10) at (-0.5, 0) {$d$};
		\node [style=white] (11) at (-2, 0) {};
		\node [style=none] (15) at (1, 0.5) {\scriptsize $y$};
	\end{pgfonlayer}
	\begin{pgfonlayer}{edgelayer}
		\draw (10) to (11);
		\draw (6.center) to (10);
	\end{pgfonlayer}
\end{tikzpicture}
}
\;\;\myeq{codel}\;\;\Wunitn{\, y}
\end{equation}

\subsubsection{Biproduct Categories}\label{sec:biproduct-category}
Categories that are both cartesian and cocartesian are called biproduct categories. They feature a comonoid and a monoid structure on each object, satisfying the $\mathsf{dup}$-$\mathsf{del}$ and $\mathsf{codup}$-$\mathsf{codel}$ axioms. Note one important consequence: if we apply $\mathsf{dup}$ to $d=\Wmult$, $\mathsf{dup}$ to $d=\Wunit$, $\mathsf{codup}$ to $d=
\begin{tikzpicture}
	\begin{pgfonlayer}{nodelayer}
		\node [style=black] (1) at (1.75, 0) {};
		\node [style=none] (4) at (0.5, 0) {};
	\end{pgfonlayer}
	\begin{pgfonlayer}{edgelayer}
		\draw (1) to (4.center);
	\end{pgfonlayer}
\end{tikzpicture}
}
$, and $\mathsf{del}$ to $d=\Wmult$, we get the following:
\begin{gather*}

\InputIfFileExists{wmult-bcomult.tikz}{}{\input{./tikz/wmult-bcomult.tikz}}
\:=\:
\InputIfFileExists{2-bcomult-sym-2-wmult.tikz}{}{\input{./tikz/2-bcomult-sym-2-wmult.tikz}}
\quad\qquad 
\InputIfFileExists{wunit-bcomult.tikz}{}{\input{./tikz/wunit-bcomult.tikz}}
\:=\:
\begin{tikzpicture}
	\begin{pgfonlayer}{nodelayer}
		\node [style=none] (6) at (2, -0.5) {};
		\node [style=white] (8) at (0.75, -0.5) {};
		\node [style=none] (14) at (2, 0.5) {};
		\node [style=white] (16) at (0.75, 0.5) {};
	\end{pgfonlayer}
	\begin{pgfonlayer}{edgelayer}
		\draw (6.center) to (8);
		\draw (14.center) to (16);
	\end{pgfonlayer}
\end{tikzpicture}
}
\\[1em] 
\InputIfFileExists{wmult-bcounit.tikz}{}{\input{./tikz/wmult-bcounit.tikz}}
\:=\:
\begin{tikzpicture}
	\begin{pgfonlayer}{nodelayer}
		\node [style=none] (6) at (0.75, 0.5) {};
		\node [style=black] (8) at (2, 0.5) {};
		\node [style=none] (14) at (0.75, -0.5) {};
		\node [style=black] (16) at (2, -0.5) {};
	\end{pgfonlayer}
	\begin{pgfonlayer}{edgelayer}
		\draw (6.center) to (8);
		\draw (14.center) to (16);
	\end{pgfonlayer}
\end{tikzpicture}
}
\quad\qquad 
\begin{tikzpicture}
	\begin{pgfonlayer}{nodelayer}
		\node [style=white] (1) at (-0.75, 0) {};
		\node [style=black] (4) at (0.5, 0) {};
	\end{pgfonlayer}
	\begin{pgfonlayer}{edgelayer}
		\draw (1) to (4);
	\end{pgfonlayer}
\end{tikzpicture}
}
\:=\: 
\InputIfFileExists{empty-diag.tikz}{}{\input{./tikz/empty-diag.tikz}}

\end{gather*}
These are the defining axioms of bimonoids, as introduced in Example~\ref{ex:bimonoids}.

\subsubsection{Hypergraph Categories}\label{sec:hypergraph-categories}
Hypergraph categories further extend the capabilities of CD categories. Like for biproduct categories, they include both a monoid and a comonoid but, as we will see, these interact differently.

The term formation rules are the same as for biproduct categories, \emph{i.e.}, those symmetric monoidal categories, with the following additions:
\[\derivationRule{}{\quad\Bcomultn{x}\quad}{\scriptstyle{x\in\Sigma_0}}\qquad \derivationRule{}{\quad\Bcounitn{x}\quad}{\scriptstyle{x\in\Sigma_0}}\qquad
\derivationRule{}{\quad\Bmultn{x}\quad}{\scriptstyle{x\in\Sigma_0}}\qquad \derivationRule{}{\quad\Bunitn{x}\quad}{\scriptstyle{x\in\Sigma_0}}\]
The first two are similar to the extra generators of CD categories. The last two are their mirror image. We write them in black instead of the white generators of cocartesian and biproduct categories since they will play a different role, as we will now see.

The corresponding notion of structural equivalence is given by the laws of symmetric monoidal categories with the addition of the axioms of  special commutative Frobenius monoids (Example~\ref{ex:commutative-special-Frobenius}) summarised in~\eqref{eq:commutative-special-Frobenius}.
However, observe that we do not impose that every diagram can be (co)copied or (co)deleted, as we did for (co)cartesian categories. This is a key difference---in fact, as we will see later, these two requirements turn out to be incompatible in a rather fundamental way.

The diagrammatic language of hypergraph categories is the most permissive: it allows any set of ports (left or right) of the same  sort to be connected together via  $\Bcomultn{x}, \Bcounitn{x}, \Bmultn{x}$, and $\Bunitn{x}$. In fact, as we have seen in Example~\ref{ex:Frobenius}, any two connected\footnote{In the sense that there is a path from any two nodes.} diagrams made exclusively of these black generators, are equal if and only if they have the same number of left and right ports, a fact known as the spider theorem. For example, we can use the defining axioms of Frobenius monoids to show that the following two diagrams are equivalent:
\[
\InputIfFileExists{spider-ex.tikz}{}{\input{./tikz/spider-ex.tikz}}
\quad = \quad 
\InputIfFileExists{spider-ex-1.tikz}{}{\input{./tikz/spider-ex-1.tikz}}
\]
This means that the only relevant structure of a given connected diagram made entirely of $\Bcomultn{x}$, $\Bcounitn{x}$, $\Bmultn{x}$, and $\Bunitn{x}$ is the number of ports on the left and on the right. As a result, as we saw in Example~\ref{ex:Frobenius}, we can introduce the following black nodes as syntactic sugar for any such diagram with $m$ dangling wires on the left and $n$ on the right:
\[
\InputIfFileExists{spider-m-n.tikz}{}{\input{./tikz/spider-m-n.tikz}}
\]

Using this convenient notation, any string diagram in a hypergraph category will look like a hypergraph, as introduced in Section~\ref{sec:graphs}: boxes act as hyperedges, which may be wired together via black nodes.
\[
\InputIfFileExists{ex-hypergraph-diag.tikz}{}{\input{./tikz/ex-hypergraph-diag.tikz}}
\]
In fact, this observation is what justifies the name `hypergraph category' for these structures. Once more, two diagrams are equivalent if they connect the same ports via black nodes. For example, the following two are equivalent:
\[
\InputIfFileExists{ex-hypergraph-diag.tikz}{}{\input{./tikz/ex-hypergraph-diag.tikz}}
\quad =\quad 
\InputIfFileExists{ex-hypergraph-diag-deform.tikz}{}{\input{./tikz/ex-hypergraph-diag-deform.tikz}}
\]
We could justify this equality through a sequence of Frobenius monoid axioms and the laws of symmetric monoidal categories, but it would be very time-consuming! It suffices to check that the the connectivity of the different labelled boxes and black nodes remains the same. This is why hypergraph categories are very appealing.


These examples lead us to observe that hypergraph categories are always self-dual compact closed.\footnote{This is in fact why they were called well-supported compact closed categories, when initially studied in~\cite{Carboni1987}.} With the previous intuition, this observation is not too surprising: if we are able to connect any set of ports, we can connect any two pairs of ports. More formally, we can define cups and caps as $\sdcupx{x} := 
\InputIfFileExists{cup-frob.tikz}{}{\input{./tikz/cup-frob.tikz}}
$ and $\sdcapx{x} :=
\InputIfFileExists{cap-frob.tikz}{}{\input{./tikz/cap-frob.tikz}}
$, for any $x$ in the signature. That they satisfy the axioms of compact closed categories is a consequence of the Frobenius monoid axioms (or, more generally, of the spider theorem). We give the diagrammatic proof explicitly here, as it is instructive:
\[
\InputIfFileExists{self-dual-s-yanking.tikz}{}{\input{./tikz/self-dual-s-yanking.tikz}}
 \;:=\;
\InputIfFileExists{s-yanking-frob.tikz}{}{\input{./tikz/s-yanking-frob.tikz}}
\;\myeq{frob}\;
\InputIfFileExists{s-yanking-frob-1.tikz}{}{\input{./tikz/s-yanking-frob-1.tikz}}
\;\myeq{coun}\;
\InputIfFileExists{s-yanking-frob-2.tikz}{}{\input{./tikz/s-yanking-frob-2.tikz}}
\;\myeq{un}\;\idx{}\]
The other equation can be proved in the same way.
That the resulting compact structure is also self-dual is immediate, since the cups and caps we have defined relate any given object to itself.

\begin{remark}[A matter of perspective.]
The reader may have noticed that the additional structure of CD categories, self-dual compact closed, and hypergraph categories can also be seen as the free SMC over a theory that includes some additional generators and equations (\emph{cf.} Section~\ref{sec:eq-theories}). For example, the free CD category over the theory $(\Sigma,E)$ is definable as the free SMC over the theory formed by signature $\left(\Sigma_0, \Sigma_1 \cup \left\{ \Bcomultn{x}, \Bcounitn{x} \mid x \in \Sigma_0 \right\}\right)$ and equations those in $E$ plus those in \eqref{eq:cd-cats-axioms}. Whether we pick one perspective or the other depends on which structure we want to see as built-in and which we want to see as domain-specific in the considered application.
\end{remark}

\subsubsection{Closed monoidal categories} \label{sec:closed}

In monoidal categories that are \emph{closed}, the set of string diagrams $v\to w$ can be seen as an object $v\multimap w$ of the category itself. Closed monoidal categories arise naturally in applications where it makes sense to consider \emph{higher-order} functions: processes that can take functions as inputs and can output other functions. Objects of the form $v\multimap w$ are called \emph{exponentials}.

The existence of exponentials $v\multimap w$ for all $v,w$ is not sufficient to form a closed monoidal category. We need extra conditions that encode the behaviour of $v\multimap w$ as some sort of function space. For this, we require the existence of a family of morphisms $\mathsf{eval}_{v,w}\from v (v\multimap w)\to w$ depicted as
\[
\InputIfFileExists{exponential-unit.tikz}{}{\input{./tikz/exponential-unit.tikz}}
\]
with the following property: for every $d\from uv\to w$, there exists a unique morphism $\Lambda_u d\from u\to (v\multimap w)$, such that
\begin{equation}\label{eq:lambda-abstr}
	
\InputIfFileExists{closed-monoidal-up.tikz}{}{\input{./tikz/closed-monoidal-up.tikz}}

\end{equation}
Intuitively, $\mathsf{eval}_{v,w}$ acts like an evaluation map that applies a function $v\to w$ to a value of type $v$ and returns a $w$. In usual programming terms, $\Lambda_u d$ is a \emph{curried} version of $d$ which, given an argument of type $u$, returns a morphism $v\to w$. We can also see $\Lambda$ as a family of morphisms, sending objects $u$ and morphisms of type $uv \to w$ to morphisms of type $u\to (v\multimap w)$, for any $u, v,w$. In fact, this defines a \emph{natural} one-to-one correspondences between sets of morphisms of these types\footnote{For a refresher on the categorical concept of naturality, see Definition~\ref{def:nat-trans} in Appendix. The categorically minded reader might also recognise an adjunction (\emph{cf.} Definition~\ref{def:adjunction} in Appendix) between the functor $(-)v$, which takes the monoidal product with an object $v$, and the exponential functor $v\multimap (-)$. In fact, a closed monoidal category is succinctly defined by the existence of such adjunction. We opted for a slightly different presentation, to emphasise the shape of string diagrams determining the closed structure.
}. This construction is also known as ($\lambda$-)\emph{abstraction} in programming language theory.

It is possible to make the string diagrammatic language of closed monoidal categories even more appealing, by introducing a pictorial notation for the abstraction map $\Lambda$, represented as a box surrounding a given string diagram. For instance, given $d \colon uv \to w$, $\Lambda_u d\from u \to (v\multimap w)$ becomes
\[
\InputIfFileExists{abstraction-box.tikz}{}{\input{./tikz/abstraction-box.tikz}}
\]
The different orientation of wires $u$ and $v$ in the box signals the different role they play: intuitively, $d$ awaits an input of type $u$ on its left, in order to form a function with input of type $v$. As the case with any string diagrammatic language introduced so far, this notation finds formal justification in terms of categorical structures: it is syntactic sugar for so-called \emph{functorial boxes}~\cite{mellies2006functorial}, which capture the behaviour of $\Lambda$. We will not cover functorial boxes in detail here, though we recall briefly what they are in Appendix.

The reader may find further details about the string diagrammatic language of closed monoidal categories in~\cite[Section 3]{ghica2023hierarchical}. We do not discuss it further, given how different it is from our other examples. We conclude by linking the closed monoidal structure to other categorical structures seen in this section.

First, as the name suggest, compact closed categories (Section~\ref{sec:compact closed}) are closed monoidal. The objects $v\multimap w$ are defined as $wv^*$ and the evaluation maps $\mathsf{eval}_{v,w}$ are defined by:
\[
\InputIfFileExists{exponential-unit.tikz}{}{\input{./tikz/exponential-unit.tikz}}
 \; :=\;
\InputIfFileExists{compact-eval.tikz}{}{\input{./tikz/compact-eval.tikz}}
\]
where $v\multimap w := wv^*$ in compact closed categories.
Moreover, in compact closed categories, abstraction can be realised by bending a wire as follows:
\[
\InputIfFileExists{abstraction-box.tikz}{}{\input{./tikz/abstraction-box.tikz}}
 \;:=\; 
\InputIfFileExists{compact-abstraction.tikz}{}{\input{./tikz/compact-abstraction.tikz}}
\]
We see that the resulting diagram has type $u\to wv^*$ as required.
The ability to represent evaluation and abstraction as just wire bending is a special property of compact closed categories, which does not hold for generic closed monoidal ones.

Second, another important class of closed categories are those that are also cartesian (Section~\ref{sec:Cartesian}). \emph{Cartesian closed} categories are the semantics of choice for most functional programming languages. The cartesian structure witnesses the fact that resources (represented as variables) may be used arbitrarily many times in most programming languages, while the closed structure reflects the ability to manipulate higher-order functions. Once again, we refer the reader to~\cite{ghica2023hierarchical} for a more extensive discussion.


\subsubsection{Mix and Match}\label{sec:mix}
We have seen several cases in which categorical structures blend together to give rise to interesting combinations. But beware! Certain combinations have undesired consequences. The classic example is that of the incompatibility between cartesian and compact closed categories. This can be made precise as the following claim: a category that is both cartesian and compact closed is degenerate, in the sense that it has at most one morphism between any two objects.\footnote{Another way to phrase this is to say that the category collapses to a partially ordered set (poset). Indeed, we may regard a poset as a category whose objects are the poset elements and there is a morphism from $x$ to $y$ only when $x \leq y$. Thus each homset contains at most one morphism.} To show this,  it suffices to derive from the axioms of cartesian and compact closed categories that all morphisms between any two objects are equal. We achieve this by proving that we can disconnect any  wire, in two steps: first we show that the cup splits as follows,
\begin{equation}\label{eq:copy-cup}

\InputIfFileExists{cup-copy.tikz}{}{\input{./tikz/cup-copy.tikz}}

\end{equation}
(the reader might recognise this as an instance of Remark~\ref{rmk:cartesian-decomposition}). Then we show that the identity can be disconnected
\begin{equation}\label{id-disconnect}

\InputIfFileExists{id-disconnect.tikz}{}{\input{./tikz/id-disconnect.tikz}}
=:\;
\begin{tikzpicture}
	\begin{pgfonlayer}{nodelayer}
		\node [style=black] (4) at (1, 0) {};
		\node [style=none] (5) at (2.25, 0) {};
		\node [style=black] (6) at (0, 0) {};
		\node [style=none] (7) at (-1.25, 0) {};
	\end{pgfonlayer}
	\begin{pgfonlayer}{edgelayer}
		\draw (4) to (5.center);
		\draw (6) to (7.center);
	\end{pgfonlayer}
\end{tikzpicture}
}
\end{equation}
Finally, we can show what we wanted: for any $f\from v \to w$,
\[
\InputIfFileExists{Cartesian-compact-poset.tikz}{}{\input{./tikz/Cartesian-compact-poset.tikz}}
\]
Therefore, modulo equivalence there is at most one string diagram of type $v \to w$. Note that the incompatibility between compact closed and cartesian structure implies that hypergraph and cartesian structure are also incompatible.

Interestingly, if we weaken the cartesian compact closed structure to that of a \emph{cartesian traced} monoidal category, the resulting combination not only avoids degeneracy, but turns out to be closely related to the notion of (parameterised) fixed-point~\cite{Hasegawa97recursionfrom}. A parameterised fixed-point operator in a cartesian SMC $(\mathcal{C},\times, 1)$ takes a morphism $f\from X\times A\to X$ and produces $f^\dagger\from A \to X$. The operator $(-)^\dagger$ is then required to satisfy a certain number of intuitive axioms. For instance, $f^\dagger$ should indeed be a fixed-point of $f$, \emph{i.e.},
\begin{equation}\label{eq:fixed-point}
\diagbox{f^\dagger}{\qquad}{\qquad} =  
\InputIfFileExists{fixpoint-eq.tikz}{}{\input{./tikz/fixpoint-eq.tikz}}

\end{equation}
It is easy to see how to define such an operation in a traced category: let
\[\diagbox{f^\dagger}{\qquad}{\qquad} :=
\InputIfFileExists{fixpoint-def.tikz}{}{\input{./tikz/fixpoint-def.tikz}}
\]
The axioms the fixed-point operator is required to satisfy are  consequences of the axioms of traced monoidal categories with those of cartesian categories. For instance, it does satisfy~\eqref{eq:fixed-point}:
\begin{equation*}\label{eq:fixed-point-trace}
\scalebox{0.85}{
\InputIfFileExists{fixpoint-eq-proof.tikz}{}{\input{./tikz/fixpoint-eq-proof.tikz}}
}
\end{equation*}
where the second equality holds by the yanking and sliding axioms of the trace. Conversely, from a given fixed-point operator, we can define a trace. In fact, the notions of parameterised fixed-points and cartesian traces are equivalent~\cite[Theorem 3.1]{Hasegawa97recursionfrom}.

We conclude by mentioning another example of a useful interaction between different structures. String diagrams for braided and self-dual compact closed categories allow us to draw arbitrary \emph{knots}:
\[
\InputIfFileExists{trefoil-knot.tikz}{}{\input{./tikz/trefoil-knot.tikz}}
\]
The central result for these categories is that two different, closed (\emph{i.e.} with empty left and right boundary) diagrams are equal if and only if the corresponding knots can be topologically deformed into one another---thus, the topological notion of knot is fully captured by a few algebraic axioms. 
We see here another advantage of working with string diagrams: they give an algebraic home to topological concepts that are otherwise difficult to express in standard algebraic syntax.

\section{Semantics}\label{sec:semantics}

So far, we have thoroughly explored an arsenal of diagrammatic syntax, each kind corresponding to a specific flavour of monoidal category. We have occasionally discussed what is the intended `meaning' of these structures, that scientists have in mind when reasoning about a certain phenomenon with string diagrams. In this section, we explain how  assigning meaning to string diagrams can be made formal, as a \emph{semantics}.

Our approach to string diagram semantics draws inspiration from the study of \emph{denotational semantics} of programming languages. In programming practice, computations are not mere manipulations of symbols, devoid of content; there is a task, a mathematical object, which we intend to describe with the program. A denotational semantics specifies what that `something' is intended to be. It allows us to define the behaviour of programs in a given language rigorously, and prove more easily certain properties that they satisfy. The same language can even have different semantic interpretations. A well-chosen semantics may allow us to circumscribe more precisely the expressiveness of the language, or to rule out certain classes of behaviour.

\subsection{From Syntax to Semantics, Functorially}
\label{sec:syntax-v-semantics}

Generally speaking, a semantics is a mapping from syntax to a domain of interpretation. Categorically, this idea may be applied to string diagrams using the ingredients introduced in the previous sections. Our starting point is a symmetric monoidal theory $(\Sigma,E)$. Then string diagrams of the free symmetric monoidal category $\FreeSMC{\Sigma,E}$ over $(\Sigma,E)$ is our syntax.   

Now, in order to interpret such syntax, a domain of interpretation should mirror its basic structure. This is why we consider categories $\mathsf{Sem}$ that are symmetric monoidal for the task. A semantics for $\FreeSMC{\Sigma,E}$ will then be a mapping that preserves such structure, that is, a symmetric monoidal functor $\sem{\cdot} \colon \FreeSMC{\Sigma,E} \to \mathsf{Sem}$.

If $\FreeSMC{\Sigma,E}$ was a generic category, in order to define $\sem{\cdot}$ we would need to come up with a definition of $\sem{v}$ and $\sem{d}$ for any object $v$ and string diagram $d$ of $\FreeSMC{\Sigma,E}$. However, because $\FreeSMC{\Sigma,E}$ is freely generated by $(\Sigma,E)$, our task is simpler. In order to fully define $\sem{\cdot}$, it suffices to assign it a value only on the generating objects and operations of $\Sigma$, and make sure they satisfy the relevant equations.

More explicitly, giving a semantic interpretation of $\FreeSMC{\Sigma,E}$ in $\mathsf{Sem}$ amounts to specifying:
\begin{itemize}
\item an object $\sem{x}$ of $\mathsf{Sem}$ for each generating object $x\in \Sigma_0$;
\item a morphism $\sem{c} \colon \sem{x_1}\otimes\dots\otimes\sem{x_n} \to \sem{y_1}\otimes\dots\otimes\sem{y_m}$ of $\mathsf{Sem}$ for each generating operation $c\in \Sigma_1$ of type $x_1\dots x_n \to y_1\dots y_n$;
\end{itemize}
Moveover, this should be done in such a way that the equations of $E$ are satisfied, in the sense that $c \myeq{E} d$ implies $\sem{c} = \sem{d}$. 

Giving such an interpretation for the generators completely defines a symmetric monoidal functor $\sem{\cdot} \colon \FreeSMC{\Sigma,E} \to \mathsf{Sem}$, in a canonical way. The semantics of an arbitrary object of $\FreeSMC{\Sigma,E}$, which is a word $w = x_1\dots x_n$ of generating objects, is computed as $\sem{w} = \sem{x_1}\otimes \dots \otimes \sem{x_n}$. The semantics of an arbitrary (composite) string diagram is computed using the composition and monoidal product in the semantics, as long as the latter has the appropriate structure:
\begin{equation}\label{eq:compositionality}
\begin{gathered}
\sem{
\InputIfFileExists{horizontal-comp.tikz}{}{\input{./tikz/horizontal-comp.tikz}}
} = \sem{\diagbox{c}{u}{v}}\poi \sem{\diagbox{d}{v}{w}} 
\\
\sem{
\InputIfFileExists{vertical-comp.tikz}{}{\input{./tikz/vertical-comp.tikz}}
} = \sem{\diagbox{d_1}{v_1}{w_1}}\otimes \sem{\diagbox{d_2}{v_2}{w_2}}.
\end{gathered}
\end{equation}
Finally, because $\sem{\cdot}$ should be a symmetric monoidal functor, symmetries $^{y}_x \ \sym \ ^x_{y}$ are mapped to symmetries $\sem{x}\otimes \sem{y}\to \sem{y}\otimes \sem{x}$ of $\mathsf{Sem}$, and similarly for the identities.

In a sense, one may regard such description of $\sem{\cdot}$ as a definition of semantics by \emph{structural induction} on string diagrammatic syntax. Remarkably, it is an inductive definition where we just need to specify the base cases, and the inductive step is always given by \eqref{eq:compositionality}.
It is worth emphasising once more that this style of definition is only possible because $\FreeSMC{\Sigma,E}$ is a \emph{free} symmetric monoidal category. Our recipe for $\sem{\cdot}$ implicitly exploits the universal property of free constructions, \emph{cf.} Remark~\ref{rmk:freeness}. Also, note that \eqref{eq:compositionality} is what we commonly refer to as the property of \emph{compositionality}: the semantics of a compound diagram is entirely determined by the semantics of its elementary components. Compositionality is a crucial property in software analysis, as it makes formal reasoning feasible at a large scale. Being able to reason semantically about graphical models using decompositions based on~\eqref{eq:compositionality} is a major appeal of string diagrammatic approaches.

One last word about syntax. In the examples of semantics that we consider below, we will often remark that the domain of interpretation $\mathsf{Sem}$ has more structure than just symmetric monoidal. In particular, we will see categories that are also cartesian, hypergraph, etc. in the sense of Section~\ref{sec:thstringdiag}. In such cases, it is often interesting to consider string diagrammatic syntax that also exhibits such structure, and study structure-preserving interpretations. To do so, we can introduce $\FreeX{\Sigma,E}
$ for the free $X$-category over $(\Sigma,E)$, where $X$ stands for one of the structures considered in Section~\ref{sec:thstringdiag}, \emph{e.g.}, cartesian, hypergraph, etc. These can be defined analogously to the free symmetric monoidal category $\FreeSMC{\Sigma,E}$, except that the built-in generators are not just the symmetries $^{y}_x \ \sym \ ^x_{y}$ and identities $
\InputIfFileExists{id-x-with-frame.tikz}{}{\input{./tikz/id-x-with-frame.tikz}}
$, but also  the generators and equations specific to that structure. For instance, the free cartesian category $\Free{Cart}{\Sigma,E}$ will have additional generators $\Bcomultn{x}$ and $\Bcounitn{x}$ for each generating object $x$, satisfying all the appropriate axioms.


\subsection{Soundness and Completeness}\label{sec:soundness-completeness}


Given a semantics $\sem{\cdot} \from  \FreeSMC{\Sigma,E} \to \mathsf{Sem}$, it is often insightful to understand which string diagrams $c$ and $d$ are identified by $\sem{\cdot}$, that is, when $\sem{c}=\sem{d}$. Investigating these equalities may inform us on the behaviour of the processes represented by string diagrams, and the differences of picking one semantics over another. First, the very existence of such a functor $\sem{\cdot}$ requires that $c\myeq{E}d$  implies $\sem{c} = \sem{d}$; otherwise, $\sem{\cdot}$ would not be well-defined. Borrowing terminology from logic, in this case we say that $E$ is \emph{sound} for $\sem{\cdot}$. Furthermore $E$ is said to be \emph{complete} if the reverse implication holds. Compared to soundness, completeness is typically much harder to prove, and often relies on identifying a `canonical shape' for the morphisms of $\mathsf{Sem}$. When we have that $c\eqE{E}d$  if and only if $\sem{c} = \sem{d}$, we say that the theory $E$ is sound and complete, and call it an \emph{axiomatisation} of the target SMC $\mathsf{Sem}$. In categorical terms, $E$ is sound and complete if $\sem{\cdot}$ is a faithful symmetric monoidal functor. The reader will often read that a theory is ``complete for $\Sem$'', rather than $\sem{\cdot}$, when the functor $\sem{\cdot}$ is clear from context. Another question that is often relevant is how expressive a diagrammatic language is. That means, we may investigate which behaviours lie in the image of $\sem{\cdot}$, and whether this image may be characterised by some property. In particular, if this image is (equivalent to) the whole of $\mathsf{Sem}$, we say that $\sem{\cdot}$ is \emph{full}.

\subsection{Examples}
\label{sec:examples-semantics}

We now cover useful examples of SMCs in which we can interpret different flavours of string diagrams. As we will see, some allow us to interpret symmetric monoidal theories with varying degrees of complexity. Some semantics have a special link with certain commonly occurring theories, in that the latter is \emph{complete} for the former. In these cases, the string diagrams for a given theory capture exactly the semantics of interest.  

Each time, we will use the same notation $\sem{\cdot}$ for the semantic functor, and only specify its domain and codomain when necessary.

\begin{example}[Functions, $\times$]\label{ex:set-Cartesian}
The category $\Set$ of sets and functions can be equipped with a monoidal structure in different ways. The cartesian product of sets is one example of a monoidal product. On objects, it is simply the set of pairs, given by $X_1\times X_2=\{(x_1,x_2)\mid x_1\in X_1\land x_2\in X_2\}$; on morphisms, it is given by $(f_1\times f_2)(x_1,x_2) = (f(x_1),f_2(x_2))$. The unit for the product is the singleton set $1 = \{\bullet\}$ (any singleton set will do).  It is straightforward to check that these satisfy the axioms of monoidal categories from~\eqref{fig:smc-axioms}. As an exercise, let us prove the interchange law:
\begin{align*}
\big((f_1\times f_2)\poi (g_1\times g_2)\big)(x_1,x_2) &= (g_1\times g_2) (f_1(x_1),f_2(x_2))
\\
& = \big(g_1(f_1(x_1)), g_2(f_2(x_2))\big)
\\
&=\big((f_1\poi g_1)(x_1),(f_2\poi g_2)(x_2)\big)
\\
& =\big((f_1\poi g_1)\times (f_2\poi g_2)\big)(x_1,x_2).
\end{align*}
(Side note: strictly speaking, using pairs for `$\times$' does not define an associative monoidal product, because $(X_1\times X_2)\times X_3$ is not \emph{equal} to $X_1\times (X_2\times X_3)$, but merely isomorphic to it. See Remarks~\ref{rmk:strictness} and~\ref{rmk:rel-not-strict}.)
Furthermore, $(Set,\times)$ is a \emph{symmetric} monoidal category, with symmetry given by the function $\sigma_X^Y\from X\times Y\to Y\times X$ defined by $\sigma_X^Y(x,y) = (y,x)$. 

Since $\Set$ is a symmetric monoidal category, we can use it to interpret string diagrams from a symmetric monoidal theory $(\Sigma, E)$. Formally, this amounts to defining a symmetric monoidal functor 
\[\sem{\cdot}\from \FreeSMC{\Sigma,E}\to\Set\]
As explained in Section~\ref{sec:syntax-v-semantics}, this places significant constraints on $\sem{\cdot}$: 
\begin{enumerate}
\item \emph{Monoidal functoriality} means that $\sem{v_1v_2} = \sem{v_1}\times \sem{v_2}$ for all $v_1,v_2\in\Sigma_0^*$, that the identity wire $\idx{v}$ over any object $v\in\Sigma_0^*$ is sent to the identity map over $\sem{v}$, and that the two compositions are preserved by $\sem{\cdot}$:
\begin{equation}\label{eq:functoriality-set}
\begin{gathered}
\sem{
\InputIfFileExists{horizontal-comp.tikz}{}{\input{./tikz/horizontal-comp.tikz}}
} = \sem{\diagbox{d}{v}{w}}\circ \sem{\diagbox{c}{u}{v}} 
\\
\sem{
\InputIfFileExists{vertical-comp.tikz}{}{\input{./tikz/vertical-comp.tikz}}
} = \sem{\diagbox{d_1}{v_1}{\;w_1}}\times \sem{\diagbox{d_2}{v_2}{\;w_2}}
\end{gathered}
\end{equation}
In addition, $\sem{\cdot}$ is also a \emph{symmetric} monoidal functor, so we necessarily have $\sem{\sym^v_w}=\sigma^X_Y$, for $\sem{v} = X$, $\sem{w}=Y$.

\item Since $\FreeSMC{\Sigma,E}$ is \emph{free},  to fully specify such a monoidal functor $\sem{\cdot}\from \FreeSMC{\Sigma}\to\Set$, it suffices to assign a set to each element of $\Sigma_0$ and a function $\sem{c}\from \sem{v}\to\sem{w}$ for each operation $c\from v\to w$ in $\Sigma_1$, such that they verify the axioms in $E$.  
\end{enumerate}
We see from points 1. and 2. above that the semantics of an arbitrary string diagram $d$ can be computed from the semantics of the generating operations of $\Sigma$ and how they are composed together to form $d$, using~\ref{eq:functoriality-set}. 

This symmetric monoidal category also has the structure to interpret string diagrams for \emph{cartesian} categories (Section~\ref{sec:Cartesian}). In fact, there is only one such structure. For an object $v$ of a given signature, the comonoid structure $\Bcomultn{v}, \Bcounitn{v}$ is given by the following copy $\Delta\from \sem{v}\to \sem{v}\times \sem{v}$ and discarding $!\from \sem{v}\to 1$ maps, defined respectively by:
\[\sem{\Bcomultn{v}}(x)= \Delta(x)=(x,x)\qquad\qquad \sem{\Bcounitn{v}}(x) = !(x) = \bullet\]
One can easily check that these satisfy the axioms of commutative comonoids~\eqref{ex:comonoids} and that any function satisfies the equations $\mathsf{dup}$ and $\mathsf{del}$ from~\eqref{eq:copy-delete}. To build a bit more intuition, let us verify $\mathsf{dup}$, for example. For any $x\in{\sem{v}}$, we have
\begin{align*}
\sem{
\InputIfFileExists{f-bcomult.tikz}{}{\input{./tikz/f-bcomult.tikz}}
}(x) & = \left(\sem{\Bcomult}\circ \sem{f}\right)(x)  = \sem{\Bcomult}\big(\sem{f}(x)\big)
\\
& = \big(\sem{f}(x),\sem{f}(x)\big) = \big(\sem{f}\times\sem{f}\big)(x) 
\\
&= \sem{
\InputIfFileExists{bcomult-2-f.tikz}{}{\input{./tikz/bcomult-2-f.tikz}}
}(x)
\end{align*}
We encourage the reader to verify the other axioms as an exercise.

Note that there is only one map $X\to 1$ for any set $X$, namely the discarding map $!_X$, given by $!_X(x) = \bullet$. In conjunction with the counitality axiom of the comonoid structure, this condition forces the interpretation of the cartesian structure to be the one we have given---there are no other possible choices. 
\end{example}

\begin{remark}[Models of algebraic theories and cartesian categories]\label{rmk:algebraic-Cartesian}
We have seen in Remark~\ref{rmk:cartesian-decomposition} that there is a close syntactic correspondence between cartesian categories and algebraic theories. It is natural to wonder weather the correspondence carries over to the semantic side. This is indeed the case: symmetric monoidal functors out of the free cartesian category over a given theory into the SMC $(\Set,\times)$ are models (in the usual algebraic sense) of the corresponding algebraic theory. For example, to specify such a functor for the cartesian theory of monoids involves choosing a carrier set $X$ and functions $\sem{m} : X\times X\to X$, $\sem{e} : 1\to X$ of the appropriate arity that satisfy the relevant axioms, which is precisely a model of the algebraic theory of monoids. 
\end{remark}

\begin{example}[Relations,$\times$]\label{ex:relational-sem}
The category $\Rel$ has as objects, sets, and as morphisms $R\from X\to Y$, binary relations, \emph{i.e.} subsets $R\subseteq X\times Y$. The composition of two relations $R\from X \to Y$ and $S\from Y\to Z$ is defined by $R\poi S = \{(x,z)\mid \exists y (x,y)\in R\land (y,z)\in S\}$.  The cartesian product $X\times Y$ further defines a monoidal product on $\Rel$, with unit the singleton set $1= \{\bullet\}$. Furthermore, $\Rel$ is \emph{symmetric} monoidal, with the symmetry $X\times Y\to Y\times X$ given by
the relation $\{((x,y), (y,x))\mid x\in X, y\in Y\}$. It is easy to verify that these satisfy all the laws of symmetric monoidal categories. 
Even though this monoidal product is the same as in $\Set$ on objects, the properties of the two SMCs are very different.

Once again, given the free symmetric monoidal category $\FreeSMC{\Sigma,E}$ over some theory $(\Sigma,E)$, specifying a symmetric monoidal functor $\sem{\cdot}\from \FreeSMC{\Sigma,E}\to\Rel$ means assigning a set to each element of $\Sigma_0$ and a relation $\sem{c}\subseteq \sem{v}\times\sem{w}$ for each $c\from v\to w$ in $\Sigma_1$ such that the axioms of $E$ are satisfied in $\Rel$. Here, monoidal functoriality, aka compositionality, means that, in particular:
\[\sem{
\InputIfFileExists{horizontal-comp.tikz}{}{\input{./tikz/horizontal-comp.tikz}}
} = \left\{(x,z)\mid\exists y\left((x,y)\in\sem{c}\land (y,z)\in\sem{d}\right)\right\}\]
\[\sem{\,
\InputIfFileExists{vertical-comp.tikz}{}{\input{./tikz/vertical-comp.tikz}}
} = \left\{((x_1,x_2),(y_1,y_2))\mid (x_1,x_2)\in\sem{d_1}\land (x_2,y_2)\in\sem{d_2}\right\}\]

One can moreover interpret string diagrams for \emph{self-dual compact closed} categories (Section~\ref{sec:self-dual-compact}) into $\Rel$, by choosing a relation for the cups $\sdcupx{v}$ and caps $\sdcapx{v}$ on every generating object $v$ of our signature, such that axiom~\eqref{eq:snake} is satisfied. Once more, there are many possible choices, but the following interpretation is a common one: 
\[
\sem{\sdcupx{v}\,} = \{(\bullet,(x,x))\mid x\in \sem{v}\}
\qquad 
\sem{\,\sdcapx{v}} = \{((x,x), \bullet)\mid x\in \sem{v}\}
\]
It is clear that these two relations satisfy the defining axiom~\eqref{eq:snake} of (self-dual) compact closed categories.

In fact, we can go even further: $\Rel$ can interpret string diagrams for \emph{hypergraph} categories (Section~\ref{sec:hypergraph-categories}). For this we need to choose a special, commutative Frobenius monoid to which we map $\Bcomultn{v},\Bcounitn{v}, \Bunitn{v},\Bmultn{v}$, for every generating object $v$ of our chosen signature. There are many possibilities. One that occurs often in the literature is an extension of the comonoid structure chosen for functions in Example~\ref{ex:set-Cartesian}. We take the diagonal relation as comultiplication and the projection as counit, with the multiplication and unit given by the converse relations. Formally:
\begin{equation}\label{eq:rel-hypergraph}
\begin{gathered}
\sem{\Bcomultn{v}} = \{(x,(x,x))\mid x\in \sem{v}\}
\quad
\sem{\Bcounitn{v}}= \{(x,\bullet)\mid x\in \sem{v}\}
\\
\sem{\Bmultn{v}} =\{((x,x),x)\mid x\in \sem{v}\} 
\quad
\sem{\Bunitn{v}} = \{(\bullet, x)\mid x\in \sem{v}\}
\end{gathered}
\end{equation}
Let us check (one side of) the Frobenius law, to see how this works in more detail. In the following, all $x$s belong to $\sem{v}$ for some $v$, which we omit:
\begin{align*}
& \big((x_1,x_2),(x_1',x_2')\big)\in \sem{
\InputIfFileExists{copy-Frobenius-left.tikz}{}{\input{./tikz/copy-Frobenius-left.tikz}}
} \\
& \Iff \big((x_1,x_2),(x_1',x_2')\big)\in \left(\sem{\idone}\times \sem{
\InputIfFileExists{bcomult.tikz}{}{\input{./tikz/bcomult.tikz}}
}\right)\poi \left(\sem{
\InputIfFileExists{bmult.tikz}{}{\input{./tikz/bmult.tikz}}
}\times\sem{\idone}\right)
\\
& \Iff \exists x_1'',x_2'', x_3''\big((x_1,x_2),(x_1'',x_2'',x_3'')\big)\in \left(\sem{\idone}\times \sem{
\InputIfFileExists{bcomult.tikz}{}{\input{./tikz/bcomult.tikz}}
}\right) 
\\ &\qquad \land \big((x_1'',x_2'',x_3''), (x_1',x_2')\big)\in \left(\sem{
\InputIfFileExists{bmult.tikz}{}{\input{./tikz/bmult.tikz}}
}\times \sem{\idone}\right)
\\
& \Iff \exists x_1''\exists x_2''\exists x_3''\big[\big((x_1 = x_1'')\land (x_2 = x_2''=x_3'')\big)
\\ &\qquad \land \big((x_1''=x_2''=x_1')\land (x_3''=x_2')\big)\big]
\\
& \Iff x_1=x_2=x_1'=x_2'
\\
& \Iff \exists x\big[(x_1=x_2=x)\land (x=x_1'=x_2')\big]
\\
& \Iff \big((x_1,x_2),(x_1',x_2')\big)\in \sem{
\InputIfFileExists{bmult-bcomult.tikz}{}{\input{./tikz/bmult-bcomult.tikz}}
}
\end{align*}
In practice, one rarely reasons this way about string diagrams for relations. There is a much more intuitive way: if we think of each wire of the diagram as being labelled by a variable, then connected networks of black nodes force all variables labelling its left and right legs to be equal. With this in mind, one may observe that any connected network of black nodes forces all the corresponding variables to be the same. This can be seen as a semantic rendition of the spider theorem (covered in Example~\ref{ex:Frobenius})! The special case we have shown above falls out as a corollary.

Functions can also be seen as relations via their \emph{graph}: 
\[
\mathsf{Graph}(f)=\{(x,y)\mid y = f(x)\}
\]
Moreover, the composite (as relations) of two functional relations is the graph of the composite of the two corresponding functions, that is, $\mathsf{Graph}(g\circ f) =$ $\mathsf{Graph}(f)\poi \mathsf{Graph}(g)$. In other words, $\mathsf{Graph}$ defines a functor $\mathsf{Graph}\from \Set\to\Rel$. We call relations that are the graph of some function, \emph{functional}.

Using $\Bcomult$, $\Bcounit$ as above, we can interpret string diagrams for cartesian categories in $\Rel$, since, as we have just seen, it contains $\Set$. However, not all interpretations of a signature that includes $\Bcomult$, $\Bcounit$ will satisfy the axioms of cartesian categories, unlike in $\Set$. In fact, in $\Rel$, it is possible to characterise functional relations purely by how they interact with $\Bcomult$, $\Bcounit$: they are precisely those that satisfy the $\mathsf{dup}$ and $\mathsf{del}$ axioms in \eqref{eq:copy-delete}, as in the example of $\Set$ above. Indeed, a relation $f$ satisfies $\mathsf{dup}$ if and only if it is single-valued, and it satisfies $\mathsf{del}$ if and only if it is total. This is a useful characterisation that often comes up in the literature.

Finally, as we have mentioned, there are other choices of special, commutative Frobenius monoids in this SMC. The interested reader will find a full classification of all such choices in~\cite{pavlovic2009quantum}. 
\end{example}

\begin{remark}[Not strict?]\label{rmk:rel-not-strict}
The observant reader may have noticed an issue with the previous examples: the SMCs of functions and relations are not strict. This is because taking the cartesian product is not strictly associative, \emph{i.e.} the set $(X\times Y)\times Z$ is not \emph{equal} to the set $X\times (Y\times Z)$. However, because of the coherence theorem for SMCs (Remark~\ref{rmk:strictness}) it is harmless to pretend that they are---and again, this is why we can draw string diagrams in this category. If the reader is still uncomfortable with this idea, we invite them to give an equivalent presentation of the same SMC that does not rely on taking pairs, but \emph{tuples} of arbitrary length. This SMC would then be the \emph{strictification} of $\Set$ or $\Rel$, and nothing of importance would be lost.

Similarly, we required the semantic functor $\sem{\cdot}$ to be strict. To be fully formal, for many examples, we should allow $\sem{w_1w_2}$ to merely be isomorphic to $\sem{w_1}\otimes\sem{w_2}$. However, for all intents and purposes, we can act as if $\sem{\cdot}$ was strict, with codomain the strictification of the semantics we have in mind, as we do here.
\end{remark}

\begin{example}[Functions, +]\label{ex:set+}
The cartesian product is only one among several possible choices of monoidal structures that one can impose on the category of sets and functions. In fact, we can turn it into an SMC in at least one other interesting way: instead of taking the monoidal product to be the cartesian product of sets, we consider the disjoint sum, defined as $X_1+X_2 := ((X_1\times \{1\})\cup (X_2\times \{2\}))$ on objects, and given by $f_1+f_2 = (f_1+f_2)(x,i) = f_i(x)$ on maps. The unit for this monoidal product is the empty set. Moreover it is also symmetric monoidal, with symmetry $\varsigma^X_Y \from X+Y\to Y+X$ given by $\varsigma(z,1) = (z,2)$ and $\varsigma(z,2) = (z,1)$. 

If we can no longer interpret diagrams for cartesian categories in this SMC, we can however interpret those for \emph{cocartesian} categories (Section~\ref{sec:cocartesian}). For each generating object $v$ of our chosen signature, we can interpret the commutative monoid operations $\Wmultn{v}$ and $\Wunitn{v}$ as the following maps:
\[
\sem{\Wmultn{v}}(x,i)= x \;\text{for } x\in\sem{v} \text{ and } i=1,2 
\qquad\quad
\sem{\Wunitn{v}} = \varnothing
\]
It is a straightforward exercise to check that these are associative, unital and comutative. Moreover, every map (of the appropriate type) satisfies the $\mathsf{codup}$ and $\mathsf{codel}$ axioms with respect to $\Wmultn{v}$ and $\Wunitn{v}$. 

Amazingly, \emph{every} map between finite sets can be represented using this syntax. We only need to give ourselves a single generating object $\bullet$ in our signature, which we interpret as the singleton set $\sem{\bullet} = 1$. Then we simply write $\Wmultn{\bullet}$ and $\Wunitn{\bullet}$  as $\Wmult$ and $\Wunit$ since object labels are redundant in this context. Given a map $f:X\to Y$, for $X$ and $Y$ two finite sets, we first fix some ordering of $X$ and $Y$. In this way, we can identify them with finite sets of the form $\{0,\dots, n\}$. This allows us to encode finite sets as sequences of wires in the diagrammatic setting (we will assume a similar encoding for several other examples below). With this encoding fixed, we can represent any map $f\from X\to Y$: use as many $\Wmult$ as necessary to connect all elements $x$ in the domain to the single $y = f(x)$ in the codomain to which $f$ sends them; those elements of $Y$ that are not in the image of $f$ are each connected to a $\Wunit$. Here are a few examples of the translation:
\[
\InputIfFileExists{ex-function-1.tikz}{}{\input{./tikz/ex-function-1.tikz}}
\qquad 
\InputIfFileExists{ex-function-2.tikz}{}{\input{./tikz/ex-function-2.tikz}}
\quad 
\InputIfFileExists{ex-function-3.tikz}{}{\input{./tikz/ex-function-3.tikz}}
\]
where the first is given by $f(0)=f(1)=f(2)=1$, the second by $g(0)=g(2)=0$ and $g(1)=g(3)=1$, and the last is the unique map from the empty set to $\{0,1\}$.

Notice that there are several ways of drawing the same function, depending on how we choose to arrange the different $\Wmult$ and $\Wunit$. The following diagrams all represent $f\from \{0,1,2\}\to\{0,1\}$ above:
\[
\InputIfFileExists{ex-function-1-1.tikz}{}{\input{./tikz/ex-function-1-1.tikz}}
\]
In a cocartesian category, all these diagrams are equal, as $\Wmult$ and $\Wunit$ form a commutative monoid. This is the first instance of a monoidal theory we encounter that fully characterises the chosen semantics: \emph{the free cocartesian category over a single object} (and no morphisms) is equivalent to the SMC $(\fSet,+)$ of finite sets and functions, with the disjoint sum as monoidal product. In other words, the symmetric monoidal theory of a commutative monoid is \emph{complete} for this semantics (in the sense explained in Section~\ref{sec:soundness-completeness}). This also means that $\FreeSMC{\Sigma,E}$, the free symmetric monoidal category over $\Sigma =\left(\bullet, \{\Wmult,\Wunit\}\right)$ and where $E$ is the theory of commutative monoids (Example~\ref{ex:monoids}), is equivalent to $(\fSet,+)$. Any two diagrams made of $\Wmult$ and $\Wunit$ that denote the same map can be shown equal using only the axioms of commutative monoids. 

The proof of this fact is typical for this kind of completeness result: it works by showing how, given an arbitrary $\Sigma$-diagram $d$, we can rewrite it to some normal form, using only the equations of commutative monoids. The chosen normal form is one from which the corresponding relation can be recovered uniquely: we can choose for example to eliminate all $\Wunit$ connected to a $\Wmult$ using unitality ($\mathsf{un}$) and to associate all connected $\Wmult$ to the top using associativity ($\mathsf{as}$). Then, any two diagrams have the same normal form if and only if they are interpreted as the same map. The fact that any diagram can be rewritten to a normal form using only the axioms above, is typically proven by induction, considering each individual cases, much like normalisation proofs in programming language theory, or cut elimination proofs in logic. Also, much like these, they tend to be quite tedious and combinatorial, so we do not reproduce it here.
\end{example}

\begin{example}[Bijections, $+$]
If we restrict the previous example to bijections  (one-to-one and onto functions), we obtain the simplest example of a SMC---call it $\Bij$. String diagrams for $\Bij$ are simply permutations of the wires! If we restrict further to finite ordinals, the resulting SMC is equivalent to $\FreeSMC{\Sigma}$, \emph{the free SMC} over the signature $\Sigma = (\{\bullet\}, \varnothing)$, the SMC of permutations we have already encountered in Example~\ref{ex:free-smc-single-object}.
\end{example}

\begin{example}[Relations, $+$]\label{ex:rel+}
As for functions, the disjoint sum gives another interesting monoidal product on relations. On objects, it remains the same: $X_1+X_2 := ((X_1\times \{1\})\cup (X_2\times \{2\})$. On morphisms, $R_1+R_2$ is given by $((x,i),(y,i))\in R_1+R_2$ if and only if $(x,y)\in R_i$ for some $i\in\{1,2\}$. Once again, the unit is the empty set and the symmetry is the graph of the corresponding function $\varsigma$ seen above. In this case, we cannot interpret string diagrams for compact closed nor hypergraph categories.

However, with the disjoint sum as monoidal product, we can interpret string diagrams for cartesian as well as cocartesian categories in $(\Rel,+)$. For each generating object $v$ of a given signature, the monoid of the cocartesian structure is given by the graph of the relations that give the cocartesian structure to $(\Set,+)$:
\[
\sem{\Wmultn{v}}= \{((x,i), x)\mid x\in \sem{v}, i=1,2\}
\qquad \sem{\Wunitn{v}} = \varnothing
\]
For the cartesian structure on with copying and deleting relations given by the converse of the above relations:
\[
\sem{\Bcomultn{v}}= \{(x, (x,i))\mid x\in \sem{v}, i=1,2\}
\qquad \sem{\Bcounitn{v}} = \varnothing
\]
One can check that $\Bcomultn{v},\Bcounitn{v}$ form a commutative comonoid for any interpretation of $v$, and that they can copy and delete any relation, \emph{i.e.}, that they satisfy the $\mathsf{dup}$ and $\mathsf{del}$ axioms from~\eqref{eq:copy-delete}. This means that we can interpret string diagrams for \emph{biproduct} categories (Section~\ref{sec:biproduct-category}) in $(\Rel,+)$.

Intuitively, we can think of the diagrams $\Wmult,\Wunit,\Bcounit,\Bcomult$ in this category as directing the flow of a single token that travels around the wires. The intuition here is that the $\Wmult$ transfers to the right wire the token that comes through any one of its left wires, $\Wunit$ is able to generate a token, $\Bcomult$ is a non-deterministic fork and $\Bcounit$ a dead end.
As we have already said, the relations for $\Wmult$ and $\Wunit$ are simply the graphs of the functions defined in the category of sets and functions. This makes sense: functions can only direct the token deterministically from left to right. Hence, they lack the nondeterministic $\Bcomult$ and $\Bcounit$.

Note that the particle intuition for diagrams in $\Rel$ with the disjoint sum as monoidal product is quite different from the intuition for diagrams in $\Rel$ with the cartesian product, where the variables for all wires are set to compatible values globally, all at once. For this reason, diagrams in biproduct categories are sometimes called \emph{particle-style}, while those of self-dual compact closed categories are said to be \emph{wave-style}. A more systematic discussion of this perspective can be found in~\cite{abramsky2005retracing}.

As in the previous example, we can represent \emph{any} relation between finite sets, using only the signature $\Sigma =\left( \{\bullet\}, \left\{\Bcomult, \Bcounit,\Wunit,\Wmult\right\}\right)$, where we set $\sem{\bullet} = 1$ and interpret $\Bcomult, \Bcounit,\Wunit,\Wmult$ as the relations above.
In our interpretation, a relation $R \from X\to Y$ corresponds to a diagram $d$ with $|X|$ wires on the left and $|Y|$ wires on the right. The $j$-th port on the left is connected to the $i$-th port on the right exactly when $(i,j)\in R$. For example,
the relation $R\from \{0,1,2\}\to \{0,1\}$ given by $\{(0,0), (0,1), (2,1)\}$,
can be represented by the following diagram:
\begin{equation}\label{ex:relation}

\InputIfFileExists{ex-relation.tikz}{}{\input{./tikz/ex-relation.tikz}}

\end{equation}
We see that this representation extends that of functions, by adding the possibility of connecting one wire on the left to several (or none) on the right. This is precisely the difference between functions and relations, between determinism and non-determinism, reflected in the diagrams.

Note that a relation can also be seen as a matrix with Boolean coefficients. The relationship between the string diagrams above and matrices (over arbitrary semirings) will be explained in Example~\ref{ex:matrices-product}.

Not only do these string diagrams allow us to represent any relation between finite sets, we can also produce an axiomatisation of this SMC, \emph{i.e.} a sound and complete equational theory for the chosen semantics. To do this, we simply quotient $\Sigma$-diagrams by the axioms of an idempotent, commutative bimonoid:
\begin{equation}\label{eq:axioms-idpt-bimonoid}
\begin{gathered}
 
\InputIfFileExists{copy-associative.tikz}{}{\input{./tikz/copy-associative.tikz}}
 = 
\InputIfFileExists{copy-associative-1.tikz}{}{\input{./tikz/copy-associative-1.tikz}}
\qquad   
\InputIfFileExists{copy-unital-left.tikz}{}{\input{./tikz/copy-unital-left.tikz}}
=\idx{} =\;
\InputIfFileExists{copy-unital-right.tikz}{}{\input{./tikz/copy-unital-right.tikz}}

 \\ 
 
\InputIfFileExists{copy-commutative.tikz}{}{\input{./tikz/copy-commutative.tikz}}
=\; \Bcomult
 \\
  
\InputIfFileExists{wmult-associative.tikz}{}{\input{./tikz/wmult-associative.tikz}}
 = 
\InputIfFileExists{wmult-associative-1.tikz}{}{\input{./tikz/wmult-associative-1.tikz}}
\qquad   
\InputIfFileExists{wmult-unital-left.tikz}{}{\input{./tikz/wmult-unital-left.tikz}}
=\idx{} =\;
\InputIfFileExists{wmult-unital-right.tikz}{}{\input{./tikz/wmult-unital-right.tikz}}

  \\ 
  
\InputIfFileExists{wmult-commutative.tikz}{}{\input{./tikz/wmult-commutative.tikz}}
=\; \Wmult
 \\

\InputIfFileExists{wmult-bcomult.tikz}{}{\input{./tikz/wmult-bcomult.tikz}}
\:=\:
\InputIfFileExists{2-bcomult-sym-2-wmult.tikz}{}{\input{./tikz/2-bcomult-sym-2-wmult.tikz}}
\quad\qquad 
\InputIfFileExists{wunit-bcomult.tikz}{}{\input{./tikz/wunit-bcomult.tikz}}
\:=\:
\begin{tikzpicture}
	\begin{pgfonlayer}{nodelayer}
		\node [style=none] (6) at (2, -0.5) {};
		\node [style=white] (8) at (0.75, -0.5) {};
		\node [style=none] (14) at (2, 0.5) {};
		\node [style=white] (16) at (0.75, 0.5) {};
	\end{pgfonlayer}
	\begin{pgfonlayer}{edgelayer}
		\draw (6.center) to (8);
		\draw (14.center) to (16);
	\end{pgfonlayer}
\end{tikzpicture}
}
\qquad 
\InputIfFileExists{wmult-bcounit.tikz}{}{\input{./tikz/wmult-bcounit.tikz}}
\:=\:
\begin{tikzpicture}
	\begin{pgfonlayer}{nodelayer}
		\node [style=none] (6) at (0.75, 0.5) {};
		\node [style=black] (8) at (2, 0.5) {};
		\node [style=none] (14) at (0.75, -0.5) {};
		\node [style=black] (16) at (2, -0.5) {};
	\end{pgfonlayer}
	\begin{pgfonlayer}{edgelayer}
		\draw (6.center) to (8);
		\draw (14.center) to (16);
	\end{pgfonlayer}
\end{tikzpicture}
}

\\ 
\begin{tikzpicture}
	\begin{pgfonlayer}{nodelayer}
		\node [style=white] (1) at (-0.75, 0) {};
		\node [style=black] (4) at (0.5, 0) {};
	\end{pgfonlayer}
	\begin{pgfonlayer}{edgelayer}
		\draw (1) to (4);
	\end{pgfonlayer}
\end{tikzpicture}
}
\:=\:\idzero  \qquad

\InputIfFileExists{bcomult-wmult.tikz}{}{\input{./tikz/bcomult-wmult.tikz}}
= \idx{}
\end{gathered}
\end{equation}
This equational theory turns out to be complete for our choice of $\sem{\cdot}$. In other words, any two diagrams $c,d$ made from $\Bcomult, \Bcounit$, $\Wunit,\Wmult$ are equal modulo the axioms in~\eqref{eq:axioms-idpt-bimonoid} if and only if $\sem{c} = \sem{d}$, \emph{i.e.}, if and only if they denote the same relation. 

\end{example}
\begin{remark}\label{rmk:between-maps-and-relations}
We just saw that going straight from functions to relations (with the disjoint sum as product) amounts to adding $\Bcomult$ and $\Bcounit$ to $\Wmult$ and $\Wunit$. These two examples fit into a hierarchy of expressiveness, from bijections to relations:
\[
\InputIfFileExists{prop-hierarchy.tikz}{}{\input{./tikz/prop-hierarchy.tikz}}
\]
Of course, any subset of the generators $\Bcomult,\Bcounit, \Wunit,\Wmult$ gives a well-defined sub-SMC of ($\Rel$,$+$). We have not included all $2^4$ of them as they do not all correspond to well-known mathematical notions.

We should also note that the ways in which these theories are combined define distributive laws, a topic we mentioned in Remark~\ref{rmk:distributive-law}.
\end{remark}

\begin{example}[Spans, $\times$]\label{ex:spans}
The category $\Span{\Set}$ has sets as objects and, as morphisms $X\to Y$, pairs of maps $f:A\to X, g:A\to Y$ with the same set $A$ as domain. We will write spans as $\spn{X}{f}{A}{g}{Y}$. One way to think about spans is as \emph{witnessed} or \emph{proof-relevant} relations. In other words, they keep track of the way in which two elements are related: an element $a$ of the apex $A$ can be thought of as a witness or a proof of the fact that $(f(a),g(a))$ are related by the span. Thus, the difference with relations is that there may be several ways in which two elements from $X$ and $Y$ are related by the same span  $(A,f,g)$; if $f(a)=f(a')=x$ and $g(a)=g(a')=y$, then $(x,y)$ are related by two different witnesses $a$ and $a'$. 

The composition of two spans is obtained by computing what is called \emph{the pullback} of $g$ and $p$ below and composing the resulting outer two functions on each side:
\begin{equation*}
  \xymatrix{
    && A\times_Y B \ar[dl]_{\pi_1}\ar[dr]^{\pi_2}\\
    & A \ar[dl]_{f}\ar[dr]^{g} & & B\ar[dl]_{p} \ar[dr]^{q}\\
    X\,  & & Y & & \, Z
  }
\end{equation*}
where $A\times_Y B := \{(a,b) \mid (g(a) = p(b)\}$ and $\pi_1,\pi_2$ are the two projections onto $A$ and $B$. Thus, the composition of $\spn{X}{f}{A}{g}{Y}$ followed by $\spn{Y}{p}{B}{q}{Z}$ is $\spn{X}{f\circ \pi_1}{A\times_Y B}{q\circ \pi_2}{Y}$. For a set $X$, the identity span is $\spn{X}{id_X}{X}{id_X}{X}$. As given, this operation is not strictly associative or unital. To make $\Span{\Set}$ into a bona fide category, we need to identify all isomorphic spans: two spans $\spn{X}{f}{A}{g}{Y}$ and $\spn{Y}{p}{B}{q}{Z}$ are isomorphic when there is a bijection $h: A \to B$ such that $p\circ h = f$ and $q\circ h = g$.

$\Span{\Set}$ can be made into a symmetric monoidal category with the cartesian product of sets: on objects $X_1\times X_2$ is the usual set of pairs of elements of $X_1$ and $X_2$, on morphisms $(\spn{X_1}{f_1}{A_1}{g_1}{Y_1})\otimes (\spn{X_2}{f_2}{A_2}{g_2}{Y_2}) = (\spn{X_1\times X_2}{f_1\times f_2}{A_1\times A_2}{g_1\times g_2}{Y_1\times Y_2})$. With the singleton set as unit and the symmetry as $\spn{X\times Y}{id}{X\times Y}{\sigma_X^Y}{Y\times X}$ where $\sigma_X^Y(x,y) = (y,x)$ as before, this equips $\Span{\Set}$ with a symmetric monoidal structure.

Furthermore, like relations, we can interpret string diagrams for hypergraph categories in the SMC of spans. There are many possible choices of where to map the Frobenius monoid $\Bcomultn{v}, \Bcounitn{v}$, $\Bunitn{v}, \Bmultn{v}$ for a given generating object $v$ of a given signature. However, there is one evident choice, dictated by the presence of finite products: 
\begin{equation}\label{eq:span-hypergraph}
\begin{gathered}
\sem{\Bcomultn{v}} = \spn{\sem{v}}{id}{\sem{v}}{\Delta}{\sem{v}\times \sem{v}}\quad \sem{\Bcounitn{v}}= \spn{\sem{v}}{id}{\sem{v}}{!}{1}\\
\sem{\Bmultn{v}} = \spn{\sem{v}\times \sem{v}}{\Delta}{\sem{v}}{id}{\sem{v}}\quad \sem{\Bunitn{v}} = \spn{1}{!}{\sem{v}}{id}{\sem{v}}
\end{gathered}
\end{equation}
Here $\Delta$ is the usual diagonal map, defined by $\Delta(x) = (x,x)$, and $!$ is the unique map $\sem{v}\to 1$. The proof that these satisfy the axioms of commutative Frobenius monoids can be computed much like for relations, with the added complexity that one has to keep track of the witnesses in each apex.

Notice that if we forget the apex of each of these, and only keep track of the pairs that they relate, the resulting relations are exactly those that give $\Rel$ its hypergraph structure, in~\eqref{eq:rel-hypergraph}. This stems from a more general fact about spans and relations. If spans (of sets or any category with categorical products) of type $X\to Y$ can be seen as maps $A\to  X\times Y$, relations are precisely \emph{injective} (or monomorphic, in the general categorical setting) spans. One can always obtain a relation from a span $A\to X\times Y$ by first factorising the map into a surjective map (epimorphism) followed by an injective (monomorphism) one, and keeping only the latter. The interested reader will find the construction of $\Rel$ from $\Span{\Set}$ explained in more detail in~\cite{Gadducci1998,Zanasi16}.
\end{example}

\begin{example}[Spans,+]\label{ex:spans+}
 As for functions and relations, spans can also be made into a symmetric monoidal category with the \emph{disjoint sum} as monoidal product. On objects, it is defined in the same way, and on spans, it is given by taking the disjoint sum of each pair of legs of the two spans. 
 
With this monoidal product, we can use the same signature as in Example~\ref{ex:rel+} to represent any span of finite sets. Let $\Sigma = (\{\bullet\}$, $\{\Bcomult$, $\Bcounit$, $\Wmult$, $\Wunit\})$ with the following slightly modified interpretation: $\sem{\bullet} = 1$ and (omitting the only object label)
\[
\sem{\Bcomultn{}}= \spn{1}{\nabla}{1+1}{id}{1+1}
\qquad \sem{\Bcounitn{}} = \spn{1}{0}{\varnothing}{id}{\varnothing}
\]
\[
\sem{\Wmultn{}}= \spn{1+1}{id}{1+1}{\nabla}{1}
\qquad \sem{\Wunitn{}} = \spn{\varnothing}{id}{\varnothing}{0}{1}
\]
where $\nabla:1+1\to 1$ is defined by $\nabla(x,i)=x$ and $0:\varnothing\to 1$ is the only map from the empty set.
These also satisfy the axioms for string diagrams of biproduct categories, \emph{cf.} Section~\ref{sec:biproduct-category}. In fact, it is \emph{the free biproduct category over a single object} and no additional morphism~\cite[5.3]{Lack2004a}.

As for relations, we can use this signature to represent any span of finite sets with $+$ as monoidal product: for the span $\spn{\{0,\dots,n\}}{f}{A}{g}{\{0,\dots,n\}}$, there is a path from the $i$-th wire on the left to the $j$-th one on the right in the corresponding diagram for each element of $\{a\in A\mid f(a)=j, g(a)=i\}$. For example,
the span $\spn{\{0,1,2\}}{f}{\{0,1,2,3\}}{g}{\{0,1\}}$, with $f(0)=f(1)=f(2)=0$, $f(3)=2$ and $g(0)=0,g(1)=g(2)=g(3)=1$,
can be represented by the following diagram:
\[
\InputIfFileExists{ex-span.tikz}{}{\input{./tikz/ex-span.tikz}}
\]
Notice that this diagram denotes the same \emph{relation} as in~\eqref{ex:relation}, but the two represent different spans.

Another way to understand the correspondence is to observe that spans of finite sets can be seen as matrices with coefficients in $\N$. Because we identify isomorphic spans, the specific label of each witness in the apex plays no role. In this sense a span just keeps track of \emph{how many ways} two elements in each of its legs are related.
More precisely, given the span $\spn{\{0,\dots,n\}}{f}{A}{g}{\{0,\dots,n\}}$, we can represent it as an $m\times n$ matrix whose $(i,j)$-th coefficient is the cardinality of $\{a\in A\mid f(a)=j, g(a)=i\}$.  The diagrammatic calculus for matrices will be explained in more details in Example~\ref{ex:matrices-product} below. This perspective is also developed in~\cite{Bruni01somealgebraic}, where the authors study some of the algebraic properties of the SMC of spans and their dual, \emph{cospans}, which we introduce next.
\end{example}

\begin{example}[Cospans]\label{ex:cospans}
Cospans, as their name indicates, are formed by inverting the arrows in the definition of spans. Let $\Cospan{\Set}$ be the category with sets as objects and morphisms $X\to Y$ given by pairs of maps $f:X\to A$ and $g: Y\to A$ with the same set $A$ as codomain, which we write as $\cospn{X}{f}{A}{g}{Y}$.
The composition of two cospans is obtained by computing what is called \emph{the pushout} of $g$ and $p$ below and composing the resulting outer two functions on each side:
\begin{equation*}
  \xymatrix{
    && A+_Y B \\
    & A \ar[ur]_{\iota_1} & & B \ar[ul]^{\iota_2} \\
    X\ar[ur]_{f} & & Y\ar[ul]^{g}\ar[ur]_{p} & & \, Z \ar[ul]^{q}
  }
\end{equation*}
where $A+_Y B =\big(\{(a,1) \mid a\in A\}\cup \{(b,2) \mid b\in B\}\big)/\sim$ where $\sim$ is the equivalence relation defined by $(a,1)\sim (b,2)$ if and only if $a=g(y)$ and $b=p(y)$ for some $y\in Y$, and $\iota_1,\iota_2$ are the obvious inclusion maps of $A$ and $B$ into $A+_YB$. Then, the composition of $\cospn{X}{f}{A}{g}{Y}$ with $\cospn{Y}{p}{B}{q}{Z}$ is $\cospn{X}{\iota_1\circ f}{A\times_Y B}{\iota_2\circ q}{Y}$. For a set $X$, the identity cospan is $\cospn{X}{id_X}{X}{id_X}{X}$. As for spans, this operation is not strictly associative or unital. To make $\Cospan{\Set}$ into a bona fide category, we need to identify all isomorphic cospans: two cospans $\cospn{X}{f}{A}{g}{Y}$ and $\cospn{Y}{p}{B}{q}{Z}$ are isomorphic when there is a bijection $h: A \to B$ that makes the two resulting triangles commute.

Like for spans, we can equip cospans with the structure of a SMC---this time with the disjoint sum as monoidal product. Take $X_1+X_2$ to be the monoidal product on objects and, on morphisms, $(\cospn{X_1}{f_1}{A_1}{g_1}{Y_1})\otimes (\cospn{X_2}{f_2}{A_2}{g_2}{Y_2}) = \cospn{X_1+X_2}{f_1+f_2}{A_1+A_2}{g_1+g_2}{Y_1+Y_2}$ where $(f_1+f_2)(x,i) = f_i(x)$; the empty set is the unit of this monoidal product and the symmetry is given by $\cospn{X+Y}{id}{X+Y}{\iota_2+\iota_1}{Y+X}$, where $\iota_1 : X\hookrightarrow X$, $\iota_2 : Y\hookrightarrow X$ are the injections into the first and second component given respectively by $\iota_1(x) = (x,1)$ and $\iota_2(y) = (y,2)$.

Once again, we can interpret Frobenius monoids into the SMC of cospans and thus draw string diagrams for hypergraph categories. For any generating object $v$ of a chosen signature, let
\begin{equation}
\begin{gathered}
\sem{\Bcomultn{v}} = \cospn{\sem{v}}{id}{\sem{v}}{\nabla}{\sem{v}+\sem{v}}\qquad \sem{\Bcounitn{v}} = \cospn{\sem{v}}{id}{\sem{v}}{0}{\varnothing}
\\
\sem{\Bmultn{v}} = \cospn{\sem{v}+\sem{v}}{\nabla}{\sem{v}}{id}{\sem{v}} \qquad \sem{\Bunitn{v}} = \cospn{\varnothing}{0}{\sem{v}}{id}{0}
\end{gathered}
\end{equation}
where $\nabla: \sem{v}+\sem{v}\to \sem{v}$ is defined as before by $\nabla(x,i) = x$ and $0 : \varnothing \to \sem{v}$ is the unique map from the empty set to $\sem{v}$. Notice the similarity (and differences) with~\eqref{eq:span-hypergraph}.

In fact, there is more than a coincidental relationship between cospans and hypergraph categories: $(\Cospan{\fSet},+)$, the SMC of cospans restricted to finite sets, is equivalent to \emph{free hypergraph category on a single object} and no morphisms~\cite[5.4]{Lack2004a}, that is, on the signature $\Sigma = (\{\bullet\}, \varnothing)$. Another way to say the same thing is that $(\Cospan{\fSet},+)$ is equivalent to $\FreeSMC{\scFrob}$, the free SMC over the theory of a special commutative Frobenius monoid (\emph{cf.} Example~\ref{ex:commutative-special-Frobenius}). 

This means in particular that any cospan between finite sets (seen again as ordinals $\{0,\dots, n\}$) can be represented as a diagram using only $\Bcomult,\Bcounit,\Bunit,\Bmult$ (where we omit the single object label $\bullet$ once again) and $\sem{\bullet} = 1$. To build some intuition for this correspondence, take for example the pair $f:\{0,1,2,3\}\to\{0,1,2\}$ and $g: \{0,1\}\to \{0,1,2\}$, given by $f(0)=f(2)=0$, $f(1)=f(3)=1$ and $g(0)=g(1)=1$; this cospan can be depicted as:
\begin{equation}\label{eq:ex-cospan}

\InputIfFileExists{ex-cospan.tikz}{}{\input{./tikz/ex-cospan.tikz}}
\qquad \text{ or, using spider  notation }\quad 
\InputIfFileExists{ex-cospan-spider.tikz}{}{\input{./tikz/ex-cospan-spider.tikz}}

\end{equation}
The rule of thumb is easy to formulate: each element of the apex of the cospan---here, $\{0,1,2\}$---corresponds to one connected network of black generators, and a boundary point is connected to a black dot if it is mapped to the corresponding apex element by (one of the legs of) the cospan. There is only one way of forming such a network from the generators modulo the axioms of special commutative Frobenius monoids, by the spider theorem, a result we saw in Example~\ref{ex:Frobenius}.

Completeness means that string diagrams in $\FreeSMC{\scFrob}$ are equal if and only if they denote the same cospan. The proof of this fact is essentially the spider theorem from Example~\ref{ex:commutative-special-Frobenius}. This theorem gives a normal form from which we can uniquely read the corresponding cospan: any diagram of the free hypergraph category on a single object is fully and uniquely characterised by the number of disconnected components (spiders) and to which of these each boundary point is connected. This is the same as defining a cospan! 

Finally, many hypergraph categories can be seen as categories of cospans equipped with additional structure~\cite{fong2015decorated}. Interestingly, not \emph{all} hypergraph categories can be described in this way. For that, we need the notion of \emph{corelation}, which we cover in the next example.

\end{example}

\begin{example}[Corelations]\label{ex:corelations}
After seeing the last few examples, it is natural to wonder: spans are to relations as cospans are to \emph{what}? The answer is \emph{equivalence relations}, also known as \emph{corelations} in this context~\cite{CF}. It turns out that we can organise equivalence relations into a SMC. In fact, they can be organised into a hypergraph category.

A corelation $C\from X\to Y$ is an equivalence relation (\emph{i.e.} a reflexive, symmetric and transitive relation) over $X+Y$. Given two corelations $C\from X\to Y$ and $D\from Y\to Z$, their composition $C\poi D \from X\to Z$ is defined by glueing together equivalence classes from $C$ and $D$ along shared elements. To define it formally, we temporarily rename $C.D$ the usual composition of relations (\emph{cf.} Example~\ref{ex:relational-sem}) and let $R^*$ be the transitive closure of a relation $R$; then $C\poi D $ is the restriction of $C\cup D\cup (C.D)^*$ to elements of $X+Z$. Intuitively, two elements $a$ and $b$ are in the same equivalence class of $C\poi D$ if there exists some sequence of elements of $X+Y+Z$, that are equivalent either according to $C$ or $D$, starting with $a$ and ending with $b$.
The disjoint sum of sets can be extended to corelations to give a monoidal product. It is moreover symmetric with symmetry given by $\sem{\sym^v_w} = \{\{(x,1),(x,2)\}x\in \sem{v}\}\cup \{\{(y,1),(y,2)\} y \in\sem{w}\}$.
Once more, we can interpret the Frobenius monoids that define hypergraph categories in this SMC---for a generator $v$ of our signature, let
\begin{equation}
\begin{gathered}
\sem{\Bcomultn{v}} = \{\{(x,1),((x,1),2),((x,2),2)\}\mid x\in \sem{v}\}
\\ 
\sem{\Bcounitn{v}} = \sem{v}
\\
\sem{\Bmultn{v}} = \{\{((x,1),1),((x,2),1),(x,2)\}\mid x\in \sem{v}\} 
\\ 
\sem{\Bunitn{v}} = \sem{v}
\end{gathered}
\end{equation}
In plain English, $\Bcomultn{v}$ (resp. $\Bmultn{v}$) is mapped to the equivalence relation over $\sem{v}+(\sem{v}+\sem{v})$ (resp. $(\sem{v}+\sem{v})+\sem{v}$) that identifies all occurrences of $x\in \sem{v}$ in the different components of the disjoint  sum.

As we did for cospans, it is easy to represent any corelation between finite sets as a diagram using only $\Bcomult,\Bcounit,\Bunit,\Bmult$. For example, the corelation $\{0,1,2,3\}\to \{0,1\}$ given by the two equivalence classes $\{\{(0,1),(2,1)\},\{(1,1),(3,1), (0,2),(1,2)\}\}$ over the disjoint sum $\{0,1,2,3\}+\{0,1\}$, can be depicted by any of the following string diagrams, using spider notation:
\[
\InputIfFileExists{ex-cospan-spider.tikz}{}{\input{./tikz/ex-cospan-spider.tikz}}
\qquad\quad 
\InputIfFileExists{ex-corelation-spider.tikz}{}{\input{./tikz/ex-corelation-spider.tikz}}
\]
Observe that the first is the same diagram as in~\eqref{eq:ex-cospan}. The individual black dot, which represents an element of the apex of the cospan that was not in the image of any of the two leg maps, is missing in the second diagram. These two string diagrams represent the same corelation, since an isolated black dot represents an empty equivalence class: $\sem{\bullet}: =\sem{
\begin{tikzpicture}
	\begin{pgfonlayer}{nodelayer}
		\node [style=black] (1) at (-2, 0) {};
		\node [style=black] (4) at (-0.75, 0) {};
	\end{pgfonlayer}
	\begin{pgfonlayer}{edgelayer}
		\draw (1) to (4);
	\end{pgfonlayer}
\end{tikzpicture}
}
}=\varnothing$.  In diagrammatic terms, this means that we can always remove networks of black generators that are not connected to any boundary points, using the fact that $
\begin{tikzpicture}
	\begin{pgfonlayer}{nodelayer}
		\node [style=black] (1) at (-2, 0) {};
		\node [style=black] (4) at (-0.75, 0) {};
	\end{pgfonlayer}
	\begin{pgfonlayer}{edgelayer}
		\draw (1) to (4);
	\end{pgfonlayer}
\end{tikzpicture}
}
 = \idzero$. However, the two string diagrams represent different cospans. This is an instance of a more general observation: at the semantic level, the only difference between corelations and cospans is that the former do not allow empty equivalence classes. 

By the spider theorem, a network of black generators is fully characterised by its number of legs. Thus,  there is only one such network, up to the laws of special commutative Frobenius monoid: the single dot $\bullet := 
\begin{tikzpicture}
	\begin{pgfonlayer}{nodelayer}
		\node [style=black] (1) at (-2, 0) {};
		\node [style=black] (4) at (-0.75, 0) {};
	\end{pgfonlayer}
	\begin{pgfonlayer}{edgelayer}
		\draw (1) to (4);
	\end{pgfonlayer}
\end{tikzpicture}
}
$. Putting all of the above together with the completeness result for cospans, we can get a similar completeness result for corelations: for this, we need only add a single axiom to remove isolated dots to the theory of special commutative Frobenius monoids: 
\[
\begin{tikzpicture}
	\begin{pgfonlayer}{nodelayer}
		\node [style=black] (1) at (-2, 0) {};
		\node [style=black] (4) at (-0.75, 0) {};
	\end{pgfonlayer}
	\begin{pgfonlayer}{edgelayer}
		\draw (1) to (4);
	\end{pgfonlayer}
\end{tikzpicture}
}
 = \idzero\]
The resulting theory is known as the theory of \emph{extraspecial} commutative Frobenius monoids \cite[Theorem 1.1]{CF}.

Recall that relations can be seen as jointly injective spans, \emph{i.e.} injective maps $R\hookrightarrow X\times Y$. Dually, corelations $C\from X\to Y$ are jointly \emph{surjective} cospans, \emph{i.e.} surjective maps $X+Y\twoheadrightarrow S$. We have already seen that, given a span $A\rightarrow X\times Y$, one can extract a relation $R$ by factorising it into $A\twoheadrightarrow R\hookrightarrow X\times Y$, a surjective map followed by an injective map. Similarly, one can obtain a corelation from a cospan by keeping only the surjective map in the factorisation of the corresponding map $X+Y\to S$. As we have just seen, diagrammatically, this corresponds to removing isolated black dots. In category theory, the factorisation of $\Set$ maps into a surjective map followed by an injective one can be abstracted into a notion called a \emph{factorisation system}. It turns out that corelations can be defined for different factorisation systems than the surjective-injective one. Moreover, their apex can be decorated with additional structure. In fact, these two generalisations are so powerful that every hypergraph category can be constructed as a category of \emph{decorated corelations}~\cite{Fong16,fong2018decorated}.
\end{example}

\begin{example}[Linear maps, $\otimes$]\label{ex:linear-maps-tensor}
The category $\fVect$ of finite-dimensional vector spaces (over some chosen field $\Field$) and linear maps is also a symmetric monoidal category, in at least two different ways. This example deals with the tensor product, while the next one considers the direct product.

We will not go over the rigorous definition of the tensor product of vector spaces here; suffices to say that $X_1\otimes X_2$ can be defined as a quotient of the free vector space over $X_1\times X_2$ that make $\otimes$ bilinear. On morphisms, it is uniquely specified as the linear map $(f_1\otimes f_2)$ that satisfies $(f_1\otimes f_2)(u_1\otimes u_2) = f(x_1)\otimes f_2(x_2)$. This defines a SMC, with unit the field $\Field$ itself, since $X\otimes \Field \cong X$, and symmetry the map fully characterised by $\sigma(x_1\otimes x_2) =x_2\otimes x_1$.

Crucially, this SMC is \emph{not} cartesian: like in $\Rel$, not all interpretations of a theory containing a commutative comonoid structure $\Bcomultn{v}$,$\Bcounitn{v}$ for each generating object of the signature, satisfy the axioms of cartesian categories ($\mathsf{dup}$ and $\mathsf{del}$). The intuition here is that, when we choose an interpretation of the comultiplication operation $\Bcomultn{v}$ over $\sem{v}$, we also choose some set of vectors $x\in \sem{v}$ that this operation copies, \emph{i.e.} that verify:
\begin{equation}\label{eq:copyable-states}
\sem{
\InputIfFileExists{v-bcomult.tikz}{}{\input{./tikz/v-bcomult.tikz}}
\;} = \sem{\;
\InputIfFileExists{2-v.tikz}{}{\input{./tikz/2-v.tikz}}
\;\;}
\end{equation}
But then, given two such copyable vectors $x_1,x_2$, consider their sum $x= x_1 + x_2$---we should have
\begin{align*}
\sem{
\InputIfFileExists{u-bcomult.tikz}{}{\input{./tikz/u-bcomult.tikz}}
\;} & = \sem{
\InputIfFileExists{v1-bcomult.tikz}{}{\input{./tikz/v1-bcomult.tikz}}
\;}+\sem{
\InputIfFileExists{v2-bcomult.tikz}{}{\input{./tikz/v2-bcomult.tikz}}
\;} = \sem{\:
\InputIfFileExists{v1xv1.tikz}{}{\input{./tikz/v1xv1.tikz}}
\;}+\sem{\:
\InputIfFileExists{v2xv2.tikz}{}{\input{./tikz/v2xv2.tikz}}
\;}
\\
& = x_1\otimes x_1 + x_2\otimes x_2
\end{align*}
On the other hand
\begin{align*}
\sem{\:
\InputIfFileExists{uxu.tikz}{}{\input{./tikz/uxu.tikz}}
\;} &= x\otimes x = (x_1+x_2)\otimes (x_1+x_2) 
\\
&= x_1\otimes x_1 + x_1\otimes x_2 + x_2\otimes x_1 + x_2\otimes x_2
\end{align*}
Thus, no linear map can copy all elements of a given vector space, as is required for the comonoid structure of a cartesian category.

In fact, we can interpret string diagrams for \emph{compact closed} categories in $(\fVect,\otimes)$ (recall that the requirement of cartesian-ness and compact closed-ness are incompatible in the sense explained in Section~\ref{sec:thstringdiag}). For a given generating object $v$, its dual $v^*$ is interpreted as the algebraic dual of $\sem{v}$ in the usual sense, \emph{i.e.}, as $\sem{v}^*$, the space of linear maps $\sem{v}\to \Field$. Then, the cap on a vector space $\sem{v}$ is the unique linear map $\sem{v}^*\otimes \sem{v}\to \Field$ that satisfies $\sem{\,\capx{v}}(f\otimes x) = f(x)$ (also known as the evaluation map). The cup is its  adjoint: to describe it explicitly, we need to pick a basis $\{e_i\}_i$ of $\sem{v}$ and a dual basis $\{f_i\}_i$ of $\sem{v}^*$ in the sense that $f_i(e_j) =1$ if $i=j$ and $0$ otherwise; $\sem{\cupx{v}\,}$ is then the map $\Field \to \sem{v}\times \sem{v}^*$ given by extending $1\mapsto \sum_i e_i\otimes f_i$ by linearity.
In summary, using the bases $\{e_i\}_i$ and $\{f_i\}_i$ for both cups and caps, we have:
\[
\sem{\cupx{v}\,}(k) = \sum_i k (e_i\otimes f_i)\quad \sem{\,\capx{v}}\left(\sum_{i,j} \lambda_{i,j} (e_i\otimes f_j)\right) = \sum_i \lambda_{i} f_i(e_i)
\]
However, observe that the resulting maps are independent of the specific choice of bases. With these expressions, we can verify the yanking equation for $\cupx{v}$ and $\capx{v}$. Let $u$ be some element of $\sem{v}$ such that $x:=\sum_i \lambda_i e_i$; we have:
\begin{align*}
\sem{
\InputIfFileExists{s-yanking-X.tikz}{}{\input{./tikz/s-yanking-X.tikz}}
}(x) & = \left(\sem{\cupx{}} \otimes \sem{\idright}\right)\poi \left(\sem{\idright}\otimes \sem{\capx{}}\right)(x)
\\
& = \left(\sem{\idright}\otimes \sem{\capx{}}\right)\left(\left(\sum_i (e_i\otimes f_i)\right)\otimes x\right)
\\
& = \sum_i f_i(x) e_i
\\
& = \sum_i f_i\left(\sum_{j}\lambda_j e_j\right) e_i
\\
& = \sum_i \sum_{j}\lambda_j f_j(e_j) e_i
\\
& = \sum_i \lambda_i e_i =: x
\end{align*}
In this SMC, the elements of the vector space $\sem{v}$ are precisely the morphisms $\Field \to \sem{v}$.

As we have seen, every compact closed category is also traced. In fact, the name (partial) \emph{trace} comes from linear algebra, where the trace $\Tr f$ of a linear map $f:\R^n\to\R^n$ is the sum of the diagonal coefficients of its matrix representation in any basis. Thus, we expect that,
\[\sem{
\InputIfFileExists{trace-linear-map.tikz}{}{\input{./tikz/trace-linear-map.tikz}}
} = \Tr f = \Tr A = \sum_i a_{ii}\]
where $A = (a_{ij})$ is the matrix that represents the action of $f$ on some chosen basis.
It is a nice exercise to show that this is indeed the case. It is then immediate to derive certain well-known properties of the trace in linear algebra, such as $\Tr(AB) = \Tr(BA)$ or, more generally, that it is invariant under circular shifts.

Another important feature is that commutative and special Frobenius monoid in $(\fVect,\otimes)$ correspond to a choice of a basis for the supporting vector space~\cite[Section 6]{Coecke2012a}. Even if there is no linear copying map for all the elements of a vector space, we have seen above that, when we choose a comultiplication operation $\Bcomultn{v}$ over $\sem{v}$ we also choose some set of elements $x\in \sem{v}$ that $\Bcomultn{v}$ copies. Conversely, given any basis, we can define a comonoid operation that copies its elements, \emph{i.e.}, whose comultiplication and counit are defined respectively by extending the following maps by linearity: $e_i\mapsto e_i\otimes e_i$ and $e_1\mapsto 1$. What about the monoid? A monoid in $(\fVect,\otimes)$ is more commonly known as an \emph{algebra}. Any basis defines not only a comonoid but an algebra given by extending the comparison map $e_i\otimes e_j\mapsto \delta^i_j e_i$ by linearity. Not only that, the corresponding monoid-comonoid pair satisfies the Frobenius axioms and define a special and commutative Frobenius monoid. Conversely, the copyable states of any commutative and special Frobenius monoid form a basis of $\sem{v}$. The last direction is more difficult to prove, and we will not do so here. Instead, we refer the interested reader to the lectures notes of Vicary and Heunen, who deal with a related case in detail~\cite[Chapter 5]{HeunenVicaryBook} and use string diagrams throughout.

A historical note: one of the earliest instances of string diagrams are \emph{Penrose graphical notation}~\cite{penrose1971applications} for working with tensors, which are precisely string diagrams for $(\fVect,\otimes)$, later systematised and generalised in~\cite{JOYAL199155}.
\end{example}

\begin{example}[Matrices, $\oplus$]\label{ex:matrices-product}
Another possible monoidal product is given by the direct sum of vector spaces $X_1\oplus X_2$ on objects and by $(f_1\oplus f_2)(x_1, x_2) = (f_1(x_1), f_2(x_2))$ on morphisms. The unit of the product is the vector space $\{0\}\cong \mathbb{K}^0$ and, with the symmetry given by $\sigma_X^Y(x,y) = (y,x)$, the resulting structure is a SMC. It is well-known that isomorphic finite-dimensional vector spaces are uniquely identified by their dimension. Therefore, in the same way that we identified finite sets with finite ordinals, we will restrict our attention to the subcategory of  $\fVect$ whose objects are $\mathbb{K}^n$ for some  $n\in\N$. We can go even further: given a linear map $\mathbb{K}^m\to\mathbb{K}^n$, we can identify it with its representation in the canonical bases of $\mathbb{K}^m$ and $\mathbb{K}^n$. We call $\Mat{\Field}$ the category whose objects are natural numbers (representing the dimension of a vector space) and morphisms $m\to n$ are $n\times m$ matrices (notice the reversal). Nothing is lost, since $\Mat{\Field}$ and $\fVect$ are equivalent.

The SMC $(\Mat{\Field},\oplus)$ can interpret diagrams for cartesian categories: given any object $v$ of some signature, the canonical comonoid structure over the vector space $\sem{v} = \mathbb{K}^n$ is given by
\[\sem{\Bcomultn{v}}(x)=(x,x)\qquad\qquad \sem{\Bcounitn{v}}(x) = \bullet\]
Since linear maps are maps with extra structure, this comonoid is inherited from $\Set$ (\emph{cf.} Example~\ref{ex:set-Cartesian}) and the proof that it satisfies the axioms of comonoid as well as the copying and deleting axioms $\mathsf{dup}$ and $\mathsf{del}$, is similar.
We can also interpret diagrams for cocartesian categories in $\Mat{\Field}$: for any object $v$, the monoid structure is given by addition and zero:
\[\sem{\Wmultn{v}}(x_1,x_2)=x_1+x_2\qquad\qquad \sem{\Wunitn{v}} = 0\]
Note that the comonoid and monoid do not interact to form a Frobenius monoid but a \emph{bimonoid} (\emph{cf.} Section~\ref{ex:bimonoids}). In fact, we have even more structure: these string diagrams satisfy the axioms of \emph{biproduct categories} (Section~\ref{sec:biproduct-category}). This means that, maps do not only satisfy the $\mathsf{dup}$ and $\mathsf{del}$ axioms of cartesian categories, but the dual axioms of cocartesian categories $\mathsf{codup}$ and $\mathsf{codel}$. In semantic terms, these last two axioms are simply implied by linearity: all maps preserve addition. Note that this structure is very similar to that of relations with the disjoint sum as monoidal product (\emph{cf.} Example~\ref{ex:rel+}), the chief difference being that the bimonoid is not idempotent for matrices over a field. The close similarity between the two cases comes from the fact that relations can be seen as matrices, not over field, but over the semiring of the Booleans.

With the bimonoid above, we are very close to being able to express all matrices diagrammatically. As before, we will use the signature $\Sigma =(\{\bullet\}, $ $\{\Bcomult$, $\Bcounit$, $\Wmult$, $\Wunit\}\cup \{\scalar{a}\mid a\in \Field\}$. We interpret the single generating object as $\sem{\bullet} = \mathbb{K}$. Contrary to the case of relations, we cannot express arbitrary matrices with just $\Bcomult$, $\Bcounit$, $\Wmult$, $\Wunit$; this is why we have added a new generating operation $\scalar{a}$ for each $a\in\Field$, intended to represent scalar multiplication and interpreted correspondingly: 
\[\sem{\scalar{a}}(x) = ax\] 
Before explaining the encoding of matrices, there are few special cases of $\scalar{a}$ that we should mention: multiplying by $1$ is the same as the identity, so $\sem{\scalar{1}} = \sem{\idone}$, and the result of multiplying by zero is always zero, so $\sem{\scalar{0}} = \sem{
\InputIfFileExists{bcounit-wunit.tikz}{}{\input{./tikz/bcounit-wunit.tikz}}
}$.

Putting all these ingredients together, we are now ready to represent matrices. An $n\times m$ matrix $A = (a_{ij})$ corresponds to a diagram $d$ with $m$ wires on the left and $n$ wires on the right---the left ports can be interpreted as the columns and the right ports as the rows of $A$. The left $j$-th port is connected to the $i$-th port on the right through an $a$-weighted wire whenever coefficient $a_{ij}$ is a scalar $a\in \Field$. When coefficient $a_{ij}$ is $0$, they are disconnected. In addition, given that $\sem{\scalar{a}} = \sem{\idone}$, we can simply draw the connection as a plain wire when $a_{ij}=1$ and since $\sem{\scalar{0}} = \sem{
\InputIfFileExists{bcounit-wunit.tikz}{}{\input{./tikz/bcounit-wunit.tikz}}
}$ we can also omit a connecting wire when $a_{ij}=0$. Conversely, given a diagram, we recover the matrix by summing weighted paths from left to right ports. For example, the matrix
\begin{equation*}
  \label{eq:matrix-diagram}
  A =
 \begin{pmatrix}
  a & 0 & 0 \\
  b & 0 & 1
 \end{pmatrix}
\end{equation*}
can be represented by any of the following diagrams, which are all semantically equal (\emph{i.e.}, represent the same matrix):
\begin{equation*}
\begin{gathered}

\InputIfFileExists{ex-matrix-coeffs-1.tikz}{}{\input{./tikz/ex-matrix-coeffs-1.tikz}}
 \:\text{ or }\quad 
\InputIfFileExists{ex-matrix-coeffs.tikz}{}{\input{./tikz/ex-matrix-coeffs.tikz}}
 \:\text{ or }
\\
\quad
\InputIfFileExists{ex-matrix.tikz}{}{\input{./tikz/ex-matrix.tikz}}

\end{gathered}
\end{equation*}
The dotted boxes in the diagram on the top left represent the columns of the corresponding matrix.

Amazingly, we can then quotient the diagrammatic syntax by an equational theory that makes these three equal. More generally, we can give an axiomatisation of $(\Mat{\Field},\oplus)$. 
The equational theory is very similar to that of relations with the disjoint union. It has all axioms of~\eqref{eq:axioms-idpt-bimonoid} \emph{except} the last one, namely $
\InputIfFileExists{bcomult-wmult.tikz}{}{\input{./tikz/bcomult-wmult.tikz}}
= \idx{}$ (which encodes $x+x=x$, a specific feature of the Boolean semiring). Furthermore, we need axioms that encode the additive and multiplicative structure of $\Field$, namely:
\begin{equation}\label{eq:scalar-semiring-axioms}
\begin{gathered}

\InputIfFileExists{bcomult-r+s-wmult.tikz}{}{\input{./tikz/bcomult-r+s-wmult.tikz}}
=\scalar{r+s}\qquad \quad 
\InputIfFileExists{bcounit-wunit.tikz}{}{\input{./tikz/bcounit-wunit.tikz}}
= \scalar{0}
\\

\begin{tikzpicture}
	\begin{pgfonlayer}{nodelayer}
		\node [style=reg] (1) at (-1, 0) {$r$};
		\node [style=none] (4) at (-2.75, 0) {};
		\node [style=none] (9) at (2.75, 0) {};
		\node [style=reg] (10) at (1, 0) {$s$};
	\end{pgfonlayer}
	\begin{pgfonlayer}{edgelayer}
		\draw (1) to (4.center);
		\draw (1) to (10);
		\draw (10) to (9.center);
	\end{pgfonlayer}
\end{tikzpicture}
}
=\scalar{rs}\qquad \quad \idx{}= \scalar{1}
\end{gathered}
\end{equation}
Finally, we need to make sure that the scalars can be copied and deleted and that scalar multiplication distributes over addition; we can obtain these from the usual $\mathsf{dup}$-$\mathsf{del}$ and $\mathsf{codup}$-$\mathsf{codel}$ for scalars:
\begin{equation}\label{eq:scalar-copy-del}
\begin{gathered}

\InputIfFileExists{r-bcomult.tikz}{}{\input{./tikz/r-bcomult.tikz}}
=
\InputIfFileExists{bcomult-r+r.tikz}{}{\input{./tikz/bcomult-r+r.tikz}}
\qquad \quad 
\begin{tikzpicture}
	\begin{pgfonlayer}{nodelayer}
		\node [style=black] (1) at (-1.5, 0) {};
		\node [style=reg] (4) at (-2.75, 0) {$r$};
		\node [style=none] (14) at (-4, 0) {};
	\end{pgfonlayer}
	\begin{pgfonlayer}{edgelayer}
		\draw (1) to (4);
		\draw (14.center) to (4);
	\end{pgfonlayer}
\end{tikzpicture}
}
 = \Bcounit
\\

\InputIfFileExists{r+r-wmult.tikz}{}{\input{./tikz/r+r-wmult.tikz}}
=
\InputIfFileExists{wmult-r.tikz}{}{\input{./tikz/wmult-r.tikz}}
\qquad \quad 
\begin{tikzpicture}
	\begin{pgfonlayer}{nodelayer}
		\node [style=white] (1) at (-4, 0) {};
		\node [style=reg] (4) at (-2.75, 0) {$r$};
		\node [style=none] (14) at (-1.5, 0) {};
	\end{pgfonlayer}
	\begin{pgfonlayer}{edgelayer}
		\draw (1) to (4);
		\draw (14.center) to (4);
	\end{pgfonlayer}
\end{tikzpicture}
}
 = \Wunit
\end{gathered}
\end{equation}
Taken with the axioms of~\eqref{eq:axioms-idpt-bimonoid} \emph{minus} the last one, the axioms listed in~\eqref{eq:scalar-semiring-axioms}-\eqref{eq:scalar-copy-del} give a complete theory for matrices over $\Field$: diagrams modulo these equations are equal if and only if they denote the same matrix.



Note that everything we have claimed in this example would have worked as well with an arbitrary semiring $\Rig$, instead of a field: we would just need to consider matrices with coefficients in $\Rig$ and have generating operations $\scalar{a}$ for all $a\in\Rig$. From this general result, combined with the equivalence between spans and matrices with coefficients in $\N$, we can derive the following corollary: the free biproduct category over a single generating object (and no morphism) is an axiomatisation of the SMC $(\Span{\fSet},+)$ (Example~\ref{ex:spans+}).
\end{example}

\begin{example}[Linear relations, $\times$]\label{ex:linrel}
In the last two examples, we have considered linear \emph{maps} with different monoidal products. It is possible to extend the notion of linearity to \emph{relations}: given two vector spaces $X$ and $Y$, a linear relation $X\to Y$ is a linear subspace of $X\oplus Y$, \emph{i.e.} a subset of the direct sum that is closed under linear combinations. The composition (as relations) of two linear relations is still a linear relation (exercise), and the identity relation is linear. Therefore, linear relations can be organised into a category. We call $\LinRel$, the category whose objects are natural numbers and morphisms $m\to n$ are linear relation $\Field^m\to \Field^n$. With the direct sum, $\LinRel$ become a SMC, with unit and symmetry the same as those of $\Mat{\Field}$ (\emph{cf.} previous example).

This SMC has a very rich structure. Firstly, just like any function can be seen as a relation, any linear map $f$ can be seen as a linear relation $\mathsf{Graph}(f)$, by taking its graph: $\mathsf{Graph}(f):=\{(x,y)\mid y=f(x)\}$. Thus, we can also interpret the diagrams that allowed us to depict linear maps/matrices diagrammatically in $\LinRel$: taking $\Bcomultn{},\Bcounitn{},\Wunitn{},\Wmult{}$, where each wire represent a single generating object $\bullet$, interpreted as $\sem{\bullet} = \Field$. Their interpretation as relations is given by the graph of the corresponding maps:
\begin{align*}
\sem{\Bcomultn{}} = \{(x,(x,x))\mid x\in \Field\}\qquad \sem{\Bcounitn{}} = \{(x,\bullet)\mid x\in \Field\}\\
\sem{\Wmultn{}} = \{((x_1,x_2), x_1+x_2)\mid x_1,x_2\in \Field\}\qquad \sem{\Wunitn{}} = \{(0,\bullet)\}
\end{align*}
Interestingly, the converse of these relations are also linear; to depict them, we add to our signature the mirror image of the corresponding diagrams: $\Bmultn{},\Bunitn{},\Wcounitn{},\Wcomultn{}$, with semantics given by
\begin{align*}
\sem{\Bmultn{}} = \{((x,x),x)\mid x\in \Field\}\qquad \sem{\Bunitn{}} = \{(\bullet,x)\mid x\in \Field\}\\
\sem{\Wcomultn{}} = \{(x_1+x_2, (x_1,x_2))\mid x_1,x_2\in \Field\}\qquad \sem{\Wcounitn{}} = \{(\bullet,0)\}\\
\end{align*}
Notice that these are the same relations as the first two, with the pairs flipped. We can do this for any map $f$: let $\mathsf{coGraph}(f):=\{(y,x)\mid f(x)=y\}$. If $f$ is linear, its cograph will also be a linear relation. This duality will translate into a pleasant symmetry of the equational theory, which we now cover. 


There is a complete axiomatisation of $\LinRel$ with the direct sum, which is sometimes called the theory of \emph{Interacting Hopf algebras}, or IH for short. The reader will find the complete theory and further details in~\cite{BonchiSZ-JPAA}. We discuss its most salient features below, less formally.

As for Example~\ref{ex:rel+}, the image by $\sem{\cdot}$ of diagrams made from $\Bcomult$, $\Bcounit$, $\Wunit$, $\Wmult$ are precisely those relations that are the graph of some linear map (aka a matrix). For these, the equational theory is the same as in that example: essentially, a commutative bimonoid, with additional axioms to encode scalar multiplication and addition. The nice thing is that their colour-swap $\Wcomult,\Wcounit,\Bunit,\Bmult$ satisfy exactly the same axioms. These two facts take care of all interactions between the black and white generators. We also need to specify how diagrams of the same colour interact: $\Bcomult,\Bcounit,\Bunit,\Bmult$ and $\Wcomult,\Wcounit,\Wunit,\Wmult$ both form extraspecial commutative Frobenius monoids (Example~\ref{ex:Frobenius})! The remaining  axioms specify the behaviour of scalars $\scalar{a}$, which may now encounter their mirrored version:
\[
\InputIfFileExists{scalar-inverse.tikz}{}{\input{./tikz/scalar-inverse.tikz}}
\;\;\text{ for } a\neq 0\]
where
\[
\InputIfFileExists{scalar-inverse-def.tikz}{}{\input{./tikz/scalar-inverse-def.tikz}}
\]
These axioms force the mirrored version of $\scalar{a}$ to be division by $a$, which we have over any field, as long as $a$ is non-zero. 
The two cups and caps are also related in an obvious way:
\[
\InputIfFileExists{ih-cup-cap.tikz}{}{\input{./tikz/ih-cup-cap.tikz}}
\]
From these axioms, most of linear algebra can be reformulated, with subspaces and linear maps on an equal (diagrammatic) footing.

As an elementary illustration of the basic principles of diagrammatic reasoning in linear algebra, let us look at systems of linear equations. The idea is simple: a system of linear equations in the form $Ax=0$ can be expressed by simply plugging $\Wcounit$ into the right side of a diagram that encodes the matrix $A$ (Example~\ref{ex:matrices-product}), \emph{i.e.}, by the diagram $
\begin{tikzpicture}
	\begin{pgfonlayer}{nodelayer}
		\node [style=white] (69) at (2, 0) {};
		\node [style=basic box] (70) at (0, 0) {$A$};
		\node [style=none] (71) at (-2, 0) {};
	\end{pgfonlayer}
	\begin{pgfonlayer}{edgelayer}
		\draw (71.center) to (70);
		\draw (70) to (69);
	\end{pgfonlayer}
\end{tikzpicture}
}
$. For example,
\[  \begin{pmatrix} a & 1 & 0\\ b & 0 & 1\end{pmatrix} \begin{pmatrix} x_1\\ x_2\\x_3\end{pmatrix} = \begin{pmatrix} 0\\ 0\end{pmatrix}\quad \text{ becomes }\quad  
\InputIfFileExists{ex-kernel.tikz}{}{\input{./tikz/ex-kernel.tikz}}
\]
Computing a basis of the set of solutions then involves rewriting the diagram into a form from which any solution can be generated easily:
\[
\InputIfFileExists{ex-kernel-computation.tikz}{}{\input{./tikz/ex-kernel-computation.tikz}}
\]
Here we find that the kernel of $A$ has dimension $1$, with basis vector, \emph{e.g.}, $\begin{pmatrix} 1 & -a & -b\end{pmatrix}^T$.
\end{example}
\begin{remark}
If we identify the complete equational theory of a particular structure in a semantic model, we can seek it in different models, and thereby identify seemingly unrelated algebraic objects as instances of the same abstract structure. For example, we have seen many different interpretations of Frobenius monoids or bimonoids in different models (\emph{i.e.} in different symmetric monoidal categories). Another common example is that of  groups and Hopf algebras, both instances of bimonoids in different symmetric monoidal categories (sets and functions with the cartesian product for the former, and vector spaces and linear maps with the tensor product for the latter). Even a complicated theory such as IH occurs in other contexts than that of linear relations. Indeed, IH can be interpreted in in the category of vector spaces \emph{with the tensor product} as monoidal product, where its models are closely related to the notion of complementary observables in quantum physics~\cite{interactingObservables,duncan2016interacting}.
\end{remark}

\begin{example}[Monotone relations, $\times$]\label{ex:monotone-relations}
So far all the examples of compact closed categories we have covered (spans, relations, cospans, corelations) also have the structure of hypergraph categories. Of course, there are compact closed categories where the compact structure does not come from some chosen Frobenius monoid. The category of \emph{monotone relations} is one such example. It has pre-ordered sets as objects (that is, sets equipped with a reflexive and transitive binary relation); its morphisms $(X,\preceq)\to (Y,\leq)$ are relations $R\subseteq X\times Y$ that preserve the order in the following sense: if $(x,y)\in R$ and $x'\preceq x$, $y\leq y'$ then $(x',y')\in R$. The composition of monotone relation is the same as the usual composition of relations (recalled in Example~\ref{ex:relational-sem}). Since the composition of two monotone relations is monotone, they form a category, with identity on each object $(X,\preceq)$ given by the pre-order relation $\preceq\subseteq X\times X$ itself.

As for plain relations, the cartesian product defines a monoidal product: $(X,\preceq)\times(Y,\leq) := (X\times Y, \preceq\times \leq)$ with $(x,y)\preceq\times \leq(x',y')$ if and only if $x\preceq x'$ and $y\leq y'$. The unit is still the singleton set $1=\{\bullet\}$ with the only possible pre-order.

As anticipated, this category can interpret the string diagrams of compact closed categories. For this, given an object $v$ of a given signature, we need to define its dual $v^*$: if $\sem{v} = (X,\preceq)$ then $\sem{v^*} = (X,\succeq)$, the same underlying set with the opposite pre-order relation $\succeq :=\preceq^{op}$. The cups and caps on each object are then given by
\[\sem{\cupx{v}\,}= \{(\bullet, (x',x))\mid x\preceq x'\}\qquad  \sem{\,\capx{v}}= \{( (x,x'),\bullet)\mid x'\preceq x\}\]
Let us check one of the defining equations of compact closed categories:
\[
\InputIfFileExists{s-yanking-X.tikz}{}{\input{./tikz/s-yanking-X.tikz}}
 =\idxright{}\]
The left-hand side of this equation has the following semantics:
\begin{align*}
&\sem{
\InputIfFileExists{s-yanking-X.tikz}{}{\input{./tikz/s-yanking-X.tikz}}
} = \left(\sem{\cupx{}} \times \sem{\idxright{}}\right)\poi \left(\sem{\idxright{}}\times \sem{\capx{}}\right)
\\
& = \big(\{(\bullet, (x,x'))\mid x\preceq x'\}\times \preceq\big)\poi \big(\preceq \times \{( (x,x'),\bullet)\mid x\preceq x'\}\big)
\\
& = \big(\{(x,(x_1,x_2,x_3))\mid x_2\preceq x_1\land x\preceq x_3\}\big)\poi \\
& \qquad \big(\{(x_1,(x_2,x_3),x')\mid x_1\preceq x'\land x_3\preceq x_2 \}\big)
\\
& = \left\{(x,x')\mid \exists x_1\exists x_2 \exists x_3\big[x\preceq x_3 \land x_3\preceq x_2\land  x_2\preceq x_1\land x_1\preceq x' \big]\right\}
\\
& = \{(x,x')\mid x\preceq x'\}
\end{align*}
where the last step hold by transitivity of $\preceq$. This is clearly the same relation as $\preceq$ itself, which is the identity on $(X,\preceq)$, as we wanted.

Finally, in this SMC, every partial order is equipped with an interesting monoid and comonoid structure: given the signature $(\{\bullet\}$,$
\InputIfFileExists{lr-copy.tikz}{}{\input{./tikz/lr-copy.tikz}}
$, $
\begin{tikzpicture}
	\begin{pgfonlayer}{nodelayer}
		\node [style=black] (37) at (0.75, 0) {};
		\node [style=none] (43) at (0.25, 0) {};
		\node [style=none] (44) at (-0.5, 0) {};
	\end{pgfonlayer}
	\begin{pgfonlayer}{edgelayer}
		\draw (43.center) to (37);
		\draw [->] (44.center) to (43.center);
	\end{pgfonlayer}
\end{tikzpicture}
}
$,$
\begin{tikzpicture}
	\begin{pgfonlayer}{nodelayer}
		\node [style=black] (37) at (-0.5, 0) {};
		\node [style=none] (43) at (0.25, 0) {};
		\node [style=none] (44) at (0.75, 0) {};
	\end{pgfonlayer}
	\begin{pgfonlayer}{edgelayer}
		\draw [->] (37) to (43.center);
		\draw (44.center) to (43.center);
	\end{pgfonlayer}
\end{tikzpicture}
}
$,$
\InputIfFileExists{lr-merge.tikz}{}{\input{./tikz/lr-merge.tikz}}
)$ we can interpret its generators as follows: let $\sem{\bullet}=(X,\preceq)$ and 
\begin{equation}\label{eq:comonoid-monrel}
\begin{gathered}
\sem{
\InputIfFileExists{lr-copy.tikz}{}{\input{./tikz/lr-copy.tikz}}
} = \left\{(x,(x'_1,x'_2)\mid x\leq x'_1\text{ and } x\leq x'_2)\right\}
\\
\sem{
\begin{tikzpicture}
	\begin{pgfonlayer}{nodelayer}
		\node [style=black] (37) at (0.75, 0) {};
		\node [style=none] (43) at (0.25, 0) {};
		\node [style=none] (44) at (-0.5, 0) {};
	\end{pgfonlayer}
	\begin{pgfonlayer}{edgelayer}
		\draw (43.center) to (37);
		\draw [->] (44.center) to (43.center);
	\end{pgfonlayer}
\end{tikzpicture}
}
} = \{(x,\bullet)\mid x\in X\}
\end{gathered}
\end{equation}
\begin{equation}
\label{eq:monoid-monrel}
\begin{gathered}
\sem{
\InputIfFileExists{lr-merge.tikz}{}{\input{./tikz/lr-merge.tikz}}
} = \left\{((x_1,x_2),x')\mid x_1\leq x'\text{ and } x_2\leq x')\right\}
\\ 
\sem{
\begin{tikzpicture}
	\begin{pgfonlayer}{nodelayer}
		\node [style=black] (37) at (-0.5, 0) {};
		\node [style=none] (43) at (0.25, 0) {};
		\node [style=none] (44) at (0.75, 0) {};
	\end{pgfonlayer}
	\begin{pgfonlayer}{edgelayer}
		\draw [->] (37) to (43.center);
		\draw (44.center) to (43.center);
	\end{pgfonlayer}
\end{tikzpicture}
}
} = \{(\bullet,x)\mid x\in X\}
\end{gathered}
\end{equation}
That these satisfy the monoid and comonoid axioms respectively is a simple exercise.
Note that they are very similar to their standard relational cousins from Example~\ref{ex:relational-sem}. In fact, the former can be seen as the latter, composed with the partial order relation on each wire: for example, if we momentarily reinterpret $\sem{\cdot}$ as a mapping into $\Rel$ (since monotone relations are, after all, relations), we have that
\[\sem{
\InputIfFileExists{lr-copy.tikz}{}{\input{./tikz/lr-copy.tikz}}
} = \sem{\idxright{}}\poi\sem{
\InputIfFileExists{bcomult.tikz}{}{\input{./tikz/bcomult.tikz}}
}\poi\sem{
\InputIfFileExists{leqxleq.tikz}{}{\input{./tikz/leqxleq.tikz}}
}\]
since $\sem{\idxright{}} = \{(x,x')\mid x\preceq x'\}$.

It should be noted that $
\InputIfFileExists{lr-copy.tikz}{}{\input{./tikz/lr-copy.tikz}}
, 
\begin{tikzpicture}
	\begin{pgfonlayer}{nodelayer}
		\node [style=black] (37) at (0.75, 0) {};
		\node [style=none] (43) at (0.25, 0) {};
		\node [style=none] (44) at (-0.5, 0) {};
	\end{pgfonlayer}
	\begin{pgfonlayer}{edgelayer}
		\draw (43.center) to (37);
		\draw [->] (44.center) to (43.center);
	\end{pgfonlayer}
\end{tikzpicture}
}
,
\begin{tikzpicture}
	\begin{pgfonlayer}{nodelayer}
		\node [style=black] (37) at (-0.5, 0) {};
		\node [style=none] (43) at (0.25, 0) {};
		\node [style=none] (44) at (0.75, 0) {};
	\end{pgfonlayer}
	\begin{pgfonlayer}{edgelayer}
		\draw [->] (37) to (43.center);
		\draw (44.center) to (43.center);
	\end{pgfonlayer}
\end{tikzpicture}
}
,
\InputIfFileExists{lr-merge.tikz}{}{\input{./tikz/lr-merge.tikz}}
$ interpreted in this way do not necessarily form a Frobenius monoid, nor do they give rise to a (co)cartesian structure. Their interaction is still interesting, and can be axiomatised, but doing so requires turning to theories with \emph{inequalities} rather than just equalities. We will look at these briefly in Section~\ref{sec:inequalities} and at monotone relations again in Example~\ref{ex:Cartesian-bicategories}.
\end{example}

\section{Other Trends in String Diagram Theory}\label{sec:other-trends}


\subsection{Rewriting}\label{sec:rewriting} When reasoning about programs, \emph{reductions} of a program $p$ into another one $q$ are important objects of study: such reduction may witness for instance the evaluation of $p$ on a certain input (akin to $\beta$-reduction  in the $\lambda$-calculus~\cite{Barendregt-lambda}), or more generically its transformation into a simpler program $q$. When considering programs as terms of an algebraic theory, reductions are typically formalised as \emph{rewriting steps}: we may apply a \emph{rewrite rule} $l \Rightarrow r$ inside $p$ if the term $l$ appears as a subterm of $p$, in which case we say that the rule has a \emph{matching} in $p$; if there is such a matching, then the outcome $q$ of the rewriting step is the term $p[r/l]$ obtained by replacing $r$ for $l$ in $p$.

When it comes to string diagrams, rewriting presents additional challenges, that we do not experience on terms. The crux of the matter is matching: as string diagrams are invariant under certain topological transformations---crossing of wires, shifting of boxes, etc.---we would like matchings to exist or not regardless of which graphical presentation we choose for our string diagram. For example, consider the rule
\[

\InputIfFileExists{monoid-unitality-left.tikz}{}{\input{./tikz/monoid-unitality-left.tikz}}

\]
We claim the rule has a matching in the string diagram below on the left. However, strictly speaking, the matching isolates a subterm only when we `massage' the string diagram as on the right:
\[

\InputIfFileExists{unitality-ex-deformed.tikz}{}{\input{./tikz/unitality-ex-deformed.tikz}}

\]
More formally, the point is that string diagrams are \emph{equivalence classes} of terms, modulo the laws of SMCs. We want to be able to match a rewriting rule $l \rightsquigarrow r$ on a string diagram $c$ whenever a term in the equivalence class of the string diagram $l$ appears as a subterm in the equivalence class of the string diagram $c$.

\begin{definition}
	Let $\Sigma$ be a signature, $l \rightsquigarrow r$ be a rewrite rule of $\Sigma$-terms, and $c$ be a $\Sigma$-term. We say that $c$ rewrites into $d$ modulo $E$ if $\diagbox{c}{}{} =_E \diagbox{c'}{}{}$ and 
\begin{equation}
\diagbox{c'}{}{} = 
\InputIfFileExists{rewrite-l.tikz}{}{\input{./tikz/rewrite-l.tikz}}
 \text{ and } \diagbox{d}{}{} = 
\InputIfFileExists{rewrite-r.tikz}{}{\input{./tikz/rewrite-r.tikz}}

\end{equation}
\end{definition}
The standard case is when $E$ are the laws of SMCs, but the notion can be adapted to fit other categorical structures where string diagrams occur, as those illustrated in Section~\ref{sec:thstringdiag}. This definition seems reasonable enough from a mathematical viewpoint. However, it is completely unpractical when it comes to \emph{implementing} string diagram rewriting. Exploring the space of all $\Sigma$-terms equivalent to a given one is an expensive computational task, and if done naively it may not even terminate given that, in principle, there are infinitely many equivalent terms to be checked for a matching. This is an issue especially because rewriting is the way we formally reason about string diagrams with a computer: whenever we want to apply the equations of a theory such as those considered in Section~\ref{sec:common-theories}, the first thing to do is to orient such equations to turn them into rewrite rules.

The way out of this impasse comes from the combinatorial interpretation of string diagrams,  introduced in Section~\ref{sec:graphs}. Recall that under this interpretation, equivalent $\Sigma$-terms are mapped to a \emph{single} open hypergraph. In other words, if $c$ is mapped to $\CspGraph{c}$, and $c$ and $d$ are equivalent modulo the laws of SMCs, then $\CspGraph{c}=\CspGraph{d}$. This feature makes open hypergraphs suitable data structures to reason about string diagram rewriting: if we want to rewrite with a rule $l \rightsquigarrow r$ and a string diagram $c$ as above, we do not need to bother with the many equivalent syntactic presentations of these diagrams, but just need to consider the corresponding open hypergraphs. In fact, open hypergraphs do come with their own rewriting theory, called \emph{double-pushout rewriting}~\cite{ehrig1973graph}. The fundamental result linking double-pushout rewriting and syntactic rewriting is the following:

\begin{theorem}\label{thm:rewritefrob}
	$c$ rewrites into $d$ modulo $\scFrob$ if and only if the open hypergraph $\CspGraph{c}$ rewrites into $\CspGraph{d}$ modulo double-pushout rewriting.
\end{theorem}

\begin{figure*}
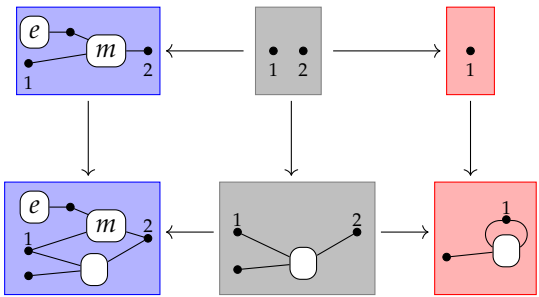

The rule $
\InputIfFileExists{monoid-unitality-left.tikz}{}{\input{./tikz/monoid-unitality-left.tikz}}
$ is interpreted as the span 
\[

\InputIfFileExists{rewrite-dpo-unitality.tikz}{}{\input{./tikz/rewrite-dpo-unitality.tikz}}
.
\]
Its application in the rewrite step $

\InputIfFileExists{monoid-unitality-ex.tikz}{}{\input{./tikz/monoid-unitality-ex.tikz}}
$ is interpreted as
\[

\InputIfFileExists{rewrite-dpo-unitality-match.tikz}{}{\input{./tikz/rewrite-dpo-unitality-match.tikz}}

\]
	\caption{Example of how string diagram rewriting is interpreted as double-pushout rewriting. Intuitively, double-pushout rewriting matches the left-hand side of the rewrite rule to a subgraph and replaces it with the right-hand side. 
	}
\end{figure*}

Theorem~\ref{thm:rewritefrob} is a consequence of the correspondence established by Theorem~\ref{thm:hypergraph-int-frob-iso} between string diagrams modulo Frobenius monoid and open hypergraphs. However, it is not completely satisfactory: we would like to interpret faithfully rewriting modulo the laws of SMCs, without the need of considering Frobenius equations too. It turns out it is still possible to obtain a correspondence, with a more restrictive notion of double-pushout rewriting, called \emph{convex}.

\begin{theorem}\label{thm:rewriteSMC}
	$c$ rewrites into $d$ modulo the laws of SMCs if and only if $\CspGraph{c}$ rewrites into $\CspGraph{d}$ modulo convex double-pushout rewriting.
\end{theorem}

We omit the details of the definition of convexity, which would require us to delve into the theory of double-pushout rewriting. Instead, we refer the interested reader to the overview of string diagram rewriting offered in~\cite{BonchiGKSZ22}.

The above two theorems settle the question of rewriting for string diagrams in symmetric monoidal categories (Theorem~\ref{thm:rewriteSMC}) and hypergraph categories (Theorem~\ref{thm:rewritefrob}). However, as we saw in Section~\ref{sec:thstringdiag}, there are other  structures for which we can draw string diagrams considered both of lower and higher complexity. Among the ones we have covered here, the question is not settled for monoidal, braided monoidal, traced monoidal, (self-dual) compact closed, and cartesian monoidal categories. It has been answered recently in~\cite{Milosavljevic-Zanasi-rewriting} for copy-delete monoidal categories\footnote{The work~\cite{Fritz-gsmonoidal}, which appeared at the same time as~\cite{Milosavljevic-Zanasi-rewriting}, also establishes the correspondence between string diagrams in copy-delete categories and suitably defined open hypergraphs, but without considering the corresponding notion of rewriting.}, and in~\cite{GhicaKaye23} for traced copy-delete categories. Finally, we point out the recent work~\cite{Alvarez-Picallo22}, which studied rewriting for monoidal closed categories.

\subsection{Higher-Dimensional Diagrams}\label{sec:higher-dimensions}

The string diagrams we have considered so far are a particular case of a more general notation for higher-categories. We highlight the connection in this section. First, we need to explain what a $n$-category is. Given the limited scope of our work, we will confine ourselves to a sketch, which should however be sufficient to grasp the link between the graphical language(s) of $n$-categories and string diagrams. The reader should also know that there are several competing definitions of $n$-categories, and that our lack of precision allows us to avoid committing to any one notion.

Very roughly, a $n$-category is a category that may have $2$-morphisms between morphisms, 3-morphisms between 2-morphisms etc. For example, in a 2-category there can be 2-morphisms between morphisms, indicated as follows:
\[\begin{tikzcd}
A \arrow[r, bend left, "f", ""{name=U, inner sep=1pt, below}]
\arrow[r, bend right, "g"{below}, ""{name=D, inner sep=1pt}]
& B
\arrow[Rightarrow, from=U, to=D, "\phi"]
\end{tikzcd}\]
Like monoidal categories, 2-categories also admit a visual representation, called \emph{surface diagrams}: objects are represented as (labelled/coloured) regions of space, morphisms as (labelled) strings or wires, and 2-morphisms as (labelled) dots.
\[
\InputIfFileExists{2-cell.tikz}{}{\input{./tikz/2-cell.tikz}}
\]
Let us see how these compose. There are two ways, just like there are two directions of composition for string diagrams:
horizontally,
\[\begin{tikzcd}
A \arrow[r, bend left, "f", ""{name=f, inner sep=1pt, below}]
\arrow[r, bend right, "g"{below}, ""{name=g, inner sep=1pt, above}]
& B \arrow[r, bend left, "p", ""{name=p, inner sep=1pt, below}]
\arrow[r, bend right, "q"{below}, ""{name=q, inner sep=1pt, above}]
& C
\arrow[Rightarrow, from=f, to=g, "\phi"]
\arrow[Rightarrow, from=p, to=q, "\psi"]
\end{tikzcd}
\qquad\quad 
\InputIfFileExists{2-cell-horizontal-comp.tikz}{}{\input{./tikz/2-cell-horizontal-comp.tikz}}
\]
and vertically
\[\begin{tikzcd}
A \arrow[r, bend left=50, "f", ""{name=f, inner sep=1pt, below}]
\arrow[r, "g"{name=g, anchor=center, fill=white}]
\arrow[r, bend right=50, "h"{below}, ""{name=h, inner sep=1pt, above}]
& B
\arrow[Rightarrow, from=f, to=g, "\phi", shorten <=0.5pt]
\arrow[Rightarrow, from=g, to=h, "\psi", shorten <=0.5pt]
\end{tikzcd}
\qquad\quad 
\InputIfFileExists{2-cell-vertical-comp.tikz}{}{\input{./tikz/2-cell-vertical-comp.tikz}}
\]
The identity (1-)morphism $id_A: A\to A$ can be depicted as a single-coloured region of 2D space, while the identity 2-morphism $id_f: f\Rightarrow f$ can be depicted as a plain wire separating the domain and codomain objects of $f: A\to B$:
\[
\InputIfFileExists{id-2-cell.tikz}{}{\input{./tikz/id-2-cell.tikz}}
\]
For these diagrams to make sense, horizontal and vertical compositions should satisfy additional associativity and unitality requirements, much like those of (1-)categories. In addition, for the diagrams to work, horizontal and vertical composition need to interact nicely; 2-categories should also verify a form of interchange law between horizontal and vertical composition. This law says that the two ways of decomposing the following diagram are equal:
\[
\InputIfFileExists{interchange-2-cells.tikz}{}{\input{./tikz/interchange-2-cells.tikz}}
\]
There is a surprising correspondence between certain 2-categories and monoidal categories: a monoidal category is simply a 2-category with a single object! Take the 2-morphisms to be the ordinary morphisms of the corresponding monoidal category, the 1-morphisms to be its objects, and the monoidal product to be composition of 1-morphisms. Diagrammatically, we can simply depict the single object as the (white here) background on which we draw our diagrams. In this sense, 2-categories can be thought of as typed monoidal categories (where the monoidal product cannot be applied uniformly, but has to match at the boundary 1-morphisms).

Note however that this correspondence is limited to monoidal categories, without any braiding or symmetry. To recover the ability to swap wires of symmetric monoidal categories a lot more structure is required. Intuitively this is because, strictly speaking, if all we have are two dimensions of ambient space, wires cannot cross---what would that even mean? Braidings can occur in at least three dimensions where it makes sense to ask the question of which wire went over or under which other wire. This is why we need to move from 2-categories to 3-or more-categories. This may sound complicated, but just like 2-categories have morphisms between morphisms, we can define 3-categories, which have 3-morphisms between 2-morphisms, 4-categories, which have 4-morphisms between 3-morphisms etc. Each of these come with different ways of composing $n$-morphisms. As it turns out, braided or symmetric monoidal categories, and their graphical languages can be recovered as special cases of  degenerate 3-and 4-categories, respectively. More precisely, a braided monoidal category is a 3-category with only a single object and (1-)morphism, while a symmetric monoidal category is a 4-category with a single object, 1-and 2-morphism. Phew! These correspondences are known as the \emph{periodic table} of $n$-categories ~\cite{baez1995higher}.

One last point: the category of categories is itself a 2-category in which objects are categories, (1-)morphisms are functors, and 2-morphisms are natural transformations between them. The graphical language of 2-categories can therefore be used to present key concepts in category theory. For an excellent introduction to category theory using this diagrammatic language (and an excellent introduction to the diagrammatic language of 2-categories itself) we  recommend~\cite{marsden2014category}.

\subsection{Inequalities}\label{sec:inequalities}

In the same way that we can reason equationally about string diagrams (see Section~\ref{sec:eq-theories}), it is also possible to reason with \emph{inequalities}. From the syntactic point of view, the changes are minimal: we can define theory with inequalities in the same way that we defined equational theories (equalities are recovered as two inequalities in both directions). To interpret inequalities requires a SMC with an order between the morphisms of the semantics that is coherent with the rest of the structure (composition, monoidal product etc.). More formally, we need a SMC in which the morphisms are partially ordered and for which the composition and monoidal product are monotone. This kind of structure appears naturally in the examples of relations that we have covered above, where morphisms can be ordered by inclusion: for two relations $R,S\from X\to Y$, we write $R\leq S$ if $R$ is included in $S$ as a subset of $X\times Y$. If we look at inequalities, some fascinating structure starts to emerge.

\begin{example}[Cartesian bicategories]\label{ex:Cartesian-bicategories}
Cartesian bicategories~\cite{Carboni1987} are SMCs in which we can assign to each object $v$ of our chosen signature a monoid and comonoid structure, which we draw once again as $\Bmultn{v},\Bunitn{v}$ and $\Bcomultn{v},\Bcounitn{v}$, respectively. These have to satisfy the following additional axioms\footnote{The categorically-minded reader will notice that these define an adjunction in the 2-categorical sense, between the comonoid and monoid structure, between $\Bcomult$ and $\Bmult$ on the one hand, and $\Bcounit$ and $\Bunit$ on the other.}:
\begin{equation}\label{eq:Cartesian-bicategories}
\begin{aligned}

\InputIfFileExists{bmult-bcomult.tikz}{}{\input{./tikz/bmult-bcomult.tikz}}
&\leq 
\begin{tikzpicture}
	\begin{pgfonlayer}{nodelayer}
		\node [style=none] (0) at (1.75, 0.75) {};
		\node [style=none] (2) at (1.75, -0.75) {};
		\node [style=none] (3) at (-2, 0.75) {};
		\node [style=none] (5) at (-2, -0.75) {};
	\end{pgfonlayer}
	\begin{pgfonlayer}{edgelayer}
		\draw (3.center) to (0.center);
		\draw (5.center) to (2.center);
	\end{pgfonlayer}
\end{tikzpicture}
}
\qquad\qquad \idx{}&\leq &\; 
\InputIfFileExists{copy-special.tikz}{}{\input{./tikz/copy-special.tikz}}

\\
\idx{}&\leq 
\begin{tikzpicture}
	\begin{pgfonlayer}{nodelayer}
		\node [style=black] (4) at (1, 0) {};
		\node [style=none] (5) at (2.25, 0) {};
		\node [style=black] (6) at (0, 0) {};
		\node [style=none] (7) at (-1.25, 0) {};
	\end{pgfonlayer}
	\begin{pgfonlayer}{edgelayer}
		\draw (4) to (5.center);
		\draw (6) to (7.center);
	\end{pgfonlayer}
\end{tikzpicture}
}
 \qquad\qquad 
\begin{tikzpicture}
	\begin{pgfonlayer}{nodelayer}
		\node [style=black] (0) at (0.5, -0) {};
		\node [style=black] (1) at (-0.75, -0) {};
	\end{pgfonlayer}
	\begin{pgfonlayer}{edgelayer}
		\draw (0) to (1);
	\end{pgfonlayer}
\end{tikzpicture}}
&\leq &\;\;
\InputIfFileExists{empty-diag.tikz}{}{\input{./tikz/empty-diag.tikz}}

\end{aligned}
\end{equation}
The SMC of monotone relations (Example~\ref{ex:monotone-relations}) is an example of a cartesian bicategory, with monoid-comonoid pair given by those of~\eqref{eq:monoid-monrel}-\eqref{eq:comonoid-monrel} for each pre-ordered set $\sem{v} = (X,\preceq)$. Note however that these, as we have seen, do not form a Frobenius monoid in general. In fact, one can show that this is only the case when the underlying partial order is given by equality. In other words, the objects $\sem{v}$ in the category of monotone relations for which the interpretations of $\Bmultn{v},\Bunitn{v}$ and $\Bcomultn{v},\Bcounitn{v}$ form a Frobenius monoid are just plain sets and monotone relations between them are just ordinary relations! This precisely characterises the SMC of relations within the larger SMC of monotone relations.

We can characterise other well-known structures in this SMC. For example,  we can require that there exists two more generating operations on the same object $v$, which we write as $\Wmultn{v},\Wunitn{v}$, and such that the dual of the inequalities~\eqref{eq:Cartesian-bicategories} hold:
\[

\begin{tikzpicture}
	\begin{pgfonlayer}{nodelayer}
		\node [style=none] (0) at (1.75, 0.75) {};
		\node [style=none] (2) at (1.75, -0.75) {};
		\node [style=none] (3) at (-2, 0.75) {};
		\node [style=none] (5) at (-2, -0.75) {};
	\end{pgfonlayer}
	\begin{pgfonlayer}{edgelayer}
		\draw (3.center) to (0.center);
		\draw (5.center) to (2.center);
	\end{pgfonlayer}
\end{tikzpicture}
}
\leq 
\InputIfFileExists{wmult-bcomult.tikz}{}{\input{./tikz/wmult-bcomult.tikz}}
\qquad\qquad 
\InputIfFileExists{bcomult-wmult.tikz}{}{\input{./tikz/bcomult-wmult.tikz}}
 \leq \; \idx{}
\]
\[ 

\InputIfFileExists{bcounit-wunit.tikz}{}{\input{./tikz/bcounit-wunit.tikz}}
 \leq \idx{} \qquad\qquad 
\InputIfFileExists{empty-diag.tikz}{}{\input{./tikz/empty-diag.tikz}}
 \leq \;\;
\begin{tikzpicture}
	\begin{pgfonlayer}{nodelayer}
		\node [style=white] (1) at (-0.75, 0) {};
		\node [style=black] (4) at (0.5, 0) {};
	\end{pgfonlayer}
	\begin{pgfonlayer}{edgelayer}
		\draw (1) to (4);
	\end{pgfonlayer}
\end{tikzpicture}
}

\]
A set $\sem{v}$ equipped with this structure in the SMC of monotone relations is precisely a semi-lattice whose binary meet and top can be identified with  $\Wmultn{v}$ and $\Wunitn{v}$ respectively. To get a lattice, we need to add $\Wcomultn{v},\Wcounitn{v}$ satisfying the same inequalities:
\[

\begin{tikzpicture}
	\begin{pgfonlayer}{nodelayer}
		\node [style=none] (0) at (1.75, 0.75) {};
		\node [style=none] (2) at (1.75, -0.75) {};
		\node [style=none] (3) at (-2, 0.75) {};
		\node [style=none] (5) at (-2, -0.75) {};
	\end{pgfonlayer}
	\begin{pgfonlayer}{edgelayer}
		\draw (3.center) to (0.center);
		\draw (5.center) to (2.center);
	\end{pgfonlayer}
\end{tikzpicture}
}
\:\leq\:
\InputIfFileExists{bmult-wcomult.tikz}{}{\input{./tikz/bmult-wcomult.tikz}}
\qquad\qquad 
\InputIfFileExists{wcomult-bmult.tikz}{}{\input{./tikz/wcomult-bmult.tikz}}
\leq  \idx{}
\]
\[

\begin{tikzpicture}
	\begin{pgfonlayer}{nodelayer}
		\node [style=black] (1) at (0.75, 0) {};
		\node [style=white] (4) at (-0.5, 0) {};
		\node [style=none] (6) at (2, 0) {};
		\node [style=none] (7) at (-1.75, 0) {};
	\end{pgfonlayer}
	\begin{pgfonlayer}{edgelayer}
		\draw (7.center) to (4);
		\draw (1) to (6.center);
	\end{pgfonlayer}
\end{tikzpicture}
}
 \leq\;\idx{}\qquad \qquad 
\InputIfFileExists{empty-diag.tikz}{}{\input{./tikz/empty-diag.tikz}}
 \leq \; 
\begin{tikzpicture}
	\begin{pgfonlayer}{nodelayer}
		\node [style=black] (1) at (-0.75, 0) {};
		\node [style=white] (4) at (0.5, 0) {};
	\end{pgfonlayer}
	\begin{pgfonlayer}{edgelayer}
		\draw (1) to (4);
	\end{pgfonlayer}
\end{tikzpicture}
}
 
\]
One might also wonder how $\Wmult,\Wunit$ and $\Wcomult, \Wcounit$ interact. For an arbitrary lattice, there is not much one can say. However, when the lattice is a \emph{Boolean algebra}, they form a commutative  Frobenius monoid!


\end{example}

\subsection{Relationship with Proof Nets}\label{sec:proof-nets}

Readers who have encountered proof nets before might wonder if there is a relationship with string diagrams. Given that proof nets are a graphical proof system for (multiplicative) linear logic~\cite{Girard87-linearLogic} and that the natural categorical semantics for linear logic takes place in monoidal categories~\cite{mellies2009categorical}, there should be a connection between the two. However, classical multiplicative linear logic usually requires two different monoidal products: one for the multiplicative conjunction, usually written $\otimes$, and one for the multiplicative disjunction, usually written $\parr$. These have  nontrivial interplay, axiomatised in the notion of linearly (or weakly) distributive category~\cite{cockett1997weakly}; if we also want a classical negation, they are related via the usual DeMorgan duality, and the relevant semantics given by $*$-autonomous categories~\cite{barr2006autonomous}.

The problem is that our diagrams already use two dimensions: one for the composition operation and one for the monoidal product. How should we deal with two monoidal products? There are different answers. The first (though not the first one historically) is to move one dimension up, from string diagrams in two-dimensional space, to surface diagrams in three-dimensional space. This is what the authors in~\cite{dunn2019coherence} propose.

However, if we prefer to retain the typesetting ease of a two-dimensional notation, \emph{proof nets} come to the rescue. Unlike standard string diagrams (for strict SMCs that is) proof nets include explicit generators for the two monoidal products\footnote{Note that, in the literature, proof nets are usually depicted going from top to bottom, but we prefer to maintain our convention here in order to make the link with string diagrams clearer.}
\[
\InputIfFileExists{otimes.tikz}{}{\input{./tikz/otimes.tikz}}
\qquad\qquad 
\InputIfFileExists{par.tikz}{}{\input{./tikz/par.tikz}}
\]
Like for compact closed categories, we also need cups and caps satisfying the usual snake equations, which the proof net literature tends to depict as undirected:
\[
\InputIfFileExists{cup-cap-ll.tikz}{}{\input{./tikz/cup-cap-ll.tikz}}
\]
However, not all diagrams we can draw in the free SMC over these generators are proof nets, in the sense they do not necessarily denote well-formed proofs in linear logic. For example, the following diagram is not a proof net and its conclusion ($A\otimes A^*$) is not a theorem of linear logic:
\[
\InputIfFileExists{non-proof-net.tikz}{}{\input{./tikz/non-proof-net.tikz}}
\]
This is why we need an additional criterion to distinguish correct proof nets among all the diagrams we are allowed to draw. There are several such criteria in the literature, under the name of \emph{correctness criteria}~\cite{Girard87-linearLogic,DanosReigner-multiplicativesProofNets}. We will not cover these criteria here---the reader should just know that they usually boil down to detecting some form of acyclicity in graphs derived from the string diagram.

Proof nets can also be understood as a two-dimensional shadow of the natural three-dimensional notation used to represent both monoidal products in~\cite{dunn2019coherence}. This projection comes at a cost: not all two-dimensional diagrams are shadows of a three-dimensional surface. Correctness criteria can therefore be seen as conditions guaranteeing that a proof net is the projection of some higher-dimensional surface diagram.

In the most degenerate cases (from the logic perspective), $\otimes = \parr$, and we are left with the diagrammatic language of compact closed categories (Section~\ref{sec:compact closed}).

\subsection{Software}\label{sec:software}

With the spread of diagrammatic reasoning, it is natural to wonder to what extent it may be automated, and more generally how computers can assist humans in manipulating string diagrams. There are several tools dedicated to this task, each with their specific focus and area of predilection. Here is a (non-exhaustive) list.
\begin{itemize}
\item \textsc{Cartographer}~\cite{sobocinski2019cartographer} deals with string diagrams for SMCs, the central concept of this introduction. It allows the user to specify arbitrary theories in this setting, and apply them as rewrites. String diagram rewriting is implemented as double-pushout hypergraph rewriting, following the approach of Section~\ref{sec:rewriting}.
\item \textsc{Chyp} (available at \url{https://github.com/akissinger/chyp}) is an interactive proof assistant for free SMCs over some signature and equational theory. The application works both with a conventional term syntax and with string diagrams. It also supports a hole-directed rewriting of terms (in the style of Agda programming).
\item \emph{DisCoPy}~\cite{DisCoPy} is a Python library that defines a DSL for diagrams given by either, a free monoidal category or a free SMC over some signature. Furthermore, DisCoPy allows the user to define a semantics for diagrams, that is, to define functors out of free (symmetric) monoidal categories---through these, diagrams can be evaluated to some Python programs (as linear maps, for example). Note however that the package does not act as a proof assistant for diagrammatic equational theories.
\item \texttt{homotopy.io} (the successor of a tool known as Globular~\cite{vicary2018globular})  is a more general tool that allows the user to construct finitely-generated $n$-categories. As a result, it is possible to encode string diagrams for SMCs into \texttt{homotopy.io} (using a correspondence that we have briefly covered in Section~\ref{sec:higher-dimensions}). However, the increased generality comes at the cost of significant sophistication: the user has to explicitly use the laws of SMCs in proofs, having to show the functoriality of the monoidal product by sliding two generators past each other, for example, instead of the two diagrams being equal in the internal representation.
\item \texttt{rewalt}~\cite{hadzihasanovic2022data} is a Python library for higher-dimensional diagrammatic rewriting in which it is possible to build presentations of higher and monoidal categories, among other applications to algebraic topology and algebra. For SMCs, this requires an encoding similar to what  was needed with \texttt{homotopy.io}. As a bonus, \texttt{rewalt} can generate TikZ output to embed diagrams directly into a \LaTeX \ document.
\item \emph{Quantomatic}~\cite{Kissinger_quantomatic} is one of the earliest tools, which deals with a restricted subset of signatures (initially motivated by the ZX-calculus, see the paragraph on quantum physics in Section~\ref{sec:diagrams-science}), namely signatures containing only commutative operations in compact closed categories. It allows the user to specify theories and rewriting strategies, as well as higher-order rules using so called !-boxes.
\end{itemize}
An important theoretical question for the above tools is which data structures best implement string diagrams and diagrammatic reasoning. Considering that string diagrams themselves are quotients of terms, it is a non-trivial task to represent their manipulation efficiently. This question has been explored recently in~\cite{WilsonZanasi21cost,WilsonZanasi23parallel}.

Finally, if one is exclusively interested in the typesetting of string diagrams into \LaTeX documents, we mention the TikZ library, which is especially convenient when paired with {TikZit} (\url{https://tikzit.github.io/}), a GUI editor designed to handle PGF/TikZ pictures.

\section{String Diagrams in Science: Some Applications}\label{sec:diagrams-science}

In the last few years, string diagrams have found application in several fields of science and engineering. This section is intended as a succinct overview of such applications, with the main aim of providing to the reader references for a more focussed study. Clearly, a survey of this type cannot possibly begin to cover all the relevant material, and it will necessarily be a partial account. Our perspective will be pedagogical rather than historical: we will typically point to the most recent surveys and introductory materials, when available. A comprehensive literature review, including a rigorous reconstruction of ``who did what first'', is out of our scope.

Many of the applications that we will describe share the same methodology. They involve noticing that the kind of systems and/or processes which constitute the focus of a given research area can be understood as the objects and/or morphisms of some SMC. This opens up the possibility of studying the topic from a functorial standpoint, using string diagrams as a syntax and the relevant systems and/or processes as semantics. In many cases the same approach also opens up the possibility of studying the equational properties of the resulting diagrammatic language. In some particular cases (this does not apply to all examples below), a universal set of generators can be found for the syntax and, if we are even luckier, the equational theory can be axiomatised by a complete monoidal theory.


It is fitting to start this section by applications of string diagrams in physics, since a notable ancestor of string diagrams is Penrose's pictorial notation, invented to carry out the complex tensor calculations in the differential geometry of general relativity~\cite{penrose1971applications} (see~\cite{Penrose-tensornotation} for a more recent overview).

\paragraph{Quantum Physics}\label{sec:quantum}
One of the most outstanding modern applications of diagrammatic reasoning has been to quantum computing. Mathematically, quantum systems are modelled as Hilbert spaces, where joining two systems is represented by taking the tensor product of the respective spaces, and processes acting on a system are linear (unitary) maps. Linear maps between Hilbert spaces with the tensor product as monoidal product form a SMC, and are thus amenable to a string diagrammatic study.

There are several diagrammatic calculi to reason about linear maps between qubits (represented as vector spaces of dimension $2^d$ for some natural $d$) with the tensor product as monoidal product. These generalise and formalise the circuit representation that is ubiquitous in quantum computing. The first and most well-known such calculus is the \emph{ZX-calculus}: its diagrams consist of nodes of two different colours\footnote{Traditionally, green and red, but white and gray have been used more recently, for accessibility.} called spiders, each labelled by an angle:
\[
\InputIfFileExists{z-spider.tikz}{}{\input{./tikz/z-spider.tikz}}
 \qquad 
\InputIfFileExists{x-spider.tikz}{}{\input{./tikz/x-spider.tikz}}
\]
The two colours denote one of two classical observables or measurement bases: the Z (or computational) basis and X (or Hadamard) basis respectively. The angle denotes a phase relative to this basis, \emph{i.e.} a rotation of the Bloch sphere along the axis determined by the chosen observable.

Equationally, the spiders form a special commutative Frobenius monoid and the two colours interact with each other to form bimonoid (more specifically, a Hopf algebra which is to a bimonoid what a group is to a monoid). Together with some other more complex equalities, the ZX-calculus completely axiomatises linear maps between qubits so that any semantic equality can be obtained by purely equational reasoning at the level of the diagrams themselves.

Because the ZX-calculus generalises quantum circuits, its axiomatisation provides a completely equational way to reason about those. Reasoning equationally at the level of circuits is more difficult, so the ZX-calculus provides a more compositional setting in which to study the behaviour of circuits. In fact, one of its most successful applications has been to quantum circuit synthesis and simplification. Given some measure of complexity of circuits, one can compile a given circuit to its corresponding ZX diagram and simplify it using the axioms of the calculus, with the important caveat that one needs to guarantee that a bona fide circuit can be recovered at the end. (There are many technical papers on this topic; we could not find a more accessible survey, though the general introduction~\cite{zxIntro} contains some pointers to the literature.) String diagrams have now reached the mainstream quantum computing community, as even one of the founders of the field has adopted the ZX-calculus in a recent preprint~\cite{khesin2023graphical}.

\begin{figure}
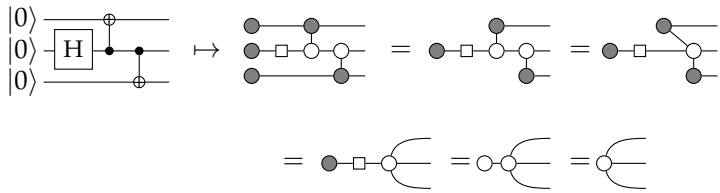

\[
\InputIfFileExists{ghz-circuit.tikz}{}{\input{./tikz/ghz-circuit.tikz}}
\;\;\mapsto\;\;
\InputIfFileExists{ghz-preparation.tikz}{}{\input{./tikz/ghz-preparation.tikz}}
\]
\caption{\emph{The quantum circuit on the left prepares the GHZ state $\ket{000}+\ket{111}$; the ZX-calculus derivation is a diagrammatic proof of this (unlabelled spiders have $0$ phase and the square box is a Hadamard gate). See~\cite[Section 5]{zxIntro} for similar examples of ZX-calculus proofs.}}
\end{figure}

There are other calculi with the same target semantics---linear maps between qubits---with different sets of generators as building blocks. The \emph{ZW-calculus} was the first for which a completeness result was found and was instrumental in deriving a complete equational theory for the ZX-calculus (by translating one into the other). Its generators are further from the conventional gates of the classical quantum circuit model, but closely related to linear optical quantum circuits~\cite{de2022quantum,poor2023completeness}.  This last reference builds a calculus that unifies both ZX-and ZW-calculus and is shown complete for arbitrary dimension. The \emph{ZH-calculus} is a variation of the ZX-calculus with which it is easier to represent certain multiply-controlled logic gates, like the Toffoli or AND gates (on the computational basis), and other related operations.

For a diagrammatic introduction to quantum computing and the foundations of quantum theory, the reference textbook is~\cite{PQP}. Alternatively, \cite{HeunenVicaryBook} provides a complementary (and more categorically minded) approach to some of the same topics. A comprehensive survey of the ZX-calculus (and its cousins) for the working computer scientist can be found in~\cite{zxIntro}.
Beyond these, there is a wealth of recent developments that have taken the original work in different directions: a calculus which incorporates finite memory elements to the ZX-calculus~\cite{carette2021graphical}, several calculi for quantum linear optical circuits~\cite{de2022quantum,clement2022lov}, and more... There are also automated tools to reason about large-scale ZX diagrams: the python library PyZX which we have already mentioned is the most recent~\cite{kissinger2019pyzx}. Finally, the website \url{https://zxcalculus.com/} contains tutorials, a helpful guide to the publications in the field and even a map of ZX research community.



\paragraph{Signal Flow Graphs and Control Theory} Engineers and control theorists have long expressed causal flow of information between different components of a system by graphical means. One popular formalism is that of \emph{signal flow graphs}, sometimes called \emph{block diagrams}. They are directed graphs whose nodes represent system variables and edges functional dependencies between variables. They are used to represent networks of interconnected electronic components, amplifiers, filters etc. In signal flow graphs, cycles represent feedback between different parts of the system.

\begin{figure}
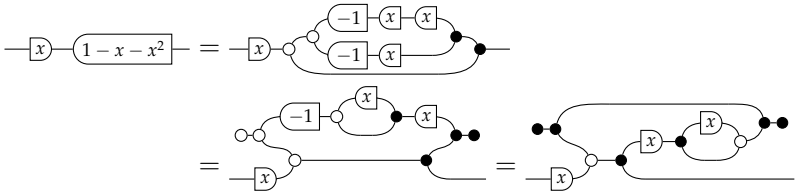

\begin{align*}

\InputIfFileExists{fibonacci-gf.tikz}{}{\input{./tikz/fibonacci-gf.tikz}}
 &= 
\InputIfFileExists{fibonacci-gf-1.tikz}{}{\input{./tikz/fibonacci-gf-1.tikz}}

\\
& =
\InputIfFileExists{fibonacci-gf-2.tikz}{}{\input{./tikz/fibonacci-gf-2.tikz}}
 = 
\InputIfFileExists{fibonacci-gf-3.tikz}{}{\input{./tikz/fibonacci-gf-3.tikz}}

\end{align*}
\caption{\emph{Steps of a derivation transforming the Fibonacci generating function $\frac{x}{1-x-x^2}$ into a signal flow graph~\cite{BonchiSZ17}. The $x$ can be interpreted as derivation in the field $\R(x)$ of rational functions or as a time delay. Note that both the specification (on the left), the signal flow graph which realises it (on the right), and the intermediate steps are all string diagrams of the same calculus, and all the steps apply laws of IH over $\R(x)$.}}
\end{figure}

Giving signal flow graphs a functorial semantics, required a change of perspective: instead of only allowing \emph{functional} dependencies between variables, we can generalise signal flow graphs to allow \emph{relational} dependencies between them, \emph{i.e.} arbitrary systems of (usually linear) equations. This is entirely consistent with the underlying physics, where laws tend to express relationships between variables, without any explicit assumption about the direction of causality. This change of perspective allowed the reinterpretation of the fundamental building blocks of signal flow graphs as relations, their connection as relation composition, and their juxtaposition as taking the cartesian product of the corresponding relations. In other words, these generalised signal flow graphs are string diagrams for a sub-category of the SMC $(\Rel,\times)$! One important such subcategory is that of vector spaces over a field and \emph{linear relations} (\emph{cf}. Example~\ref{ex:linrel}) between them. It turns out that signal flow graphs are intimately related to linear relations over the field of rational functions over $\R$. The connection was established independently in~\cite{Bonchi2015,BonchiSZ17}  and~\cite{BaezErbele-CategoriesInControl}, where the authors also give a complete equational theory, called \emph{Interacting Hopf Algebra} (IH) by the first set of authors, to reason about the behaviour of these systems entirely diagrammatically. Since these early developments, the theory IH has been employed to reason algebraically about various tasks related to signal flow graphs: we mention the realisability of rational behaviours as circuits~\cite{Bonchi2015}, semantic refinement~\cite{BonchiHPS17}, and a compositional criterion for controllability~\cite{Fong2015}.
 The interested reader should note however that there are some subtle discrepancies between this generalisation of signal flow graphs and the standard control-theoretic interpretation of their behaviour: a more accurate, but closely related semantics in terms of bi-infinite streams is given in~\cite{Fong2015}, along with a complete equational theory.

Finally, linear relations are not just important because of the relationship with signal flow graphs; more generally the diagrammatic calculus and the equational theory IH provide a playground in which a substantial amount of standard linear algebra can be reformulated entirely diagrammatically. The blog \url{graphicallinearalgebra.net} is a great introduction  to this topic, aimed at a general audience.

\paragraph{Circuit Theory}\label{sec:circuits} String diagrams are particularly compelling where they can give an algebraic foundation to existing graphical representations that are usually treated purely combinatorially. This is the case of many existing approaches to electrical or digital circuits. Despite the existence of a standard graphical representation for circuits, the string diagrammatic approach is not without challenges: taking the original graphical representation as a starting point, one needs to decompose them into a an suitable set of generators from which all other circuits can be built and, more importantly, give this syntax a functorial interpretation that assigns to each circuit its intended behaviour. In traditional introductions to electrical circuits, this last step usually appeals informally to some intuitive connection between a circuit and the set of differential equations that it specifies. String diagrams can make this connection precise and compositional. In some particular cases, it is even possible to equip the resulting syntax with a complete equational theory that axiomatises semantic equivalence of circuits.

\begin{figure}
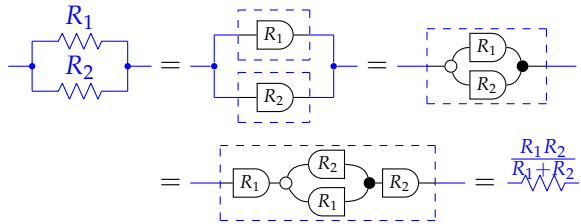

\begin{align*}
\begin{tikzpicture}[color=blue]
	\draw (-1,0) -- (0,0);
	\draw (0,0) -- (0,1);
	\node[elecdot] at (0,0) {};
	\draw (0,0) -- (0,-1);
    \draw (0,1) to[R=$R_1$] (4,1);
    \draw (0,-1) to[R=$R_2$] (4,-1);
    \draw (4,1) -- (4,-1);
    \node[elecdot] at (4,0) {};
    \draw (4,0) -- (5,0);
\end{tikzpicture}
& = 
\InputIfFileExists{parallel-resistors-unboxed.tikz}{}{\input{./tikz/parallel-resistors-unboxed.tikz}}

= 
\InputIfFileExists{parallel-resistors-unboxed-1.tikz}{}{\input{./tikz/parallel-resistors-unboxed-1.tikz}}
 
\\
& = 
\InputIfFileExists{parallel-resistors-unboxed-2.tikz}{}{\input{./tikz/parallel-resistors-unboxed-2.tikz}}
 = \begin{tikzpicture}[color=blue]
    \draw (0,0) to[R=$\frac{R_1R_2}{R_1+R_2}$] (3,0);
\end{tikzpicture}
\end{align*}
\caption{\emph{Deriving textbook properties of electrical circuits by compiling them to graphical linear algebra. The black diagrams represent the voltage-current pairs enforced by elements of the circuit~\cite{boisseau2021string}.}}\label{fig:parallel-resistors}
\end{figure}

For \emph{electrical circuits} with linear/affine behaviour (including resistors, inductors, capacitors, for example) this ambitious goal has not yet been achieved, though it is possible to compile them down to an intermediate representation in IH (or its affine extension) for which, as mentioned in the paragraph above, we do have a complete diagrammatic calculus. The case of circuits with passive components is treated in~\cite{BaezCoyaPropsNetworkTheory} while the extension to current and voltage sources (in AC regime only) is carried out in~\cite{BonchiPSZ19}. Building on this,~\cite{boisseau2021string} develop a convenient calculus that blends both circuit elements and their compilation in IH, \emph{cf.} Figure~\ref{fig:parallel-resistors}. The resulting language allows for entirely diagrammatic proofs of standard textbook results of electrical engineering such as the superposition theorem or Th\'{e}venin/Norton's theorem. In this work, the ability to reason inductively on circuits as a genuine syntax proves very useful to prove these general theorems.

For \emph{digital circuits}, there are many possible variants of interest to consider, each at their own level of abstraction (and each, with their own limitations). The simple case of acyclic circuits consisting only of logic gates, without any memory elements, reduces to boolean algebra and thus defines a cartesian monoidal category, which can be presented by the symmetric monoidal version of the algebraic theory of boolean algebras, with only minor adaptations (as explained in Remark~\ref{rmk:algebraic-Cartesian} for example)---details can be found in the pioneering~\cite{Lafont2003}.

More complex cases, involving delays or cycles are more delicate. Recently, the sequential synchronous (\emph{i.e.} where a global clock is assumed to define the time at which signals can meaningfully change) case has found a complete axiomatisation in~\cite{fullAbstractionDigitalCircuits}. Notably, this work also allows combinational (that is, without any delay) cycles in the syntax. This feature is usually avoided in traditional treatments of digital circuits because of the difficulty of handling these types of cycles compositionally. The case of asynchronous (cyclic) circuits remains elusive and an important open problem, although some work has already been done in this direction~\cite{GhicaAsynchCircuits}.



\paragraph{Probability and statistics}
Issues with the encoding of probability theory in set-theoretic measure theory have pushed researchers to develop a different, synthetic approach to probability theory. In recent years, some have sought alternative categorical foundations.

One approach studies categories of measurable spaces and Markov kernels (conditional probability measures) in an attempt to find an axiomatic setting for probability theory. This category is not only symmetric monoidal but a CD category. Moreover, in general, its morphisms satisfy only the $\mathsf{del}$ equation below:
\[
\InputIfFileExists{Lawvere-distributive-copy.tikz}{}{\input{./tikz/Lawvere-distributive-copy.tikz}}
\; \myeq{dup}\; 
\InputIfFileExists{Lawvere-distributive-copy-1.tikz}{}{\input{./tikz/Lawvere-distributive-copy-1.tikz}}
 \qquad \qquad 
\begin{tikzpicture}
	\begin{pgfonlayer}{nodelayer}
		\node [style=none] (6) at (-3.25, 0) {};
		\node [style=basic box] (10) at (-1.5, 0) {$d$};
		\node [style=black] (11) at (0, 0) {};
		\node [style=none] (15) at (-3, 0.5) {\scriptsize $v$};
	\end{pgfonlayer}
	\begin{pgfonlayer}{edgelayer}
		\draw (10) to (11);
		\draw (6.center) to (10);
	\end{pgfonlayer}
\end{tikzpicture}
}
\;\myeq{del}\;\Bcounitn{v} \]
At the semantic level, this equation simply means that conditional probabilities are measures that normalise to $1$. Crucially, not all morphisms satisfy the $\mathsf{dup}$ equation above, so that categories of Markov kernels are not cartesian; those morphisms that can be copied and do satisfy $\mathsf{dup}$ are precisely deterministic kernels, \emph{i.e.} those that, given a element of their domain, map it to a single element in their codomain with probability one. We already see that a few elementary properties of random variables can be expressed in these categories, called \emph{Markov categories}; their string diagrams for Markov categories give a graphical language to treat standard properties such as conditional independence, disintegration, almost sure properties, sufficient statistics and more. The interested reader will find~\cite{fritz2020synthetic} to be a good introduction to the topic, with applications to statistics.

\begin{figure}
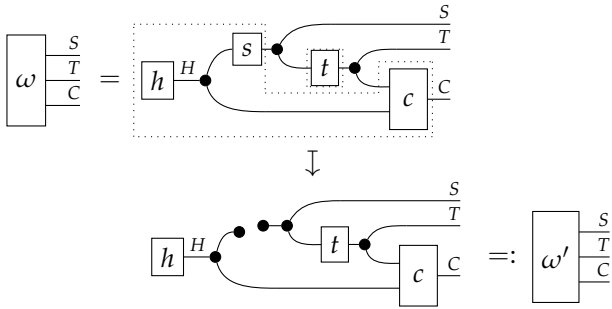

\begin{align*}

\InputIfFileExists{ex-bayesian-net-diag.tikz}{}{\input{./tikz/ex-bayesian-net-diag.tikz}}
\qquad \qquad\qquad   
\\
 \qquad\qquad \qquad 
\InputIfFileExists{ex-bayesian-net-intervention.tikz}{}{\input{./tikz/ex-bayesian-net-intervention.tikz}}

\end{align*}
\caption{\emph{In this scenario, from~\cite{jacobs2021causal}, we seek to identify the causal effect of smoking on cancer. String diagrams represent generalised Bayesian networks, encoding causal dependencies between a set of variables. The prior is $\omega$, the joint probability distribution of smoking (variable $S$), presence of tar in the lungs (variable $T$), and developing cancer (variable $C$). A tobacco company contends that, even though there is a statistical correlation between $S$ and $C$, there might be some confounding factor $H$ (perhaps genetic) which causes both smoking and cancer (decomposition of $\omega$, on the left). How can we rule out this causal scenario, when direct intervention is impossible? We see that performing a `cut' at $S$ and replacing it with the uniform distribution, as on the right, would remove any confounding influence of $H$ over $S$. In this case, we can infer from the structure of the diagram that the distribution corresponding to the resulting diagram $\omega'$ can be computed from observational data only. If, under $\omega'$, a smoker is still more likely to develop cancer, then we have demonstrated that there is a causal relationship between $S$ and $C$.}}\label{fig:causal-identifiability}
\end{figure}

Much like the existing graphical methods of Bayesian networks and related representations, string diagrams make the flow of information between different variables explicit, highlighting structural properties, such as (conditional) independence. In fact, this connection was already explored in~\cite{Fong12} which gave a functorial account of Bayesian networks. Around the same time, the authors of~\cite{coecke2012picturing} proposed a diagrammatic calculus for Bayesian inference. Since then, a surprising amount of probability theory has been recast in this synthetic mold:   a growing list of results have been reproven in this more general setting, many of which use string diagrams to streamline proofs, including zero-one laws~\cite{fritz2020zero}, de Finetti theorem~\cite{fritz2021finetti}, the ergodic decomposition theorem~\cite{moss2022category}, and more.

The same diagrams (in Markov or CD categories) also allow for a treatment of standard concepts in causal reasoning. In~\cite{jacobs2021causal}, the authors give a diagrammatic account of interventions and a sufficient criterion to identify when a given causal effect can be reliably estimated from observational data, see Figure~\ref{fig:causal-identifiability}. The recent~\cite{lorenz2023causal} extends this work to counterfactuals and shows how causal inference calculations can be carried out fully diagrammatically. Beyond its pedagogical value, one advantage of the diagrammatic approach is that it is axiomatic: as such, it is not restricted to the category of Markov kernels, but applies in all categories with the relevant structure.

\paragraph{Machine Learning with Neural Networks}\label{sec:ML} 

Having mentioned string diagrammatic treatments of Bayesian networks, it is natural to wonder about analogous studies of neural networks, another chief graphical model of machine learning. Categorical approaches to these structures are fairly recent and so far have mainly focussed on providing an abstract account of the gradient-based learning process~\cite{FongST19,CruttwellGGWZ22,Gavranovic-MSc}. See e.g.~\cite{Shiebler-MLsurvey} for an overview. Use of string diagrams to describe the network structure only cursorily appear in~\cite{FongST19}. String diagrams are heavily used in~\cite{CruttwellGGWZ22} to represent the categorical language of \emph{lenses} in the context of machine learning, but presentations by generators and equations of these diagrams are not investigated. The works on gradient-based learning with ``quantised'' versions of neural networks, such as Boolean circuits~\cite{WilsonZ20} and polynomial circuits~\cite{WilsonZ22}, adopt string diagrams in a more decisive way. These works define reverse derivatives compositionally on the diagrammatic syntax for circuits, building on the theory of reverse derivative categories~\cite{CockettCGLMPP20} and Lafont's algebraic presentation of Boolean circuits~\cite{Lafont2003}. Going forward, the expectation is that such an approach may work also for real-valued networks, once the different neural network architectures are properly understood in terms of algebraic presentations. A starting point is provided in~\cite{FongST19} for feedforward neural networks (see also the diagrammatic presentation of piecewise-linear relations found in~\cite{boisseau2022graphical}, which may be used to model networks with ReLu activation units). Another important research thread concerns automatic differentiation (AD): the work~\cite{Alvarez-Picallo23} uses rewriting of string diagrams in monoidal closed categories to describe an algorithm for AD and prove its soundness. In this context, string diagrams are appealing as they can be reasoned about as a high-level language, while at the same time exhibiting the same information of lower-level combinatorial formalisms, which in traditional AD are introduced via compilation. Finally, we recommend the webpage~\cite{BrunoWebpage} to the interested reader, as it maintains a list of papers at the intersection of category theory and machine learning. 



\paragraph{Automata Theory}\label{sec:automata}

Automata have always been represented graphically, as state-transition graphs. However, the graphical representation is usually treated as a visual aid to convey intuition, not as a formal syntax.  Kleene introduced regular expressions to give finite-state automata an algebraic syntax. Their equational theory---under the name of Kleene algebra---is now well-understood and multiple complete axiomatisations have been given, for both language and relational models.




With string diagrams however, it is possible to go directly from the operational model of automata to their equational properties, without going through a symbolic algebraic syntax~\cite{piedeleu2023finite}. This approach lets us axiomatise the behaviour of automata directly, freeing us from the necessity of compressing them down to a one-dimensional notation like regular expressions.
In addition, embracing the two-dimensional nature of automata guarantees a strong form of compositionality that the one-dimensional syntax of regular expressions does not have. In the string diagrammatic setting, automata may have multiple inputs and outputs and, as a result, can be decomposed into subcomponents that retain a meaningful interpretation.
For example, the Kleene star can be decomposed into more elementary building blocks, which come together to form a feedback loop:
\[
\InputIfFileExists{star-decomposed.tikz}{}{\input{./tikz/star-decomposed.tikz}}
\]
It should be noted that a similar insight was already present in the work of \c{S}tef\u{a}nescu~\cite{stefanescu2000network} who studied the algebraic properties of traced monoidal categories  with several additional axioms in order to capture the properties of automata, flowchart schemes, Petri nets, data-flow networks, and more.

\paragraph{Databases and Logic}\label{sec:databases} 

As we have discussed above in several places, string diagrams are a convenient syntax for relations. They are particularly well-suited to \emph{conjunctive queries}, the first-order language that contains relation symbols, equality, truth, conjunction, and existential quantification. This is a core fragment of query languages for relational databases with appealing theoretical properties, such as NP-completeness (and thus, decidability) of query inclusion. Moreover, it admits a flexible diagrammatic language, which is exactly that of cartesian bicategories (Example~\ref{ex:Cartesian-bicategories}). These were introduced in~\cite{Carboni1987} in the 1980s (without a diagrammatic syntax, most likely for typesetting reasons, though the authors give an axiomatisation that is easy to translate into string diagrams). The precise connection with conventional conjunctive query languages is worked out in~\cite{bonchi2018graphical}. More recently, the authors of~\cite{haydon2020compositional} have generalised these diagrams to all of (classical) predicate logic. This requires adding boxes that represent negation to the diagrams of~\cite{Carboni1987,bonchi2018graphical}. Remarkably, these give a categorical account of a graphical notation for first-order logic invented by the American philosopher C.S. Peirce in a series of manuscripts~\cite{peirce1974collected} dating as far back as the 19th century!


\paragraph{Computability Theory}

Computers are machines that can be programmed to exhibit a certain behaviour. The range of behaviours that they can exhibit (its processes) can be axiomatised into a monoidal category with additional properties: we require that every process the machine can perform has a name---its corresponding program---encoding the intentional content of the process. In turn, the category contains distinguished processes, called evaluators, which run 	a given a program on an input state of the machine. These  simple requirements, with the ability to copy and delete data, are what~\cite{pavlovic2013monoidal} calls a \emph{monoidal computer}. This structure is sufficient to reproduce a substantial chunk of computability (and complexity) theory using string diagrams. A textbook that does just that is~\cite{pavlovic2022categorical}.

\paragraph{Concurrency Theory}

Concurrency lends itself to graphical methods, a fact noticed early by Petri. Initially, like so many other graphical representations, Petri nets were treated monolithically, and little attention was given to their composition. Once again, it is possible to take Petri nets seriously as a diagrammatic syntax with a functorial semantics. There are several ways to do so: one can either compose Petri nets along shared state (places) or shared actions (transitions). The  former was developed in~\cite{Baez_lics20}, while the latter was initiated in~\cite{bruni2011connector,Bruni2013}. The authors have contributed to developing this last approach further, by studying and axiomatising the algebra of Petri net transitions~\cite{BHPSZ-popl19}. Significantly, the syntax is the same as that of signal flow graphs (see above)---only the semantics changes, replacing real numbers (modelling signals) with natural numbers (modelling non-negative finite resources like the tokens of Petri net). This simple change also changes the equational theory dramatically.


\paragraph{Linguistics}\label{sec:linguistics}

String diagrams have made a surprising appearance in formal linguistics and natural language processing. The core idea relies on a formal analogy between syntax and semantics of natural language. On the semantic side, vectors (in other words, arrays of numbers) are a convenient way of encoding statistical properties of words or features (\emph{word embeddings}) extracted from large amounts of text by training machine learning model. The semantics obtained in this way is often called \emph{distributional}.
On the syntactic side, various formal structures of increasing complexity have been used to study the grammatical properties of languages and explain what distinguishes a well-formed sentence from an incorrect one.

Reconciling, or rather, combining the insights of these two perspectives has been a long-standing problem. One possible approach, first proposed in~\cite{clark2008compositional}, is premised on a formal correspondence between grammatical structure and distributional semantics: both fit into monoidal categories! We have already seen before that vector spaces and linear maps form a SMC, with the tensor as monoidal product. Similarly, formal grammars can be recast as certain (non-symmetric) monoidal categories in which objects are parts-of-speech (think nouns, adjectives, transitive verbs etc.) and morphisms derivations of well-formed sentences. The analogy allows for a functorial mapping from one to the other. With this correspondence in place, it becomes possible to interpret grammatical derivations of sentences as string diagrams in the category of vector spaces and linear maps. This gives a compositional way to build the meaning of sentences from the individual meaning of words. Moreover, certain symmetric monoidal theories can model grammatical features of language whose distributional semantics is less clear. Relative pronouns, for example, have been successfully interpreted as certain Frobenius algebras~\cite{sadrzadeh2013frobenius,sadrzadeh2014frobenius}, allowing equational reasoning about the meaning of sentences that contain them. String diagrams also reveal that two a priori distinct areas of scientific enquiry can share some formal structure. 


\paragraph{Game Theory}

In classical game theory, games and their various solution concepts are usually studied monolithically. Remarkably, a compositional approach was shown possible in~\cite{ghani2018compositional}. In this paper, the authors build games from smaller pieces, called \emph{open games}. An open game is a component that chooses its next move strategically, given the state of its environment and some counterfactual reasoning about how the environment might react to its move (and what payoff it would derive from it). They form a SMC with an interesting diagrammatic syntax which contains forward and backward flowing wires (the latter represent this counterfactual reasoning, flowing from future to present). In fact, these diagrams are related to those for \emph{lenses}, which we have encountered in the paragraph on machine learning. Closed diagrams represent classical games, whose semantics is given by the appropriate equilibrium condition. Initially developed only for standard games with pure Nash equilibria as solution concept, open games have been extended to more general settings, such as Bayesian games~\cite{bolt2019bayesian} with the appropriate solution concept.

\newpage


\bibliographystyle{plain}
\bibliography{refs}

\newpage

\appendix

\section{Category Theory: the Bare Minimum}\label{sec:category-theory}

In this appendix, we recall the most basic notions of category theory: category, functor, natural transformation, adjunctions. This very brief recap is intended as reference material for some of the explanations provided in the main text. For a more pedagogical treatment, the reader should turn to an introductory textbook on the topic, such as~\cite{leinster2014basic}.

\begin{definition}\label{def:category}
A \emph{category} $\mathsf{C}$ consists of
\begin{itemize}
\item a collection of objects;
\item a collection of morphisms such that every morphism $f$ of $\mathsf{C}$ has a unique object $x$ called its domain, and a unique object $y$ called its codomain, which we then write $f\from x\to y$;
\item for every pair of morphisms $f\from x\to y$ and $g\from y \to z$, a morphism $f;g\from x\to z$ which we call the \emph{composition} of $f$ and $g$;
\item for every object $x$, a morphism $\mathsf{id} \from x\to x$ which we call the \emph{identity} on $x$;
\item such that
\begin{enumerate}
\item composition is associative, \emph{i.e.},
\[f\poi (g\poi h) = (f\poi g)\poi h\]
\item identities are the (two-sided) unit for composition, \emph{i.e.},
\[f\poi \mathsf{id}_y = f = \mathsf{id}_x\poi f\]
\end{enumerate}
\end{itemize}
\end{definition}
\begin{remark}\label{rmk:order-composition}
The order of composition in the definition above is chosen to adhere to the diagrammatic order of composition, from left to right. It is common to see the reverse-order operation, $g\circ f = f\poi g$, in particular when dealing with maps between sets.
\end{remark}
\begin{remark}\label{rmk:diagrams-category}
In the main text, the first categorical structure in which we consider string diagrams are (symmetric) monoidal categories (Definition~\ref{def:free-smc}). However, plain categories already accommodate a representation of their morphisms as string diagrams, albeit of a simpler kind. Just as in the monoidal case, one may use wires to depict (the identity on) each object, and boxes for each morphism:
\[\derivationRule{}{\quad
\InputIfFileExists{id-x-with-frame.tikz}{}{\input{./tikz/id-x-with-frame.tikz}}
\quad}{\scriptstyle{x\in \Sigma_0}}
\qquad\
\derivationRule{}{\quad\diagbox{d}{v}{w}\quad}{\scriptstyle d\in\Sigma_1}
\]
\emph{Sequential} composition of boxes with matching wires in the middle is the only allowed operation:
\[
\derivationRule{\diagbox{c}{u}{v}\quad \diagbox{d}{v}{w}}{
\InputIfFileExists{horizontal-comp.tikz}{}{\input{./tikz/horizontal-comp.tikz}}
}{}
\]
The diagrammatic notation has the benefit of absorbing the associativity and unitality laws of Definition~\ref{def:category}:
\[
\begin{gathered}
{
\InputIfFileExists{smc/sequential-associativity.tikz}{}{\input{./tikz/smc/sequential-associativity.tikz}}
 = 
\InputIfFileExists{smc/sequential-associativity-1.tikz}{}{\input{./tikz/smc/sequential-associativity-1.tikz}}
}
 \\ 
\scalebox{1}{
\InputIfFileExists{smc/unit-right.tikz}{}{\input{./tikz/smc/unit-right.tikz}}
 = \diagbox{c}{}{} = 
\InputIfFileExists{smc/unit-left.tikz}{}{\input{./tikz/smc/unit-left.tikz}}
}
 \end{gathered}
\]
\end{remark}

The appropriate notion of structure-preserving mapping between categories is called a functor.
\begin{definition}\label{def:functor}
Given two categories $\mathsf{C}$ and $\mathsf{D}$, a \emph{functor} $F\from\mathsf{C}\to \mathsf{D}$ is a map from the objects of $\mathsf{C}$ to those of $\mathsf{D}$, and a map from the morphisms of $\mathsf{C}$ to those of $\mathsf{D}$ that preserves composition and identities, \emph{i.e.},
\[F(f\poi g) = Ff\poi Fg \qquad F(\mathsf{id}_x) = \mathsf{id}_{Fx}\]
\end{definition}
Clearly, functors can also be composed like ordinary maps. With functors as morphisms, categories themselves form a category. A functor is called \emph{faithful} if it is injective on morphisms of the same type, \emph{i.e.} if $Ff = Fg$ implies that $f=g$; it is \emph{full} if it is surjective on morphisms of the same type, \emph{i.e.}, for any $g$ in $\mathsf{D}$, there exists $f$ in $\mathsf{C}$ such that $Ff = g$.

There is also a notion of mapping between functors, which we now recall.
\begin{definition}\label{def:nat-trans}
Given two functors $F\from \mathsf{C}\to\mathsf{D}$ and $G\from \mathsf{C}\to\mathsf{D}$, a \emph{natural transformation} $\eta\from F \Rightarrow G$ is a family of morphisms $\eta_x\from Fx \to Gx$ indexed by the objects of $\mathsf{C}$, such that $\eta_x \poi Gf = Ff\poi \eta_y$ for every morphism $f\from x\to y$.
\end{definition}
\begin{remark}\label{rmk:functorial-boxes}
Functors can also be represented pictorially, via \emph{functorial boxes}~\cite{mellies2006functorial}. Plainly, for a functor $F\from \mathsf{C}\to \mathsf{D}$, they are $F$-labelled boxes that frame diagrams
\[
\InputIfFileExists{F-functorial-box.tikz}{}{\input{./tikz/F-functorial-box.tikz}}
\]
Diagrams inside the box live in the category $\mathsf{C}$, and those outside in $\mathsf{D}$.
To represent functors, functorial boxes have to be functorial! This means that they satisfy the following equality:
\[
\InputIfFileExists{Fc-Fd.tikz}{}{\input{./tikz/Fc-Fd.tikz}}
 \; =\; 
\InputIfFileExists{Fcd.tikz}{}{\input{./tikz/Fcd.tikz}}
\]
This is easily seen to be the translation of the first equality in Definition~\ref{def:functor}. Note that the preservation of identities required by Definition~\ref{def:functor} is already a diagrammatic tautology, as it is absorbed by the diagrammatic notation.

We do not cover these here in detail, though we refer to their use in representing diagrams for closed monoidal categories, in Section~\ref{sec:closed}.
\end{remark}
Recall that an isomorphism is a morphism $f\from x\to y$ which has an inverse: a morphism $g\from y\to x$ such that $f\poi g = \mathsf{id}_x$ and $g\poi f = \mathsf{id}_y$. We then say that $x$ is isomorphic to $y$. The notion of isomorphism makes sense for functors and categories too. The inverse of a functor $F\from \mathsf{C}\to \mathsf{D}$ is a functor $G\from\mathsf{D}\to\mathsf{C}$ such that $ F\poi G = \mathsf{id}_{\mathsf{C}}$ and  $G\poi F = \mathsf{id}_{\mathsf{D}}$. However, it is sometimes too restrictive to ask for isomorphisms between categories. Indeed, there are categories that we would like to identify which are not isomorphic. The more general notion of equivalence of categories is often more appropriate.
\begin{definition}\label{def:equivalence}
A functor $F\from\mathsf{C}\to\mathsf{D}$  is an \emph{equivalence} of categories if there exists a functor $G\from \mathsf{D}\to\mathsf{C}$ and two natural transformations $\epsilon\from FG\Rightarrow \mathsf{id}_{\mathsf{C}}$ and $\eta \from \mathsf{id}_{\mathsf{D}} \Rightarrow GF$ whose components are isomorphisms.
\end{definition}
\begin{remark}\label{rmk:monoidal-equivalence}
In this text, we make use of \emph{monoidal equivalences} of monoidal categories. The definition is the obvious modification of Definition~\ref{def:equivalence}: it is a monoidal functor $F\from\mathsf{C}\to\mathsf{D}$ (\emph{cf.} Definition~\ref{def:symon-functor}) for which there exists a monoidal functor $G$ satisfying the conditions of the definition above. The same goes for \emph{symmetric} monoidal categories.
\end{remark}
We can even weaken further the notion of equivalence to that of \emph{adjunction}. Instead of requiring  equalities $ F\poi G = \mathsf{id}_{\mathsf{C}}$ and  $G\poi F = \mathsf{id}_{\mathsf{D}}$, we can consider functors for which we have natural transformations $G\poi F\Rightarrow \mathsf{id}_{\mathsf{C}}$ and $\mathsf{id}_{\mathsf{D}} \Rightarrow F\poi G$ that satisfy some conditions that we now recall.
\begin{definition}\label{def:adjunction}
An \emph{adjunction} between two categories consists of a pair of functors $F\from \mathsf{C}\to\mathsf{D}$ and $G\from \mathsf{D}\to\mathsf{C}$ and two natural transformations $\epsilon\from FG\Rightarrow \mathsf{id}_{\mathsf{C}}$ (the \emph{unit}) and $\eta \from \mathsf{id}_{\mathsf{D}} \Rightarrow GF$ (the \emph{counit}) such that
\[F\eta_x\poi \epsilon_{Fx}= \mathsf{id}_{Fx} \quad \text{ and } \quad \eta_{Ga}\poi  G\epsilon_{a}= \mathsf{id}_{Ga}\]
We call $F$ the \emph{left adjoint} and $G$ the \emph{right adjoint}.
\end{definition}
\end{document}